\documentstyle[axodraw,graphicx,epsf,amssymb]{aipproc2}
\newcommand{\newc}{\newcommand}
\newc{\nn}{\noindent}
\newc{\ibid}{{\it ibid}.}
\newc{\matth}{\mathsurround=0pt}
\newc{\ppbar}{\mbox{$p\overline{p}$}}
\newc{\bbbar}{\mbox{$b\overline{b}$}}
\newc{\ccbar}{\mbox{$c\overline{c}$}}
\newc{\ttbar}{\mbox{$t\overline{t}$}}
\newc{\ee}{\mbox{$ee$}}
\newc{\ti}{\times}
\newc{\er}[1]{\times \frac{m_{\tilde{#1}_R}}{100 \ {\rm GeV}}}
\newc{\rnd}{(\frac{m_{\tilde{q}}}{\tilde{\Lambda} \ {\rm GeV}})^{5/2}}
\newc{\erl}[1]{\times \frac{m_{\tilde{#1}_L}}{100 \ {\rm GeV}}}
\newc{\ero}{\sqrt{m_{\tilde{\tau}}/100 \ {\rm GeV}}}
\newc{\era}[1]{\sqrt{m_{\tilde{#1}}/100 \ {\rm GeV}}}
\newc{\zzero}{\mbox{$Z^0$}}
\newc{\pelp}{\mbox{$e^+$}}
\newc{\pelm}{\mbox{$e^-$}}
\newc{\pelpm}{\mbox{$e^{\pm}$}}
\newc{\epem}{\mbox{$e^+e^-$}}
\newc{\lplm}{\mbox{$\ell^+\ell^-$}}
\newc{\qbar}     {\mbox{$\overline{q}$}}
\newc{\gluino}   {\mbox{$\tilde{g}$}}
\newc{\squark}   {\mbox{$\tilde{q}$}}
\newc{\sqbar}    {\mbox{$\bar{\tilde{q}}$}}
\newc{\mgluino}  {\mbox{$M(\gluino)$}}
\newc{\msquark}  {\mbox{$M(\squark)$}}
\newc{\csquarkl} {\mbox{$\tilde{c}_L$}}
\newc{\mcsl}     {\mbox{$M(\csquarkl)$}}
\newc{\ssb}      {\mbox{$\squark\overline{\squark}$}}
\newc{\csquark}  {\mbox{$\tilde{c}$}}
\newcommand{\tsquark}  {\mbox{$\tilde{t}$}}
\newcommand{\stopo}    {\mbox{$\tilde{t}_1$}}

\newc{\ttbone}   {\mbox{$\tsquark_1\overline{\tsquark}_1$}}
\newc{\chione}   {\mbox{$\tilde{\chi}_{1}^{\pm}$}}
\newc{\mchione}  {\mbox{$M(\tilde{\chi}_{1}^{\pm})$}}
\newc{\mstopo}   {\mbox{$M(\tilde{t}_1)$}}
\newc{\mcone}    {\mbox{$M(\tilde{\chi}_{1}^{\pm})$}}
\newc{\none}     {\mbox{$\tilde{\chi}_{1}^0$}}
\newc{\mchio}    {\mbox{$M(\none)$}}
\newc{\lsp}      {\mbox{$\tilde{\chi}_{1}^0$}}
\newc{\mz}       {\mbox{$M_0$}}
\newc{\mo}       {\mbox{$M_{1/2}$}}
\newc{\lamp}     {\mbox{$\lambda_{121}'$}}
\newc{\tev}  {\mbox{$\;{\rm TeV}$}}
\newc{\gevc} {\mbox{$\;{\rm GeV}/c$}}
\newc{\gevcc}{\mbox{$\;{\rm GeV}/c^2$}}
\newc{\chis}{\mbox{$\chi^{2}$}}
\newc{\ifb}{\mbox{${\rm fb}^{-1}$}}
\newc{\ipb}{\mbox{${\rm pb}^{-1}$}}
\newc{\met}{\mbox{${E\!\!\!\!/_T}$}}
\newc{\intlum}{\mbox{${ \int {\cal L} \; dt}$}}
\newc{\pt}{\mbox{$p_T$}}
\newc{\et}{\mbox{$E_T$}}
\newc{\modulus}[1]{\left| #1 \right|}
\newc{\bp}{\mbox{$b'$}}
\newc{\lxy}{\mbox{$L_{xy}$}}
\newc{\dedx}{\mbox{${\rm d}E/{\rm d}x$}}
\newc{\R}{$R$}
\newc{\charginom}{M_{\tilde \chi}^{+}}
\newc{\mue}{\mu_{\tilde{e}_{iL}}}
\newc{\mud}{\mu_{\tilde{d}_{jL}}}
\newc{\barr}{\begin{eqnarray}}
\newc{\earr}{\end{eqnarray}}
\newc{\beq}{\begin{equation}}
\newc{\eeq}{\end{equation}}
\newc{\ra}{\rightarrow}
\newc{\lra}{\longrightarrow}
\newc{\lam}{\lambda}
\newc{\eps}{\epsilon}
\newc{\gev}{\,GeV}
\newc{\eq}[1]{(\ref{eq:#1})}
\newc{\eqs}[2]{(\ref{eq:#1},\ref{eq:#2})}
\newc{\etal}{{\it et al.}\ }
\newc{\eg}{{\it e.g.}\ }
\newc{\ie}{{\it i.e.}\ }
\newc{\nonum}{\nonumber}
\newc{\lab}[1]{\label{eq:#1}}
\newc{\dpr}[2]{({#1}\cdot{#2})}
\newc{\lsim}{\stackrel{<}{\sim}}
\newc{\gsim}{\stackrel{>}{\sim}}
\newc{\ol}{\overline}
\newc{\half}{\frac{1}{2}}
\newc{\rpv}{{\not\!\! R_p}}
\newc{\rpvs}{{\not R_p}}
\newc{\rp}{{${R_p}$}}
\newc{\kap}{\kappa}
\def\npb#1 #2 #3 #4 {Nucl.~Phys. B {\bf #1}, #2 (#3)#4 }
\def\plb#1 #2 #3 #4 {Phys.~Lett. B {\bf #1}, #2 (#3)#4 }
\def\prd#1 #2 #3 #4 {Phys.~Rev.  D {\bf #1}, #2 (#3)#4 }
\def\prl#1 #2 #3 #4 {Phys.~Rev.~Lett. {\bf #1}, #2 (#3)#4 }
\newc{\IJMP}[3]{Int. J. Mod. Phys. A {\bf#1},\ #2 (#3)}
\newc{\MPL}[3]{Mod. Phys. Lett. A {\bf#1},\ #2 (#3)}
\newc{\NPB}[3]{Nucl. Phys. {\bf#1},\ #2 (#3)}
\newc{\PLBold}[3]{Phys. Lett. {\bf#1},\ #2 (#3)}
\newc{\PLB}[3]{Phys. Lett. {\bf#1},\ #2 (#3)}
\newc{\PR}[3]{Phys. Rep. {\bf#1},\ #2 (#3)}
\newc{\PRD}[3]{Phys. Rev. {\bf#1}\, #2 (#3)}
\newc{\PRL}[3]{Phys. Rev. Lett. {\bf#1},\ #2 (#3)}
\newc{\PTT}[3]{Prog. Theor. Phys. {\bf#1},\ #2 (#3)}
\newc{\RMP}[3]{Rev. Mod. Phys. {\bf#1},\ #2 (#3)}
\newc{\ZPC}[3]{Z. Phys. C {\bf#1},\ #2 (#3)}
\newc{\Journal}[4]{{#1} {\bf {#2}}, {#3} ({#4})}
\newc{\PRDR}[3]{{Phys. Rev. D} {\bf #1}, Rapid Communications, #2 (#3)}
\newc{\NIMA}{{Nucl. Instrum. Methods} A}
\newc{\ptmiss}{/ \hskip-7pt p_T}
\newc{\mgut}{M_U}
\newc{\wt}{\widetilde}
\newc{\gl}{\wt g}
\newc{\mgl}{m_{\gl}}
\newc{\cnone}{\wt\chi^0_1}
\newc{\cpmone}{\wt \chi^{\pm}_1}
\newc{\mcpone}{m_{\cpone}}
\newc{\mcpmone}{m_{\cpmone}}
\newc{\mcnone}{m_{\cnone}}
\newc{\bit}{\begin{itemize}}    
\newc{\eit}{\end{itemize}}
\newc{\delgs}{\delta_{GS}}
\newc{\mth}{m_{3/2}}
\newc{\bea}{\begin{eqnarray}}   \newc{\eea}{\end{eqnarray}}
\newc{\baa}{\begin{array}}      \newc{\eaa}{\end{array}}
\newc{\thetaw}{\theta_W}
\newc{\call}{{\cal L}}
\newc{\mplanck}{M_{\rm P}}
\newc{\mpl}{\mplanck}
\newc{\dmchi}{\Delta m_{\tilde\chi}}
\newc{\cpone}{\wt \chi^+_1}
\newc{\cmone}{\wt \chi^-_1}
\newc{\gam}{\gamma}
\newc{\cntwo}{\wt\chi^0_2}
\newc{\lampp}{\lam^{\prime\prime}}
\newc{\wtil}{\widetilde}
\newc{\glsp}{$\wtil g$-LSP}
\newc{\mev}{~{\rm MeV}}
\newc{\msq}{m_{\squark}}
\newc{\anti}{\overline}
\newc{\mtil}{\widetilde m}
\newc{\mt}{m_t}
\newc{\mw}{m_W}
\newc{\wbp}{W^+}
\newc{\rts}{\sqrt{s}}
\newc{\wm}{W^-}
\newc{\fbi}{~{\rm fb}^{-1}}
\newc{\chitil}{\wt\chi}
\newc{\cmtwo}{\wt \chi^-_2}
\newc{\cnthree}{\wt\chi^0_3}
\newc{\cnfour}{\wt\chi^0_4}
\newc{\pbi}{~{\rm pb}^{-1}}
\newc{\pb}{~{\rm pb}}
\newc{\etmiss}{/ \hskip-7pt E_T}
\newc{\tanb}{\tan\beta}
\newc{\mhalf}{m_{1/2}}
\newc{\mcntwo}{m_{\cntwo}}
\newc{\mcpmtwo}{m_{\cpmtwo}}
\newc{\mcnthree}{m_{\cnthree}}
\newc{\mcnfour}{m_{\cnfour}}
\newc{\cpmtwo}{\wt \chi^{\pm}_2}
\newc{\vev}[1]{{\left\langle #1\right\rangle}}

\begin{document}
\title{Searching for R-Parity Violation at Run-II of the Tevatron}

\author{B. Allanach$^1$, H.~Baer$^2$, S.~Banerjee$^3$,
E.L. Berger$^4$, M. Chertok$^5$, F. de Campos$^6$, K. Cheung$^7$, 
A. Dedes$^8$, M.A.~Diaz$^2$,
H.~Dreiner$^8$, 
O.J.P. Eboli$^{9,10}$, J. Gunion$^7$, B.W. Harris$^4$,
J. Hewett$^{11}$, M.B. Magro$^{10}$, N.K.~Mondal$^3$,
V.S.~Narasimham$^3$, L. Navarro$^{12}$, N.~Parua$^3$, W.~Porod$^{12,13}$, 
D.A. Restrepo$^{12}$, P.~Richardson$^{14}$, T.~Rizzo$^{11}$, 
M.H. Seymour$^8$, Z. Sullivan$^4$, J.W.F. Valle$^{12}$\\\vspace{0.4cm}
Edited by: H. Dreiner$^6$}

\address{
$^1$ DAMTP, University of Cambridge, Cambridge, CB3 9EW, UK\\
$^2$ Department of Physics, Florida State University, Tallahassee,
        FL 32306, USA\\
$^3$ Tata Inst. of Fundamental Research, Bombay, India \\
$^4$ Argonne National Laboratory, Argonne, IL 60439 \\
$^5$ Texas A\&M University, College Station, TX 77843,\\
$^6$ Depart. de F\'\i sica y Qu\'\i mica, Univ. Estadual Paulista, 
       Guaratinguet\'a, Brasil \\
$^7$ Department of Physics, University of California, Davis, 
CA 95616, USA\\
$^8$Rutherford Appleton Laboratory, Chilton, Didcot OX11 0QX, U.K.\\
$^9$Instituto de F\'\i sica Te\'orica, Univ. Estadual Paulista, Sao Paulo, 
Brasil\\
$^{10}$ University of Wisconsin, Madison, WI 53706, USA\\
$^{11}$ Stanford Linear Accelerator, Stanford, CA 94309\\
$^{12}$ Departamento de F\'\i sica Te\'orica, IFIC-CSIC, Univ.~de Valencia,
            Spain \\ 
$^{13}$ Insitut f\"ur Theoretische Physik, Universit\"at Wien, Austria\\
$^{14}$ Depart.  of Theoret. Physics, University of Oxford, 
Oxford OX1 3NP, U.K.\vspace{-0.5cm}}

\maketitle

\begin{abstract}
We present an outlook for possible discovery of supersymmetry with
broken R-parity at Run II of the Tevatron. We first present a review
of the literature and an update of the experimental bounds.
In turn we then discuss the following processes:\\
1. Resonant slepton production followed by $\rpv$ decay, (a) via $LQD^c$
and (b) via $LLE^c$. \\
2. How to distinguish resonant slepton production from $Z'$ or $W'$ 
production. \\
3. Resonant slepton production followed by the decay to neutralino
LSP, which decays via $LQD^c$.\\
4. Resonant stop production followed by the decay to a chargino,
which cascades to the neutralino LSP.\\
5. Gluino pair production followed by the cascade decay to charm
squarks which decay directly via $L_1Q_2D^c_1$.\\
6. Squark pair production followed by the cascade decay to the
neutralino LSP which decays via $L_1Q_2D^c_1$.\\
7. MSSM pair production followed by the cascade decay to the LSP which
decays (a) via $LLE^c$, (b) via $LQD^c$, and (c) via $U^cD^cD^c$, 
respectively.\\
8. Top quark and top squark decays in spontaneous $\rpv$. 
\end{abstract}

\section{Introduction}
{\centerline{\large{\it{H. Dreiner}}}}
\vspace{0.7cm}

The chiral superfields of the supersymmetric Standard Model are 
\barr
L_i\;(1,2,-\half),\quad {E}_i^c\;(1,1,1),\quad H_1\;(1,2,-\half),
\quad H_2\;(1,2,\half), \nonumber \\ Q_i\;(3,2,\frac{1}{6}),\quad {
U}_i^c\;(3,1,-\frac{2}{3}),\quad {D}_i^c\;(3,1,\frac{1}{3}).  
\earr 
In parentheses we have given the quantum numbers under the Standard
Model gauge group, $G_{SM}=SU(3)\times SU(2)\times U(1)$. $i=1,2,3$ is
a generation index. The most general superpotential for this field
content is\footnote{Note that $^c$ here refers to charge conjugation
and is not a $SU(2)$ index.} 
\barr 
W&=&W_{MSSM}+W_\rpvs, \\ 
W_\rpvs&=& \half
\lam_{ijk}\eps_{ab} L_i^aL_j^b{E}_k^c +\lam'_{ijk} \eps_{ab}
L_i^aQ_j^b{D}_k^c +\lam''_{ijk} \eps_{xyz} {U}_i^{cx} {
D}_j^{cy} {D}_k^{cz} +\kap_i \eps_{ab} L_i^a H_2^b.  \lab{superpot}
\earr 
Here, $W_{MSSM}$ is the superpotential of the MSSM containing the terms
which give mass to the SM fermions. We do not further discuss it
here. $i,j,k=1,2,3$ are generation indices, $a,b=1,2$ are $SU(2)$
isospin indices, and $x,y,z=1,2,3$ are $SU(3)$ colour indices. $\lam
,\lam',\lam''$ are dimensionless Yukawa couplings. $\kap_i$ are mass
terms mixing the leptonic and Higgs doublets. The terms in $W_\rpvs$
violate the discrete and multiplicative symmetetry R-parity
\beq 
R_p=(-1)^{3B+L+2S},
\eeq
where $B$ is baryon-, $L$ is lepton-number and $S$ denotes the spin.
It is the purpose of this chapter to investigate the phenomenological
consequences of the superpotential terms $W_\rpvs$ at the Tevatron.
For a recent review on $R_p$ violation ($\rpv$) see \cite{herbi}. A
very nice detailed overview of $\rpv$ collider phenomenology is 
given in \cite{france}.

At the level of the superpotential, the terms $L_iH_2$ can be rotated
away by a field redefinition \cite{hallsuz}. When including the
soft-breaking terms this is no longer true in general. However, we
shall split our investigations into explicit $R_p$ violation,
where we neglect all contributions from the $L_iH_2$
mixing, and spontaneous $\rpv$.

\section{Explicit R-parity Violation}
\subsection{Existing Bounds}
\label{sec:bounds}
{\centerline{\large{\it{B. Allanach, A. Dedes, H. Dreiner}}}}
\vspace{0.7cm}

In Table \ref{tab:bounds}, we show a recent compilation of indirect
bounds on the $\rpv$\ Yukawa couplings \cite{dedes,gdr,herbi,bhatt}.  We
present the bounds at the 2$\sigma$ level and give two significant
figures where possible. The bounds depend on the mass of super partner
and we give the functional dependence. For more detail on how the
bounds are obtained see \cite{dedes,gdr}. We ignore strict bounds from
cosmology \cite{davidson} because they are model dependent
\cite{rosscosmo}. A discussion of the bounds on $\lam''_{ijk}$ is
given in Sect. \ref{sec:baer}.

\begin{table*}
\caption{Latest $2\sigma$ limits on the magnitudes of weak scale 
trilinear R-parity violating couplings from indirect decays and
perturbativity. We have made use of the data in \protect\cite{pdg}.  The
dependence on the relevant superparticle mass is shown
explicitly. When the perturbativity bounds are more stringent than the
empirical bounds for masses $m_{\tilde{l},\tilde{q}}$=1 TeV, then we
display them in parentheses. Where a bound without parentheses has no
explicit mass dependence shown, the mass dependence was too
complicated to detail here and a degenerate sparticle spectrum of
100~GeV is assumed.}
\label{tab:bounds}
\begin{tabular}{cccc}
%\tableline
ijk     & $\lam_{ijk}(M_Z)$\tablenote{Updated bounds from 
Ref.\cite{gdr,herbi}. Bounds on $\lam_{121}$, $\lam_{122}$, $\lam_
{123}$ have been obtained from charged current universality \cite{han}. 
Bounds on $\lam_{131}$, $\lam_{132}$, $\lam_{231}$, $\lam_{232}$ and 
$\lam_{233}$ have been derived from~\cite{han} measurements of $R_{\tau}
=\Gamma(\tau\ra e\nu\bar{\nu})/\Gamma(\tau\ra \mu\nu
\bar{\nu})$ and $R_{\tau\mu}=\Gamma(\tau\ra\mu\nu\bar{\nu})/
\Gamma(\mu\ra e\nu\bar{\nu})$ \cite{pdg}. The bound on $\lam_{133}$ 
\cite{tata} has been obtained from the experimental limit on the 
electron neutrino mass \cite{pdg}.}
        & $\lam'_{ijk}(M_Z)$\tablenote{Bounds on $\lam'_{112}$, $\lam'_{113}$,
$\lam'_{121}$, $\lam'_{122}$, and $\lam'_{123}$ have been obtained
from charged current universality \cite{han}. The bound on $\lam'_{111}$ 
has been derived from neutrino-less double beta decay \cite{beta,babu}
where $\fm =(m_{\tilde{e}}/100\ {\rm GeV})^2\ti (m_{\tilde{\chi}^0}/
100\ {\rm GeV})^{1/2}$, and on $\lam'_{131}$ from atomic parity violation 
\cite{han,apv}. This latter bound is at the 3$\sigma$ level, 
since the data disagree with the standard model at the 2.5$\sigma$ level 
\cite{apv}. The bound on $\lam'_{132}$ comes from the forward-backward 
asymmetry in $e^+e^-$ collisions~\cite{han}. Bounds on $\lam'_{133}$,
$\lam'_{233}$ have been obtained from bounds on the neutrino masses 
\cite{tata} and on $\lam'_{211}$, $\lam'_{212}$, $\lam'_{213}$ from $R_\pi=
\Gamma(\pi\ra e\nu)/\Gamma(\pi\ra \mu\nu)$ \cite{han,gdr}. Bounds on 
$\lam'_{221}$, $\lam'_{231}$ come from $\nu_\mu$ deep inelastic scattering 
\cite{han,gdr} and on $\lam'_{222}$, $\lam'_{223}$ from the D-meson decays 
\cite{gdr}, $D\ra K l \nu$. The bounds without parentheses on $\lam'_{232}$, 
$\lam'_{331}$, $\lam'_{332}$, $\lam'_{333}$ have been derived from $R_l =
\Gamma(Z\ra {\rm had})/\Gamma(Z\ra l\bar{l})$ for $m_{\tilde{q}}=100$ 
GeV~\cite{ellis} and on $\lam'_{311}$, $\lam'_{312}$, $\lam'_{313}$ from 
$R_{\tau\pi}=\Gamma(\tau\ra \pi\nu_\tau)/\Gamma(\pi\ra \mu\nu_\mu)$ 
\cite{han,gdr}. The bounds on the couplings $\lam'_{321}$, 
$\lam'_{322}$ and $\lam'_{323}$ have been derived from $D_s$ decays 
\cite{gdr}, {\it i.e.,} $R_{D_s}=\Gamma(D_s\ra \tau \nu_\tau)/\Gamma
(D_s \ra \mu\nu_\mu)$. There are also bounds on $\lam'_{3j3}$ from $R_b$
\cite{yang} but these are weak at 2$\sigma$ level and thus not displayed.}
        & $\lam''_{ijk}(M_Z)$\tablenote{The indirect bounds on $\lam''_{ijk}$
existing in the literature are on $\lam''_{112}$ from double nucleon decay
\cite{goity} ($\tilde{\Lambda}$ is a hadronic scale and it can be varied
from 0.003 to 1 GeV and $\rnd$ from $2 \times 10^{11}$ to $10^{5}$ for 
$m_{\tilde{q}}=100$ GeV)   
and on $\lam''_{113}$ from neutron oscillations~\cite{zwirner,goity}
 for $m_{\tilde{q}}$=100 GeV. For $m_{\tilde{q}}$=200 (600)
GeV the bound on $\lam''_{113}$ is 0.002 (0.1). Bounds on $\lam''_{3jk}$
have been derived from
$R_l =\Gamma(Z\ra {\rm had})/\Gamma(Z\ra l\bar{l})$ at 
$1\sigma$ for $\tilde{m}=100$ GeV \cite{bat3} and, for heavy squark masses, is
not more stringent than 
the perturbativity bound, which is displayed in the parenthesis.}
\\ \tableline 
%%%%%%%%%%%%%%%%%%%%%%%%%%%%%%%%%%%%%%%%%%%%%%%%%%%%%%%%%%%%%%%%%%%%%%
111     &    -            &$5.2\ti10^{-4}\ti\fm$&   -  \\
112     &    -            &$0.021 \er{s}$&                $10^{-15}\ti \rnd$ \\
113     &    -            &$0.021 \er{b}$&                  $10^{-4}$ \\
121     &$0.049 \er{e}$   &$0.043 \er{d}$&                $10^{-15}\ti \rnd$ \\
122     &$0.049 \er{\mu}$ &$0.043 \er{s}$&                 -\\
123     &$0.049 \er{\tau}$&$0.043 \er{b}$&        $(1.23)$\\
131     &$0.062 \er{e}$   &$0.019 \erl{t}$&        $10^{-4}$\\
132     &$0.062 \er{\mu}$ &$0.28 \erl{t}$ (1.04)&         $(1.23)$\\
133     &$0.0060\ero$     &$1.4\ti10^{-3}\era{b}$&  -\\
211     &$0.049 \er{e}$   &$0.059 \er{d}$&                 -\\
212     &$0.049 \er{\mu}$ &$0.059 \er{s}$&                 $(1.23)$\\
213     &$0.049 \er{\tau} $&$0.059 \er{b}$&       $(1.23)$\\
221     &   -             &$0.18\er{s}$ (1.12)&             $(1.23)$\\
222     &   -             &$0.21\er{s}$ (1.12)&             -\\
223     &   -             &$0.21\er{b}$ (1.12)&             $(1.23)$\\
231     &$0.070 \er{e}$   &$0.18\erl{b}$ (1.12)&            $(1.23)$\\
232     &$0.070 \er{\mu}$ &$0.56$ (1.04)&                  $(1.23)$\\
233     &$0.070 \er{\tau}$&$0.15\era{b}$&         -\\
311     &$0.062 \er{e}$   &$0.11\er{d}$ (1.12)&             -\\
312     &$0.062 \er{\mu}$ &$0.11\er{s}$ (1.12)&            $0.50$ (1.00)\\
313     &$0.0060\ero$     &$0.11\er{b}$ (1.12)&             $0.50$ (1.00)\\
321     &$0.070 \er{e}$   &$0.52\er{d}$ (1.12)&             $0.50$ (1.00)\\
322     &$0.070 \er{\mu}$ &$0.52\er{s}$ (1.12)&            -\\
323     &$0.070 \er{\tau}$&$0.52\er{b}$ (1.12)&           $0.50$ (1.00)\\
331     &  -              &$0.45$ (1.04)&           $0.50$ (1.00)\\
332     &  -              &$0.45$ (1.04)&           $0.50$ (1.00)\\
333     &  -              &$0.45$ (1.04)&          -      
\end{tabular}
\end{table*}
%%%%%%%%%%%%%%%%%%%%%%%%%%%%%%%%%%%%%%%%%%%%%%%%%%%%%%%%%%

\subsection{Outline of Signals and Summary of the Literature}
The two main differences between the standard MSSM phenomenology and
the phenomenology of explicit $\rpv$ are
\begin{enumerate}
\item Single production of supersymmetric particles is possible. For
example at a hadron collider we can have resonant slepton (charged and
neutral) production via the operators $L_iQ_j{\bar D}_k$ and resonant
squark production via the operators ${\bar U}_i{\bar D}_j{\bar D}_k$.
The lowest order Feynman diagrams are shown in Figs. \ref{fig:cross}a.
This lowers the kinematic threshold for the discovery of
supersymmetry.
\item The lightest supersymmetric particle (LSP) is not stable, and
can possibly decay in the detector. If the LSP is the lightest
neutralino it can decay, for example, via the operator $L_iQ_j{\bar
D}_k$, as shown in Fig. \ref{fig:decay}.  However, since the LSP is
not stable, the cosmological argument requiring it to be electrically
and colour neutral no longer applies \cite{cosmo}. To date there have 
been no systematic studies of non-neutralino LSPs in the context of 
$\rpv$. If the LSP is charged and has an appreciable lifetime (or is 
stable), it can be detected in a search for charged massive particles 
(CHAMPs).
\end{enumerate}
\begin{figure}[htbp]
\begin{center} 
\begin{picture}(360,60)(0,30)
\SetScale{0.7}
\ArrowLine(5,76)(60,102)
\ArrowLine(60,102)(5,128)
\DashArrowLine(90,102)(60,102){5}
\ArrowLine(90,102)(145,76)
\ArrowLine(145,128)(90,102)
\Text(80,55)[]{$\mathrm{\tilde{\chi}^{0}}$}
\Text(80,90)[]{$\mathrm{\ell^{+}}$}
\Text(25,90)[]{$\mathrm{\bar{d}}$}
\Text(25,57)[]{$\mathrm{u}$}
\Text(53,78)[]{$\mathrm{\tilde{\ell}_{L}}$}
\Vertex(60,102){1}
\Vertex(90,102){1}
\SetScale{0.7}
\ArrowLine(240,128)(185,128)
\ArrowLine(240,128)(295,128)
\ArrowLine(240,76)(185,76)
\ArrowLine(295,76)(240,76)
\DashArrowLine(240,76)(240,128){5}
\Text(190,98)[]{$\mathrm{\tilde{\chi}^{0}}$}
\Text(190,47)[]{$\mathrm{\ell^{+}}$}
\Text(153,47)[]{$\mathrm{\bar{d}}$}
\Text(153,95)[]{$\mathrm{u}$}
\Text(160,70)[]{$\mathrm{\tilde{d}_{R}}$}
\Vertex(240,128){1}
\Vertex(240,76){1}
\ArrowLine(420,128)(365,128)
\ArrowLine(475,128)(420,128)
\ArrowLine(365,76)(420,76)
\ArrowLine(420,76)(475,76)
\DashArrowLine(420,76)(420,128){5}
\Text(315,47)[]{$\mathrm{\tilde{\chi}^{0}}$}
\Text(315,98)[]{$\mathrm{\ell^{+}}$}
\Text(277,98)[]{$\mathrm{\bar{d}}$}
\Text(277,47)[]{$\mathrm{u}$}
\Text(287,70)[]{$\mathrm{\tilde{u}_{L}}$}
\Vertex(420,128){1}
\Vertex(420,76){1}
\end{picture}
\end{center}
\caption{Production of $\tilde{\chi}^{0}\ell^{+}$.\label{fig:cross}}
\begin{center} 
\begin{picture}(360,80)(0,0)
\SetScale{0.7}
\ArrowLine(5,78)(60,78)
\ArrowLine(105,105)(60,78)
\ArrowLine(84,53)(129,26)
\ArrowLine(129,80)(84,53)
\DashArrowLine(60,78)(84,53){5}
\Text(25,63)[]{$\mathrm{\tilde{\chi^{0}}}$}
\Text(55,70)[]{$\mathrm{\ell^{+}}$}
\Text(75,20)[]{$\mathrm{d}$}
\Text(75,54)[]{$\mathrm{\bar{u}}$}
\Text(45,40)[]{$\mathrm{\tilde{\ell}_{L}}$}
\Vertex(60,78){1}
\Vertex(84,53){1}
\ArrowLine(185,78)(240,78)
\ArrowLine(285,105)(240,78)
\ArrowLine(264,53)(309,26)
\ArrowLine(309,80)(264,53)
\DashArrowLine(240,78)(264,53){5}
\Text(150,63)[]{$\mathrm{\tilde{\chi}^{0}}$}
\Text(200,54)[]{$\mathrm{\ell^{+}}$}
\Text(200,20)[]{$\mathrm{d}$}
\Text(180,70)[]{$\mathrm{\bar{u}}$}
\Text(170,40)[]{$\mathrm{\tilde{u}_{L}}$}
\Vertex(240,78){1}
\Vertex(264,53){1}
\ArrowLine(365,78)(420,78)
\ArrowLine(420,78)(465,105)
\ArrowLine(489,26)(444,53)
\ArrowLine(489,80)(444,53)
\DashArrowLine(444,53)(420,78){5}
\Text(277,63)[]{$\mathrm{\tilde{\chi}^{0}}$}
\Text(330,20)[]{$\mathrm{\ell^{+}}$}
\Text(310,70)[]{$\mathrm{d}$}
\Text(330,54)[]{$\mathrm{\bar{u}}$}
\Text(300,40)[]{$\mathrm{\tilde{d}_{R}}$}
\Vertex(420,78){1}
\Vertex(444,53){1}
\end{picture}
\end{center}
\caption{LQD decay of $\mathrm{{\tilde\chi}^0}$. The neutralino is a Majorana
fermion and decays to the charge conjugate final state as well.
\label{fig:decay}}
\end{figure}
A first systematic study of R-parity violating signals at hadron collider
was presented in \cite{esmail,rosshadro}. There are in principle two kinds 
of signals given by the two kinds of supersymmetric production.
\begin{enumerate}
\item Single Sparticle Production, \eg $u+{\bar d}\ra\ell^+$.
\item Pair production of Sparticles via MSSM couplings, \eg $u+{\bar
u}\ra{\tilde q}+{\tilde {\bar q}}$.
\end{enumerate}

In both cases the production is followed either by direct R-parity
violating decays of the supersymmetric particles or by cascade decays
to the LSP, which in turn decays via $\rpv$. This latter decay can be
prompt or delayed. If it is delayed it can lead to a detached vertex
or decay outside the detector. To date only prompt decays have been
considered.  Within this report we also discuss delayed decays (see
\ref{sec:baer} and \cite{landsberg}) leading to detached
vertices. The LSP decays outside the detector for small R-parity
violating couplings, $\lam,\lam',\lam''\lsim10^{-5}$,
\cite{rosshadro,dawson}.  The resonant production is then also
suppressed and we thus retrieve the MSSM phenomenology, where
sparticles are pair-produced and the LSP escapes detection. This is
discussed in the contribution by the SUGRA working group.

In the following we summarize the possible signatures and indicate
what work has already been performed. There is at present only one
published experimental search for $\rpv$ at hadron colliders
\cite{cdfrpv}. A summary as well as the projection of this analysis for
Run II is presented in Sect. \ref{sec:chertok}. There is a
preliminary D0 analysis publically available \cite{d0rpv}. We present the
extension of this analysis to Run II in Sect.\ref{sec:banerjee}.
There {\it are} published results on searches for pair production of
leptoquarks, which can be interpreted as searches for $\rpv$ (see
Sect.\ref{sec:spartpair}). In the following, most references are
thus to theoretical work on specific signals.

\subsubsection{Single Sparticle Production}
At a hadron collider sneutrinos and charged leptons can be produced on
resonance via the operator $L_iQ_j{D}_k^c$. They can then decay
again via the same operator, another $LQ{D}^c$ operator, the
operator $L_iL_j{E}^c$ or via the MSSM gaugino interactions. Thus
we obtain the possible reactions
\barr
{\bar d}_j d_k&\ra{\tilde\nu}_i&\ra 
\left\{ \begin{array}{ccc}
\ell^+_m\ell^{'-}_n,& L_iL_m{\bar E}_n, & (a)\\
d_k{\bar d}_j,  & L_iQ_j{\bar D}_k, & (b)\\
\nu_i\chi^0_m,    & {\rm MSSM},       & (c)\\
\ell^+_i\chi^-_m, & {\rm MSSM},       & (d)
\end{array}
\right.,\quad
u_j{\bar d}_k\;\ra{\tilde\ell}_i^+\;\ra 
\left\{ \begin{array}{ccc}
\ell^+_k\nu_j,& L_iL_j{\bar E}_k, & (e)\\
u_j{\bar d}_k,  & L_iQ_j{\bar D}_k, & (f)\\
\ell^+_i\chi^0_m,    & {\rm MSSM},       & (g)\\
\nu_i\chi^+_m, & {\rm MSSM}.     & (h)
\end{array}
\right.  
\lab{sleptonprod} 
\earr 
Here we have indicated the decay products of the slepton, as well as
the coupling which leads to this decay. The decays to gauginos proceed
via MSSM gauge couplings.  Resonant slepton production was first
considered in \cite{esmail}, where the production cross sections for
the Tevatron were determined and the decays \eq{sleptonprod}a,b,e,f
were compared to the background. The specific case of scalar tau
production in \eq{sleptonprod}e was considered in \cite{spies} and
compared to more recent Tevatron data on Drell-Yan di-lepton
production. In Sect. \ref{sec:hewett} we investigate the reactions
\eq{sleptonprod}a,b,e,f in detail and compare them to exisiting
Tevatron data. Furthermore the reactions \eq{sleptonprod}a,e can look
identical in their final state to new gauge boson, $W',Z'$, production
and we also investigate how this can be disentangled in Sect.
\ref{sec:hewett}. For both cases we use present data to project the
search reach for Run II. In Sect. \ref{sec:richardson}, we
study for the first time the case of like-sign di-lepton production
via the reactions \eq{sleptonprod}g.  The focus is on a single
dominant $\rpv$ operator, so that the neutralino decays as $\chi^0
\ra\ell^\pm+2\,{\rm jets}$, as shown in Fig.\ref{fig:decay}. The 
analysis for the reaction \eq{sleptonprod}d is completely analogous.

The reactions for resonant squark production via the operators ${
U}^c{D}^c{D}^c$ are
\barr
qq' &\ra{\tilde q}''&\ra 
\left\{ \begin{array}{ccc}
(\ell,\nu)_i\,q,  & L_iQ{D}^c, & (a)\\
qq',  & {U}^c{D}^c{D}^c, & (b)\\
q\,(\chi^0_m,{\tilde g}),    & {\rm MSSM},       & (c)\\
q\,\chi^+_m, & {\rm MSSM}.       & (d)
\end{array}
\right.
\lab{squarkprod}
\earr
The two-jet production via \eq{squarkprod}b was considered in
\cite{esmail}. In special cases the reactions \eq{sleptonprod}f and
\eq{squarkprod}b can lead to single top quark production.  This has
been studied in \cite{top}. In Sect.\ref{sec:berger} we study
the reaction \eq{squarkprod}d in detail, where the produced squark is
a stop.  On resonance, the reaction \eq{squarkprod}c is kinematically
blocked because of the final state top quark. The chargino of reaction
\eq{squarkprod}d cascade decays to the lightest neutralino in a
semi-leptonic decay and the charged lepton is used for detection. The
dominant decay of the neutrlino is also suppressed because of the top
quark in the final state and thus decays outside the detector.

\subsubsection{Sparticle Pair Production}
\label{sec:spartpair}
For $\rpv$ couplings $\lam,\lam',\lam''\lsim10^{-3}$ resonance
production is no longer viable. We then expect pair production of
supersymmetric particles to dominate. The main reactions are
\beq
p{\bar p}\ra\{{\tilde g}{\tilde g},\;{\tilde g}{\tilde q},\;{\tilde
q}{\tilde q},\; {\tilde \chi}^0_i{\tilde \chi}^0_j,\; {\tilde \chi}^+_i
{\tilde \chi}_j^-,\;{\tilde \chi}^0_i{\tilde\chi}^+_j,\;{\tilde\ell}^+
{\tilde \ell}^-,\;{\tilde\nu}{\tilde \nu},\;{\tilde\ell}^+{\tilde \nu}\}
\lab{prods}
\eeq
The supersymmetric particles can undergo cascade decays to the LSP
as in the MSSM. The LSP, which we here assume to be the lightest
neutralino then decays via $\rpv$
\barr
{\tilde\chi}^0_1\ra\left\{
\begin{array}{llc}
\ell^\pm_i\ell_k^\mp \nu_j,&L_iL_j{E}_k^c,&(a)\\
\ell^\pm_i+2\,{\rm jets},&L_iQ_j{D}_k^c,&(b)\\ 
\nu+2\,{\rm jets},&L_iQ_j{D}_k^c,&(c)\\
3\,{\rm jets},& {U}_i^c{D}_j^c{D}_k^c.&(d)
\end{array}
\right.
\lab{lspdecays}
\earr
The resulting reactions were first considered in \cite{esmail} and
classified in \cite{rosshadro}. Gluino and squark pair production
were first analysed in detail and compared to CDF data in \cite{dp}.
In each case, the MSSM cascade decays were explicitly neglected, giving
branching ratio 1 for the decay to the neutralino LSP. In \cite{barger}
the production of the electroweak gauginos was studied focusing on
the reactions
\barr
p{\bar p} &\ra& \left\{ 
\begin{array}{ccl}
  \chi^\pm_1\chi^0_2 &\ra &  \left\{ 
     \begin{array}{lc}
     (\chi^0_1\ell^\pm\nu) (\chi^0_1\ell{'^+}\ell^{'-}),&(a)\\
     (\chi^0_1\ell^+\nu) (\chi^0_1\ell{'^-}\nu),&(b)
     \end{array}\right.\\
  \chi_1^\pm\chi_1^0 &\ra &\;\;\;\; (\chi^0_1\ell^\pm\nu)\; \chi^0_1,
\hspace{1.5cm}(c) \\
  \chi_1^+\chi_1^- &\ra & \;\;\;\;(\chi^0_1\ell^+\nu) 
  (\chi^0_1\ell{'^-}\nu),\hspace{0.7cm}(d) \\
\end{array}
\right.
\earr
followed by the decay of the LSP via the operator $L_iL_j{
E}_k^c$ as in \eq{lspdecays}. In this case the discovery reach is
significantly above that of the MSSM.

In Ref.\cite{baer} {\it all} the production mechanisms given in
Eq.\eq{prods} were studied simultaneously in the framework of minimal
SUGRA to see what combined contribution there would be to several
possible discovery signals: (i) quartic-leptons, (ii) tri-leptons,
(iii) opposite sign di-leptons, (iv) same-sign di-leptons, and (v)
missing transverse energy. Each signature was compared for the two
scenarios of an LSP decaying via $LL{E}^c$ and ${U}^c{D}^c{D}^c$,
respectively. In Sect.\ref{sec:chertok} we study the case of pair
production followed by cascade decay to the LSP including a detector
simulation. The LSP decays via the operator $LLE^c$. The results are
compared to the present Tevatron data and projected to a discovery
reach for Run II. In Sect.\ref{sec:banerjee} we investigate in
detail the potential at Run II for the di-lepton signature in the case
of a dominant $LQD^c$ operator. In \cite{dp2} the previous analysis of
gluino and squark pair production of \cite{dp} was extended to include
cascade decays and investigate the search reach in the gluino and
squark mass at the Tevatron. In Sect. \ref{sec:baer}, we
investigate the possibility of the LSP decaying via ${U}^c{D}^c{D}^c$
to three jets, where the chargino and neutralino mass are nearly
degenerate, thus making the cascade leptons of \cite{baer} too soft
for detection. This is the worst case scenario for supersymmetry
searches.

In February, 1997 both HERA experiments announced an anomaly in their
high $Q^2$ deep inelastic scattering data \cite{hera}. This has not
been confirmed in the later data but it has also not been excluded
yet.  The initial anomaly could naturally be explained in terms of
resonant production of squarks via $\rpv$ \cite{herarpv} or as
resonant production of leptoquarks \cite{leptos}.  It lead to an
increased interest in the searches for squarks and leptoquarks at the
Tevatron. In leptoquark pair production at the Tevatron, the
leptoquarks directly decay to a lepton and a quark.  The experimental
signature is equivalent to pair produced squarks decaying directly via
the $LQ{D}^c$ operators. This is thus a distinct signature from the
cascade decay to the LSP. We can interpret the leptoquark searches
\cite{D0-lq,CDF-lq} in terms of $\rpv$ with large $\lam'_{ijk}$ Yukawa
couplings. We would expect additional decays for the squarks to
gauginos. So the leptoquark mass bound corresponds to an excluded
region in the mass branching ratio, $BR({\tilde q}\ra\ell+q')$, plane.
The present Tevatron data \cite{D0-lq,CDF-lq,combined} excludes the
solution of the HERA anomaly where $BR({\tilde q}\ra\ell+q')=1$ and
severely constrains the case where $BR({\tilde q}\ra\ell+q')<1$
\cite{kramer}. In Sect. \ref{sec:chertok} we present a study of gluino
pair production followed by the cascade decay to charm squarks (as
motivated by the HERA anomaly) which in turn decay directly via
$L_1Q_2D^c_1$, \ie not via the LSP. We also investigate the pair
production of squarks followed by the cascade decay to the LSP which
decays via $L_1Q_2D^c_1$. This focuses particularly on the $\rpv$
interpretation of the HERA data which is not covered by the leptoquark
searches.

\vspace{0.4cm}

{\centerline {\Large {\bf PART 1: Resonant SParticle Production}}}

\section{R-parity Violating Decay of the Slepton and Identification}
\label{sec:hewett}
{\centerline{\large{\it{J. Hewett, T. Rizzo\footnote{Work supported by the 
Department of Energy, Contract DE-AC03-76SF00515}}}}}
\input{psfig}
\vspace{0.7cm}

In this section, we concentrate on the two sets of trilinear
$L$-violating terms in $W_\rpvs$, and the case of single charged or
neutral slepton production via $q\bar q^{(')}$ annihilation at the
Tevatron through the $\lam'$ couplings.  If this slepton decays to
opposite sign leptons (through the $\lam$ couplings) then an event
excess, clustered in mass, will be observed in the Drell-Yan channel
similar to that expected for a new neutral or charged gauge boson,
$Z'$ or $W'$.  In addition, both $\tilde \ell$ and $\tilde \nu$
resonances may decay hadronically via the same vertices that produced
them, leading to potentially observable peaks in the dijet invariant
mass distribution.  Thus resonant slepton production, first discussed
in Ref.\cite{esmail,bumps}, offers a unique way to explore the $\rpv$
model parameter space at hadron colliders.  It is important that
$\rpv$ also allows for other SUSY particles, such as $\tilde t$ and/or
$\tilde b$, to be exchanged in the non-resonant $t,u-$channels and
also contribute to Drell-Yan events. However, it can be easily shown
that their influence on cross sections and various distributions will
be quite small if the low energy constraints on the Yukawa couplings
are satisfied.\cite{leptos} The questions addressed in this analysis
are: ($i$) what are the mass and coupling reaches for slepton
resonance searches at the Tevatron in the Drell-Yan and dijet channels
and ($ii)$ how can slepton resonances, once discovered, be
distinguished from $Z',W'$ production.

\vspace*{-0.5cm}
\nn
\begin{figure}[htbp]
\centerline{
\psfig{figure=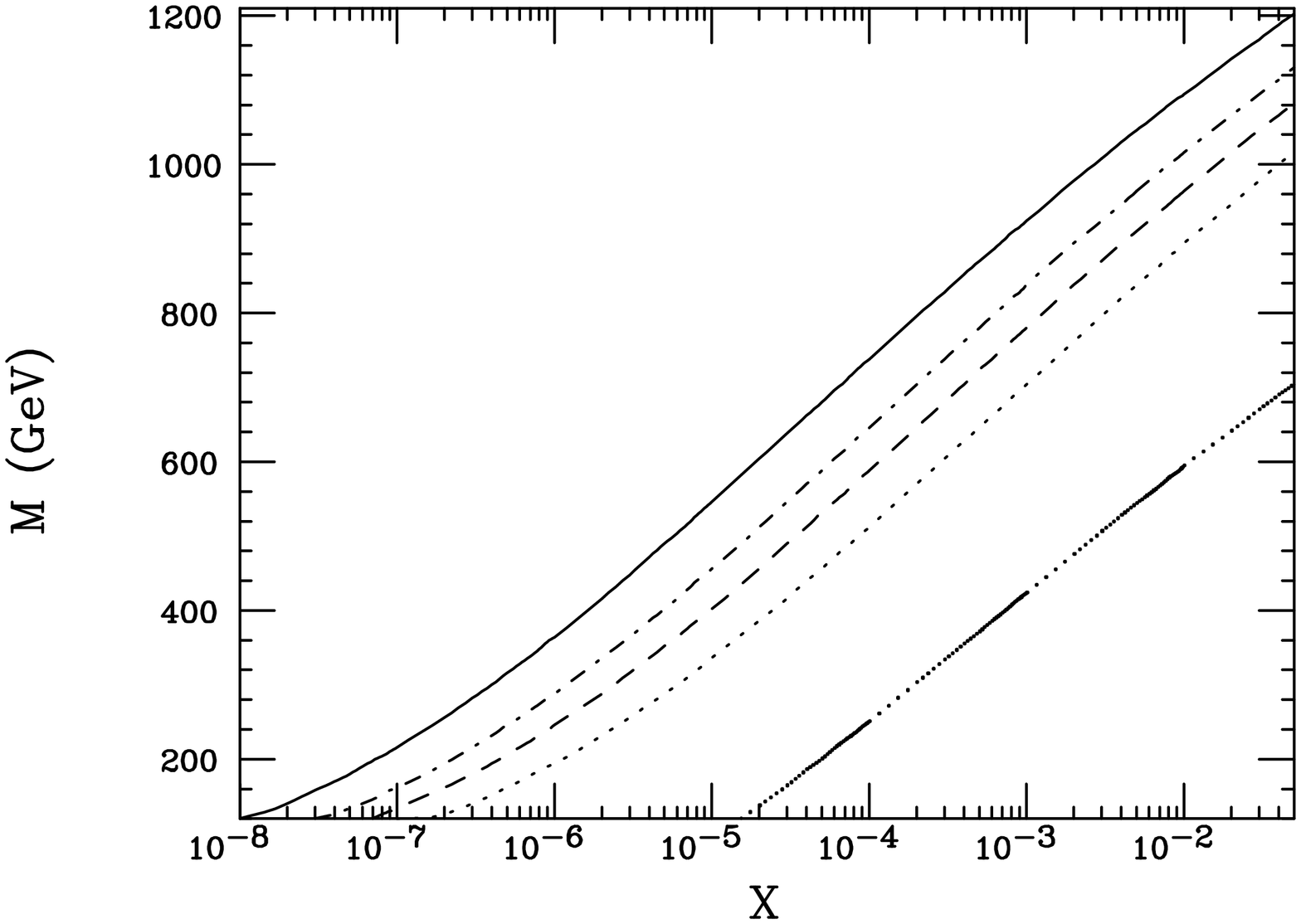,height=6cm,width=8cm,angle=-90}
\hspace*{-5mm}
\psfig{figure=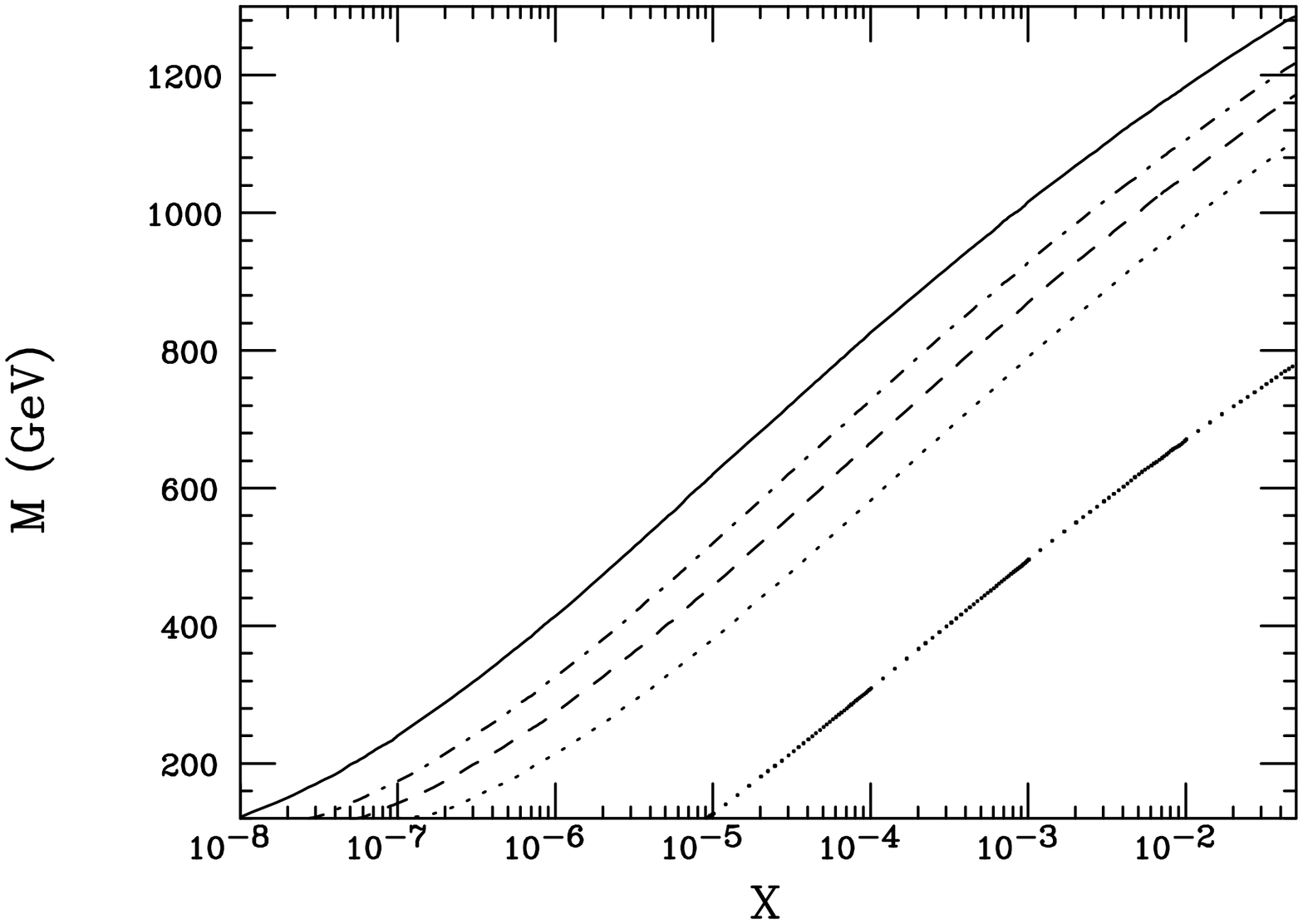,height=6cm,width=8cm,angle=-90}
}
\vspace*{-0.6cm}
\caption{Discovery regions (lying below the curves) in the mass-coupling plane 
for $\rpv$ resonances in the neutral (left) and charged (right) 
Drell-Yan channels at the Run II Tevatron. From top to bottom the curves 
correspond to integrated luminosities of 30, 10, 5 and 2 $fb^{-1}$. The 
estimated reach for Run I is given by the lowest curve. The parameter X is 
defined in the text.}
\label{tevlim}
\end{figure}
\vspace*{-.2mm}

\subsubsection{Drell-Yan Channel for Sleptons}

From the bounds in Section \ref{sec:bounds}, we see that the coupling
responsible for first generation slepton production, $\lam'_{111}$, is
much too restricted to generate a sizable cross section.  However the
bounds on the couplings for second and third generation slepton
production and decay are all of similar numerical values and allow for
a reasonably sized signal. In addition, the background free
$e^\pm\mu^\mp$ signature is also possible.

In the case of Drell-Yan production the search reach analysis is
straightforward being nearly identical to that used for new gauge
boson production, apart from acceptance issues due to the differences
between spin-0 and spin-1 resonances. Since sleptons are expected to
be narrow, the narrow width approximation is adequate and the analysis
presented in Ref. \cite{snow} can be directly followed. In addition to
the slepton mass the only other parameter in the calculation is the
product of the appropriate Yukawa couplings, $\lam'$, from the initial
state $d\bar u$ or $d\bar d$ coupling vertex, and the slepton's
leptonic branching fraction, $B_\ell$.  Denoting this product as
$X=(\lam')^2B_\ell$, we obtain the $95\%$ C.L.  search reach as a
function of $X$ in both the charged and neutral channels.  These
results are displayed in Fig. \ref{tevlim} for various values of the
integrated luminosity.  Not only is it important to notice the very
large mass reach of these colliders for sizeable values of $X\sim
10^{-3}$, but we should also observe the small $X$ reach, $X\sim
10^{-(5-7)}$ and below, for relatively small slepton masses. These
results show the rather wide opportunity available to discover slepton
resonances over extended ranges of masses and couplings at the
Tevatron. Note that for fixed values of $X$ the search reach is
greater in the charged current channel due to the higher parton
luminosities in that case.

\vspace*{-0.3cm}
\nn
\begin{figure}[htbp]
\centerline{
\psfig{figure=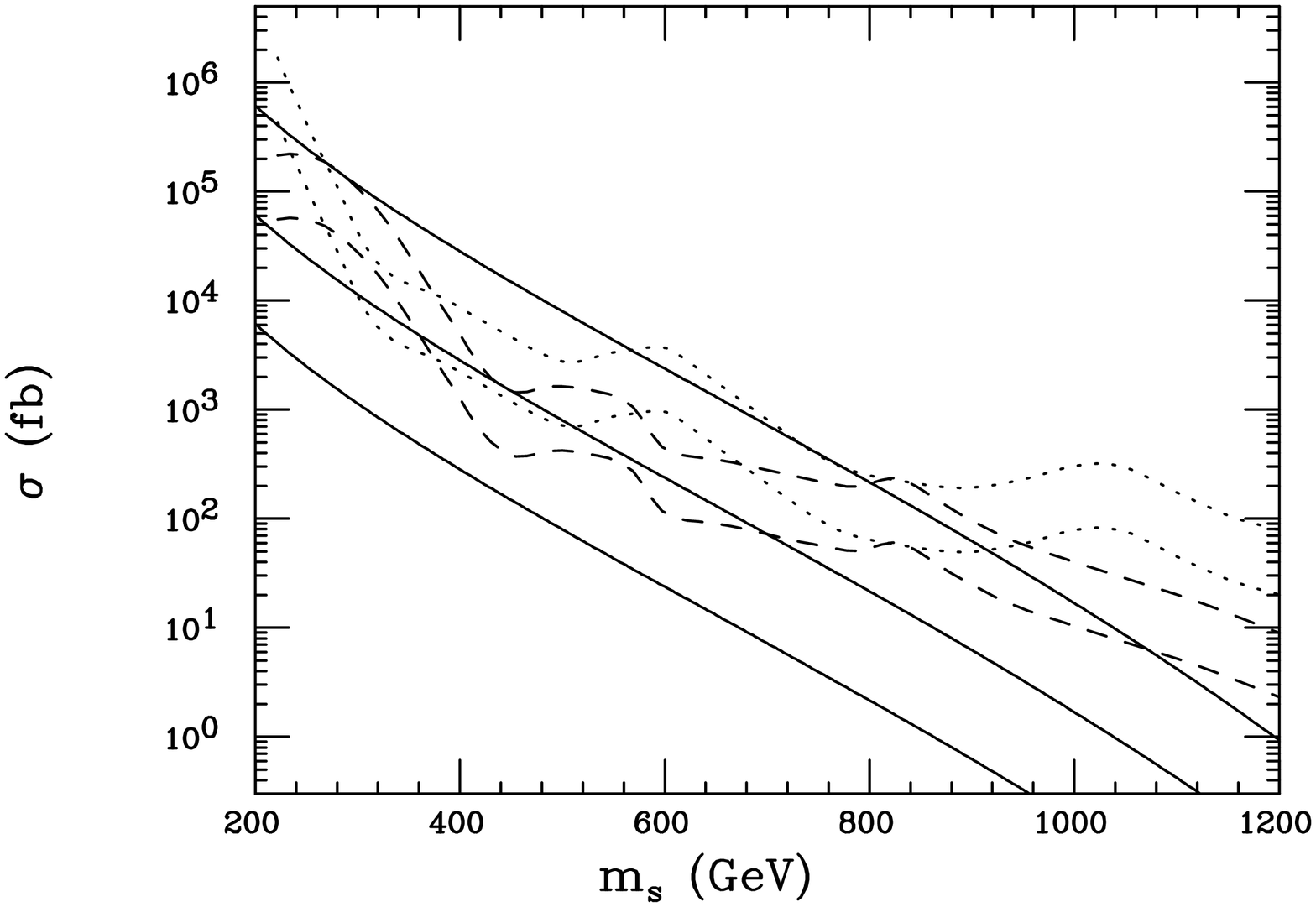,height=6cm,width=8cm,angle=-90}
\hspace*{-5mm}
\psfig{figure=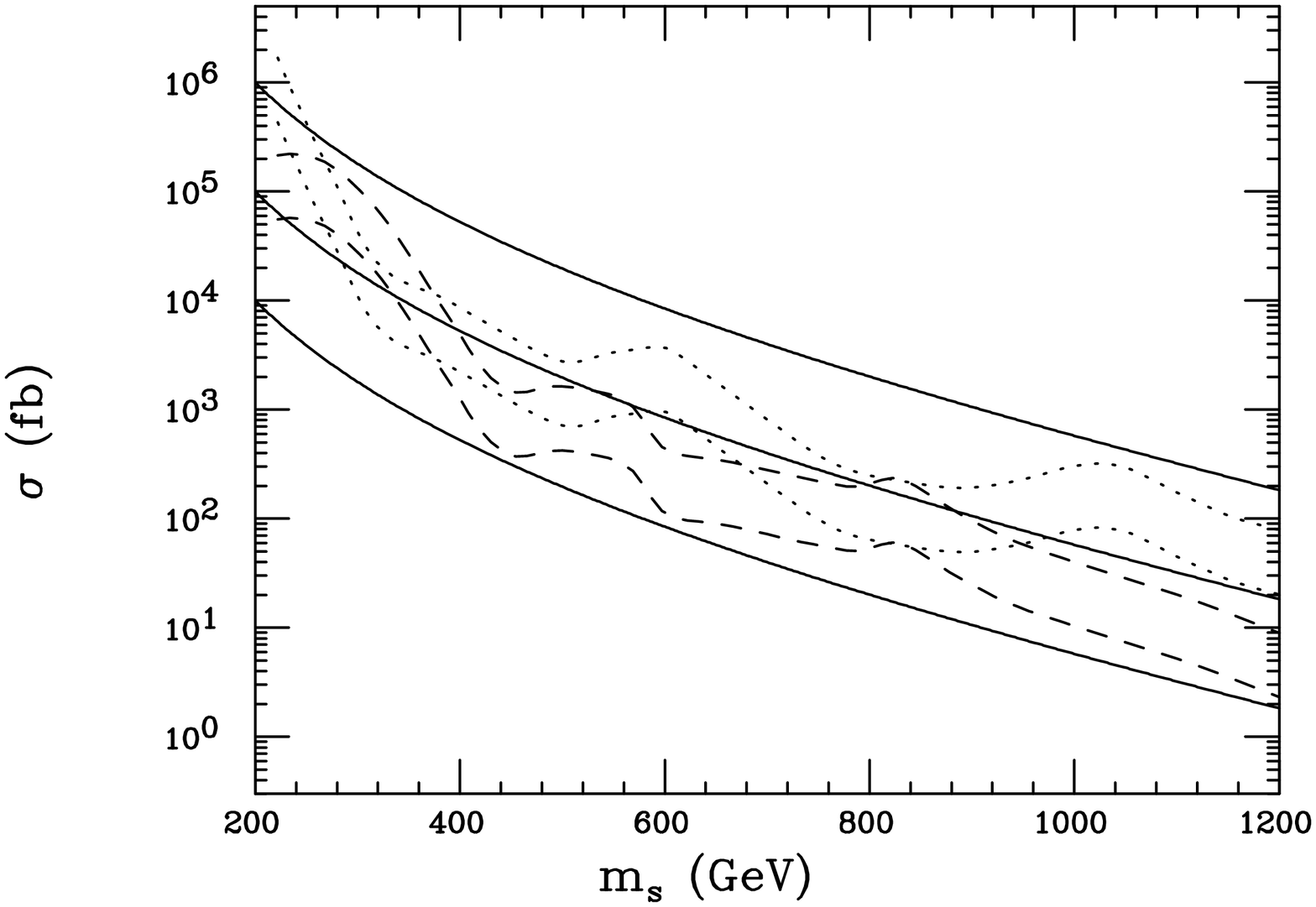,height=6cm,width=8cm,angle=-90}}
\vspace*{-0.6cm}
\caption{Cross sections for narrow dijet resonances (solid) at the 2 TeV 
Tevatron arising from $\tilde \nu$ (left) or $\tilde \ell$ (right) production 
in comparison to the anticipated search reaches of CDF (dots) and D0 
(dashes). The upper(lower) curve for each experiment assumes an integrated 
luminosity of 2(30) $fb^{-1}$. The three solid curves from top to bottom 
correspond to slepton resonance predictions for $Y$=0.1, 0.01 and 0.001, 
respectively, where $Y$ is defined in the text.}
\label{tevdijets}
\end{figure}
\vspace*{0.1mm}

\vspace*{-0.5cm}
\nn
\begin{figure}[htbp]
\centerline{
\psfig{figure=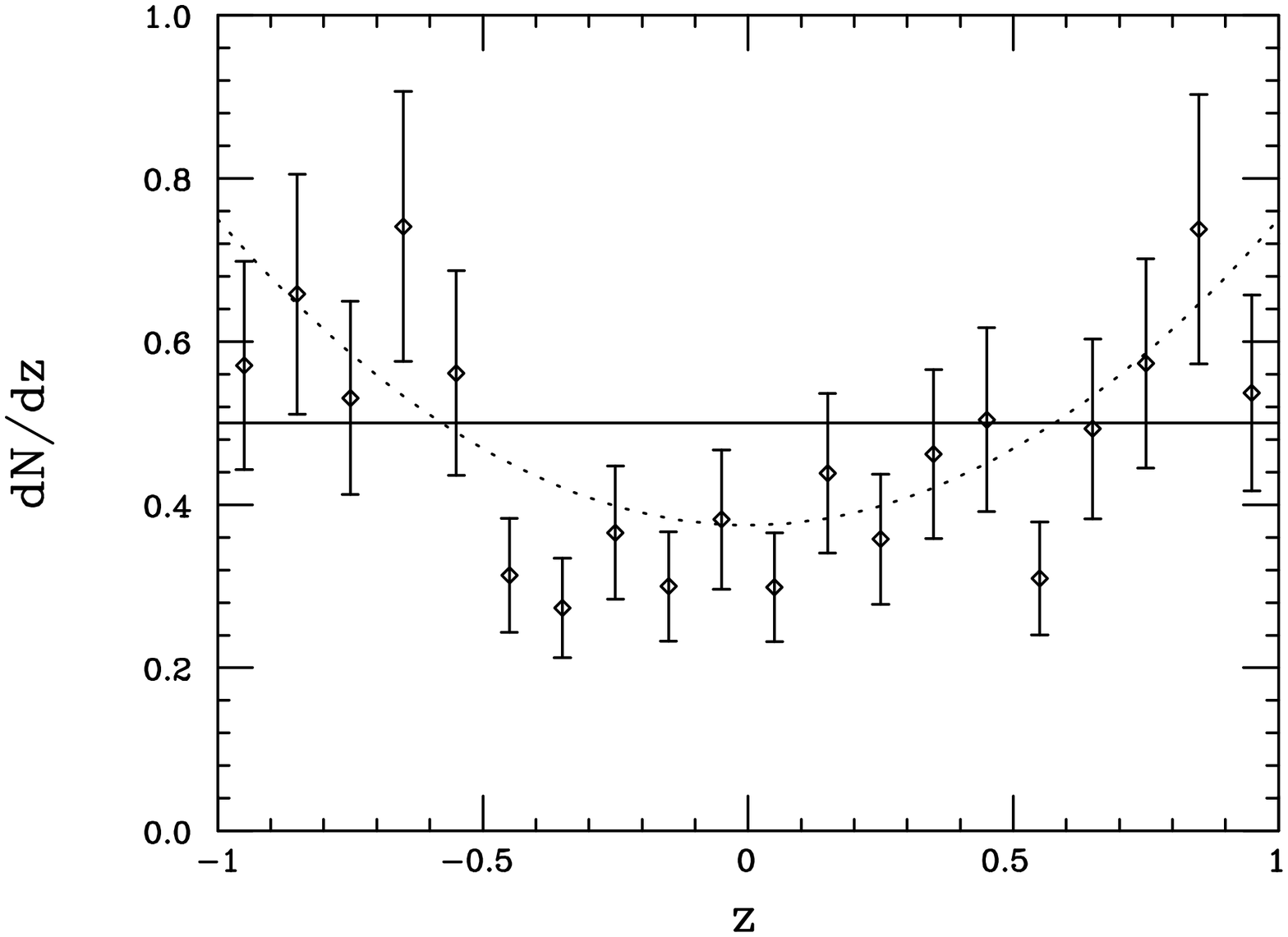,height=6cm,width=8cm,angle=-90}
\hspace*{-5mm}
\psfig{figure=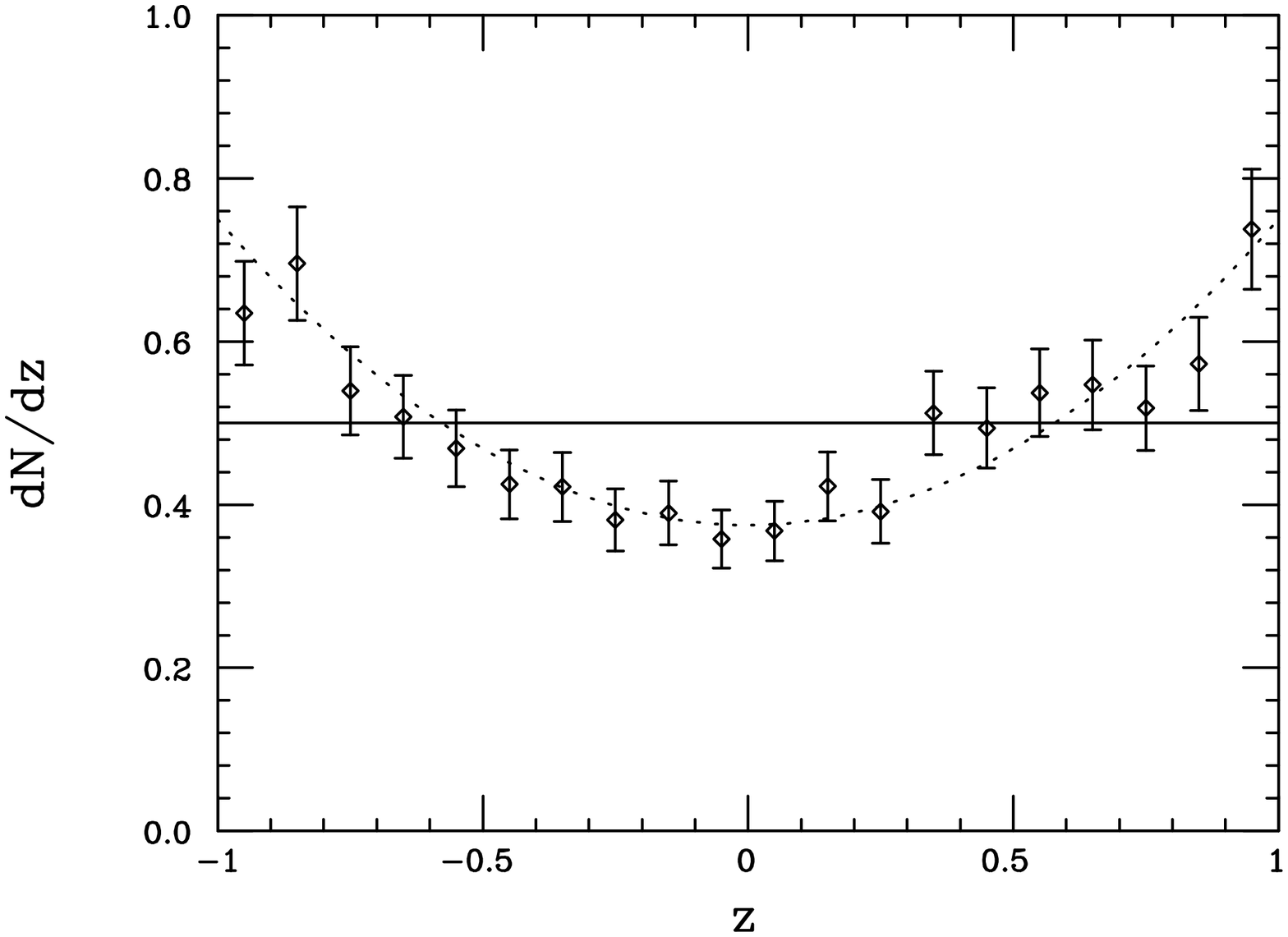,height=6cm,width=8cm,angle=-90}}
\vspace*{-0.6cm}
\caption{Comparison of the Monte Carlo generated normalized angular 
distribution for the 
leptons in $Z'$ decay (data points) with that for a $\tilde \nu$ (solid curve)
assuming a (left)400 event sample or a (right)2000 event sample; the displayed 
errors are purely statistical.  The $Z'$ is assumed to have no 
forward-backward asymmetry due to its fermionic couplings. A $1+z^2$ 
angular distribution is represented by the dotted curve.}
\label{angledist}
\end{figure}
\vspace*{0.1mm}

\subsubsection{Di-Jet Decays of the Slepton}
Since $d\bar d$ and/or $u\bar d$ annihilation are responsible for single
slepton production, slepton resonance can also decay into these same
fermion pairs. Hence $\tilde \ell$ or $\tilde \nu$ will have dijet decays 
which may appear as observable peaks above the conventional QCD
backgrounds.  Searches for such narrow dijet resonances have already
been performed\cite{dijetbumps} at the Tevatron by both CDF and D0
during Run I.  Using their results and scaling by appropriate factors
of beam energy and integrated luminosities we may estimate the
probable search reaches for CDF and D0 from Run II. Our estimates
conform to the expectations given in Ref. \cite{tev2000}. The cross
sections themselves are calculable in the narrow width approximation
in terms of the product $Y=(\lam')^2B_{2j}$, where $\lam'$ is the
Yukawa coupling and $B_{2j}$ is the dijet branching fraction. The
results are presented in Fig. \ref{tevdijets}. Here we see
that for values of $Y \gsim 0.001-0.01$ the Tevatron will have a
substantial mass reach for slepton induced dijet mass bumps during Run
II.  Note that as in the case of Drell-Yan production, larger cross
sections for fixed $Y$ occur in the charge current channel than in the
neutral current case due to the larger parton luminosities.
Unfortunately, if such a bump is observed it will not be
straightforward to identify it as a slepton resonance.

\subsubsection{Distinguishing Sleptons from $W',\,Z'$}
Our next issue concerns the ability to determine that an observed
resonance (or Jacobian peak) in the Drell-Yan channel is due to
slepton production, instead of a new gauge boson, $Z', W'$.  One clear
signature for $\tilde \nu$ production would be the observation of
$e\mu$ final states. These are not expected to occur in $Z'$ models
and would point to $\rpv$. In addition, if the $R_p$ conserving decay
modes of the slepton dominate there will be no identification problem.
If the $\rpv$ modes dominate one should first look for universality
violations, {\it e.g.}, if the resonance decays to only one of
$e^+e^-$ or $\mu^+\mu^-$ or if these two rates are substantially
different.  Most models with new gauge bosons do not have
substantially different couplings to the first two fermion
generations.

In the neutral current channel, \ie, $\tilde \nu$ or $Z'$ production,
the forward-backward asymmetry $A_{FB}$ in the leptonic decay
distributions provides a good diagnostic tool.  It is well-known that
most models of $Z'$ bosons have parity violating fermionic couplings
which would lead to $A_{FB}\not=0$. However, $\tilde \nu$, being
spin-0, would always produce a null asymmetry.  $A_{FB}$ is more
easily measured and requires less statistical power than does the
reconstruction of the complete angular distribution.  This is
important since, whereas only 10 or so background free events would
constitute a discovery, many more events, $\sim 100-200$ are required
to determine the asymmetry.  This implies that the reach for
performing this test is somewhat, if not substantially, less than the
discovery reach.  For example, the Tevatron may discover a $\tilde
\nu$ with a mass of 700 GeV for a certain value of $X$ but only for
masses below 500 GeV would there be enough statistics to extract
$A_{FB}$ for this same $X$ value.

\vspace*{-0.5cm}
\nn
\begin{figure}[htbp]
\centerline{
\psfig{figure=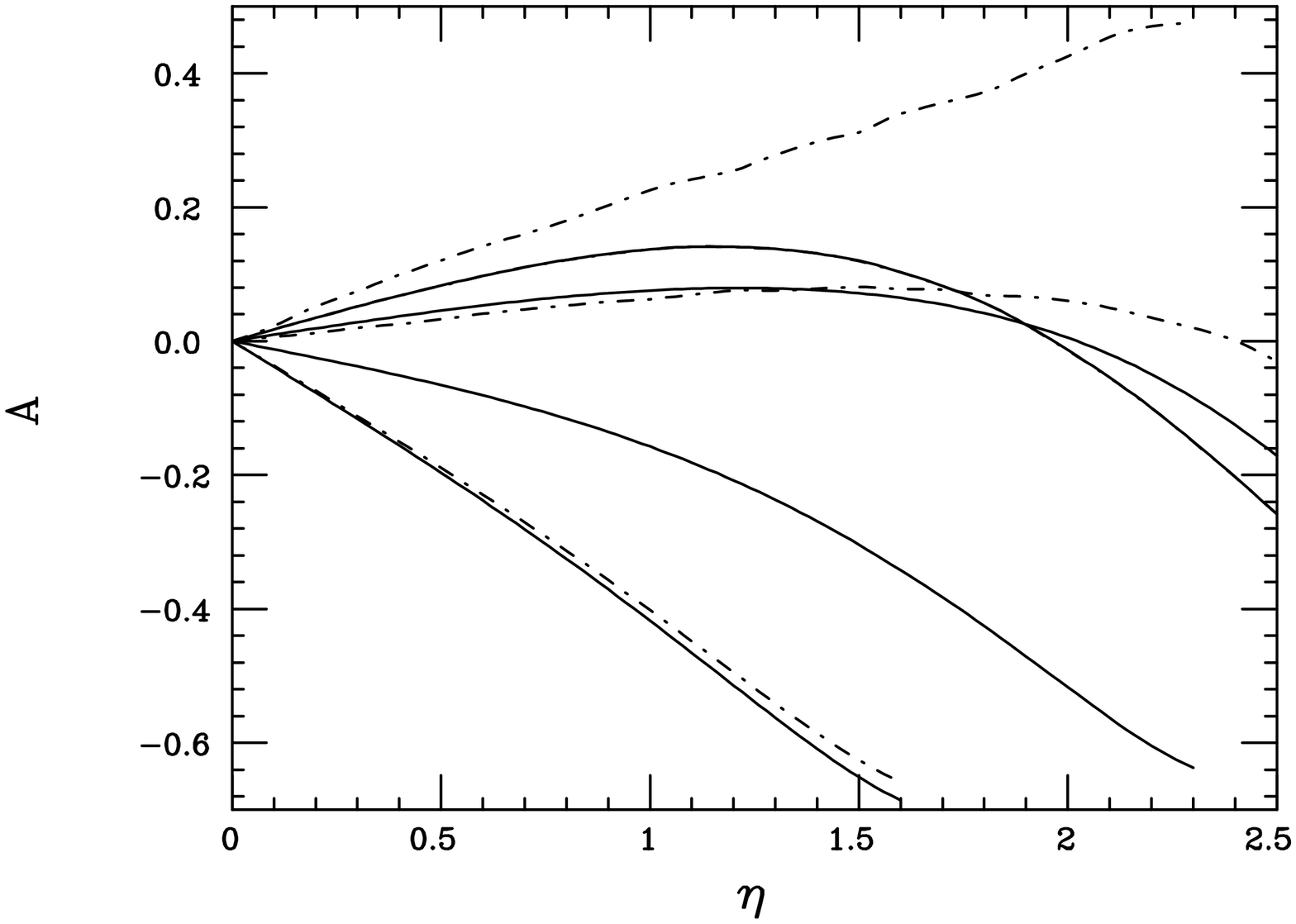,height=6cm,width=8cm,angle=-90}
\hspace*{-5mm}
\psfig{figure=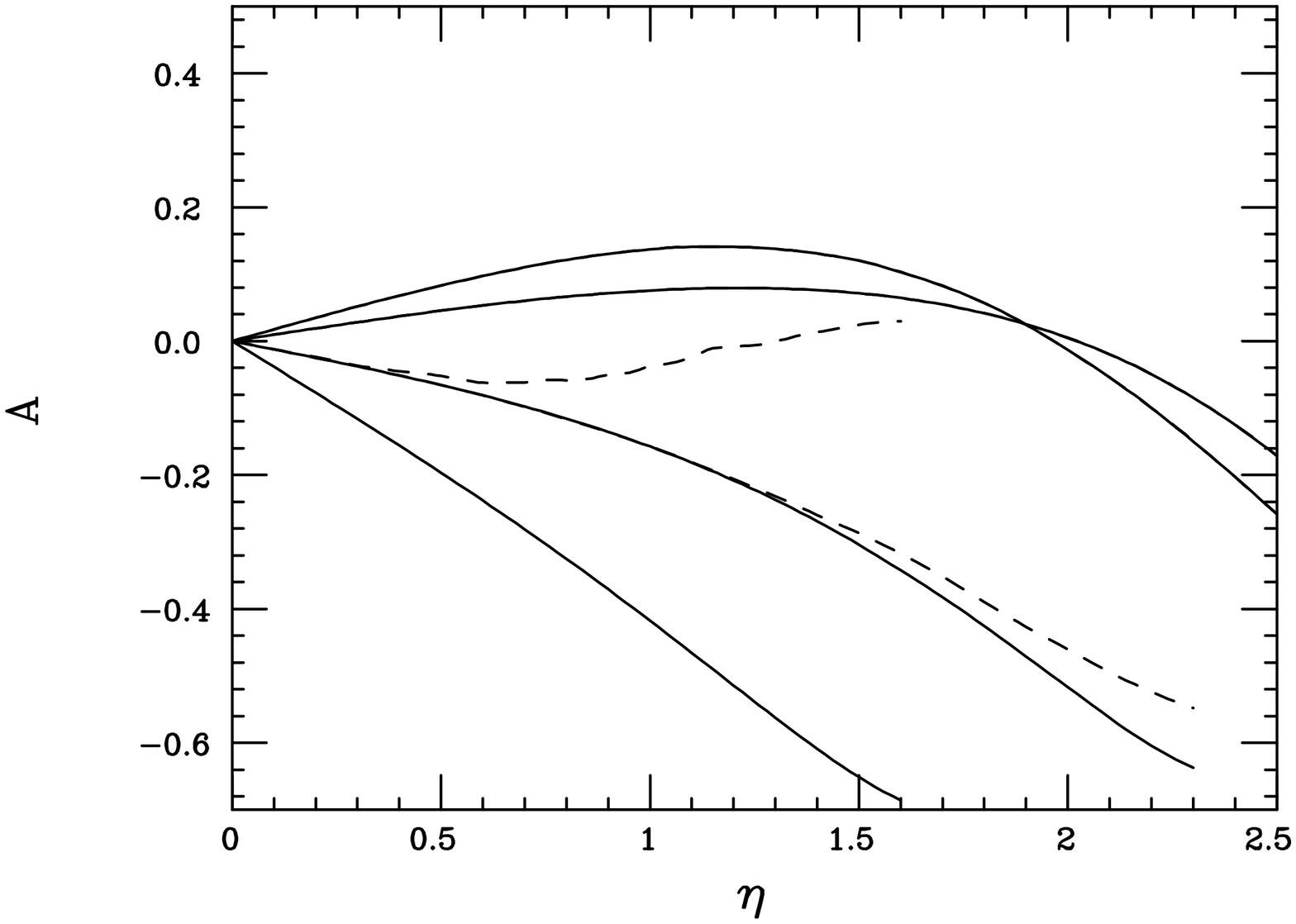,height=6cm,width=8cm,angle=-90}}
\vspace*{-0.6cm}
\caption{The lepton charge asymmetry in the charged current Drell-Yan 
production channel at the 2 TeV Tevatron for the SM (solid curves) and 
with 250(700) GeV $\tilde \ell$ exchange (the dash-dotted and dashed curves) 
assuming $\lam,\lam'=0.15$.  From top to bottom in the center of the figure, 
the SM curves correspond to $M_T$ bins of 50-100, 100-200, 200-400 and 
$>400$ GeV, respectively.}
\label{deviations}
\end{figure}
\vspace*{0.1mm}

A more complex and interesting situation arises when the $Z'$
naturally has $A_{FB}=0$ as in, \eg, some $E_6$ models\cite{jlhtgr};
in this case the on-resonance asymmetry data alone is insufficient.
If $A_{FB}$ could be measured throughout the resonance region, it
would be possible to deduce through detailed line-shape studies
whether or not the new contribution interferes with the SM amplitude
(something that does not occur in the case of spin-0 $\tilde \nu$
production).  However, finite dilepton mass resolution, especially for
the $\mu^+\mu^-$ final state, in addition to the increased required
statistics may disrupt this program.

With a plethora of statistics the complete angular distribution can be
obtained as shown in Fig.~\ref{angledist}. Here we compare Monte Carlo
generated data for a $Z'$ with a zero forward-backward asymmetry with
both the flat distribution expected in the spin-0 case, and the $\sim
1+z^2$ distribution ($z=\cos \theta$) of spin-1 exchange.  We ignore
complications due to possible acceptance losses arising from rapidity
cuts in the forward and backward directions.  Such a distribution has
been measured by CDF both on and above the SM $Z$ resonance \cite{angular}. 
This analysis indicates that of order $\sim 1000$ events are required
to make a clean measurement, which is a sample approximately 100 times
larger than that required for discovery.  Although such measurements
would be conclusive as to the identity of the spin of the resonance,
the required statistics results in a significant loss in the mass
range over which it can be performed. In the example discussed above
where the slepton search reach was 700 GeV we find that the angular
distribution could only be determined for masses below $\sim 400$ GeV
assuming the same $X$ value.

Angular distributions cannot be used to separate between possible
$\tilde \ell$ and $W'$ production in the charged current channel due
to the missing energy in the event.  However, there are two other
useful observables in this case.  First, is the transverse mass
($M_T$) distribution associated with the new Jacobian peak region
where interference with the SM amplitude would occur for $W'$, but not
slepton, production.  This test is unfortunately far more difficult to
perform than in the neutral current channel again due to the missing
energy as well as mass smearing.  A second possibility is to examine
the leptonic charge asymmetry, $A(\eta)$, for the electrons or muons
in the final state as a function of their rapidity, where
\begin{equation}
A(\eta)\equiv{dN_+/d\eta -dN_-/d\eta \over {dN_+/d\eta +dN_-/d\eta}}\,,
\end{equation}
where $N_{\pm}$ is the number of positively/negatively charged electrons of a 
given rapidity, $\eta$.  In the SM, the charge asymmetry is sensitive to 
the ratio of u-quark to d-quark parton densities and the $V-A$ nature of the
$W$ production and decay.\cite{chargew}  Since the coupling structure of the 
SM $W$ has been 
well-measured, any deviations in this asymmetry within the $M_T$ bin 
surrounding the $W$ mass have been attributed to modifications in the parton 
distributions.  Here, we 
are more interested in events with larger values of $M_T$.  Note that 
$A(-\eta)=-A(\eta)$ if $CP$ is conserved (which we assume) so that we will 
only need to deal with $\eta \geq 0$ in the following discussion. 

\vspace*{-0.5cm}
\nn
\begin{figure}[htbp]
\centerline{
\psfig{figure=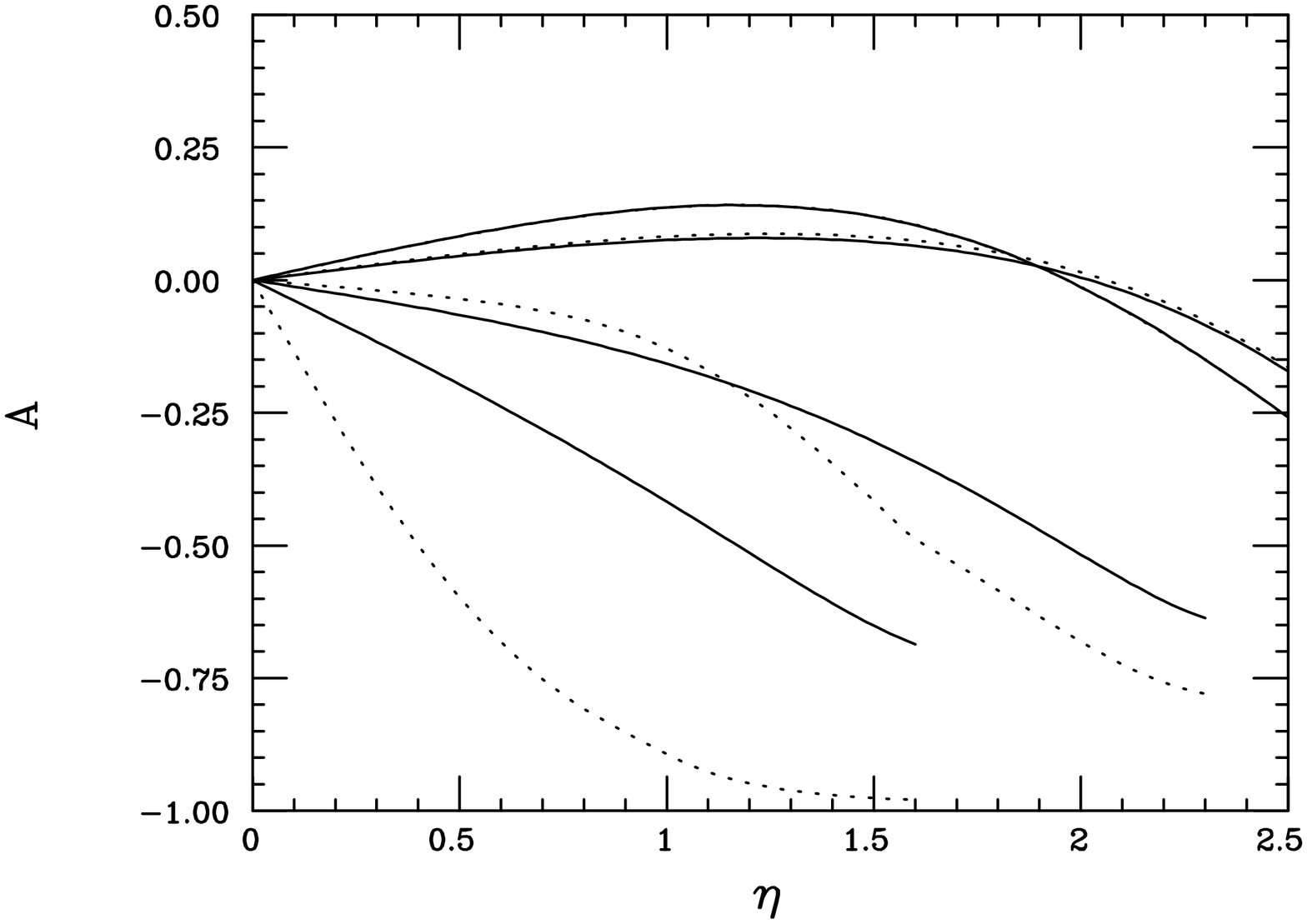,height=6cm,width=8cm,angle=-90}
\hspace*{-5mm}
\psfig{figure=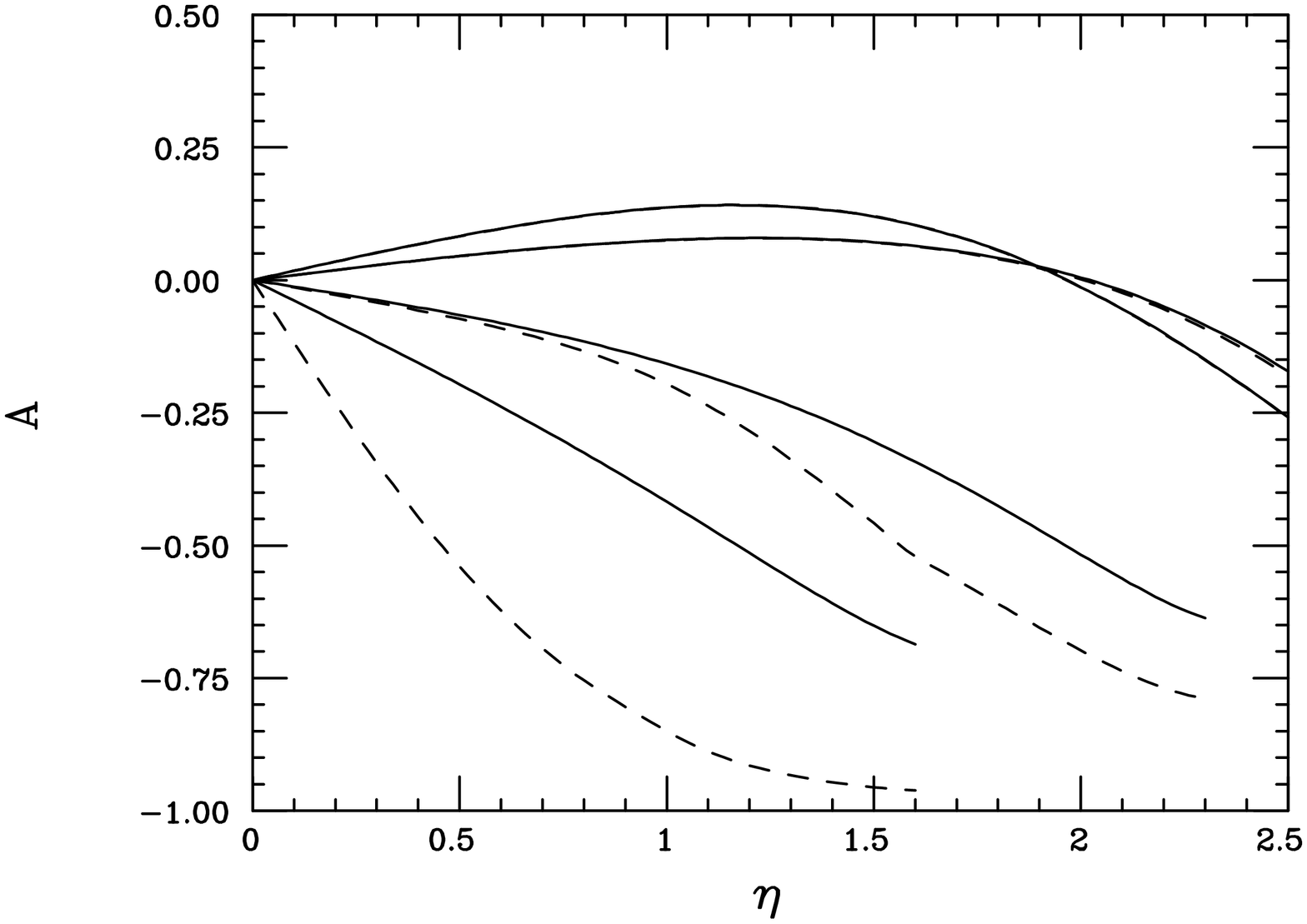,height=6cm,width=8cm,angle=-90}}
\vspace*{-0.6cm}
\caption{Same as the previous figure but now for the case of a 800 GeV 
$W'$ with purely left-handed(left panel) or 
purely right-handed(right panel) couplings.}
\label{wprime}
\end{figure}
\vspace*{0.1mm}

Figure \ref{deviations} displays the lepton charge asymmetry within
four $M_T$ bins, corresponding to $50<M_T<100$ GeV, $100<M_T<200$ GeV,
$200<M_T<400$ GeV and $400<M_T<1800$ GeV, for the SM and with a
250(700) GeV $\tilde \ell$ with, for purposes of demonstration,
$\lam,\lam'=0.15$.  In particular, the lepton charge asymmetry can be
significantly altered for larger values of $M_T$ in the bins
associated with the new Jacobian peak.  Note, however, that there is
essentially no deviation in the asymmetry in the transverse mass bin
associated with the SM $W$ peak, $50<M_T<100$ GeV, so that this $M_T$
region can still be used for determination of the parton densities.
The figure also shows that the presence of the $\tilde \ell$ tends to
drive the asymmetry to smaller absolute values as perhaps might be
expected due to the presence of a spin-0 resonance.

\vspace*{-0.4cm}
\nn
\begin{figure}[t]
\centerline{
\psfig{figure=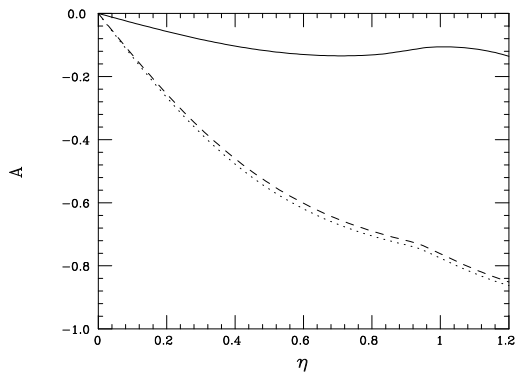,height=6cm,width=8cm,angle=-90}}
\vspace*{-0.6cm}
\caption{Direct comparison of the charge asymmetry induced by a 800 GeV 
$\tilde \ell$(solid) and a left- or right-handed $W'$(dot and dash) of the 
same mass, taking $600 < M_T <900$ GeV. 
The assumed values of the Yukawa couplings are as in the earlier figure.}
\label{dis}
\end{figure}
\vspace*{0.5mm}

Figure \ref{wprime} presents the corresponding modifications in the 
leptonic charge asymmetry 
due to an 800 GeV $W'$ with either purely left-handed(LH) or purely 
right-handed(RH) fermionic couplings.  Note that the $W'$ with purely RH 
couplings, unlike the LH $W'$, does 
not interfere with the SM amplitude, similar to the case of $\tilde \ell$ 
production.  We see that the deviation in the asymmetry due to either type of 
$W'$ is very different than that for a slepton.   In the $W'$ case, the
magnitude of the asymmetry is substantially increased for 
both left- and right-handed couplings and these two cases 
are themselves potentially 
distinguishable by using the data in the $M_T$ bin below, but not containing,
the Jacobian peak.

The $M_T$ bins we have taken in this analysis are rather broad.  We
might expect that if the width of the $M_T$ bin around the $W'$ or
slepton Jacobian peak is compressed then the purity of the resonant
contribution will be increased resulting in a better separation of the
two cases (at the price of reduced statistics).  These expectations
are realized in Fig. \ref{dis} which shows a more direct comparison of
the lepton charge asymmetries for a $\tilde \ell$ and $W'$ of the same
mass (800 GeV) taking the width of the $M_T$ bin surrounding the
Jacobian peak to be only 300 GeV.  However, the left- and
right-handed $W'$ possibilities are no longer separable since this
requires interference with the SM.

\begin{figure}[htbp]
\centerline{
\psfig{figure=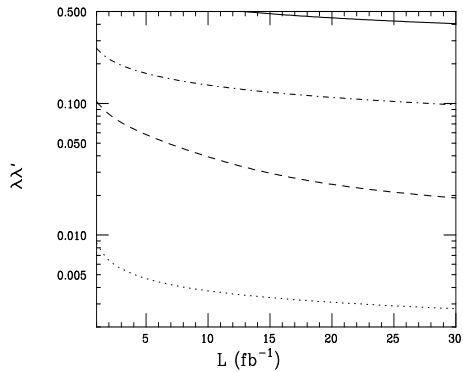,height=6cm,width=7cm,angle=-90}}
\vspace*{-0.6cm}
\caption{Search reach for $\tilde \ell$ exchange as a function of the 
integrated luminosity assuming $\Gamma/m=0.004$ for masses of 1500, 
1000, 750 and 250 GeV (from top to bottom).}
\label{asymreach}
\end{figure}

Deviations in the leptonic charge asymmetry can also be used to probe
indirectly for the exchange of $\tilde \ell$ through $\rpv$ couplings.
To demonstrate this, we fix the $\tilde \ell$ width to mass ratio to
be $\Gamma/m=0.004$ and subdivide each of the four $M_T$ bins
discussed above into rapidity intervals of $\Delta \eta=0.1$.  For a
fixed slepton mass and integrated luminosity we generate Monte Carlo
data for various values of the Yukawa couplings and perform a $\chi^2$
analysis to obtain the sensitivity to the product $\lam \lam'$.  The
$95\%$ C.L. search reach from this analysis is shown in
Fig.\ref{asymreach} where it is clear that the reach obtained in this
manner is rather modest.
\vspace*{-0.1cm}
\nn
\vspace*{-5.mm}

\subsubsection{Summary}
The above analysis shows that resonant $s$-channel slepton production
with subsequent decay into purely leptonic or dijet final states via
$\rpv$ couplings is observable in hadronic collisions over a wide
range of parameters.  If this signature is observed, we have
demonstrated that the leptonic angular distributions and the lepton
charge asymmetry can be successfully used to distinguish slepton
resonances from those associated with new gauge bosons.  This process
provides a clean and powerful probe of $\rpv$ supersymmetric parameter
space.

\section{R-Parity Conserving Decay of the Slepton}
\label{sec:richardson}
{\centerline{\large {\it {H. Dreiner, M. Seymour, P. Richardson}}}}
\vspace{0.7cm}

In this section we discuss the resonant charged slepton production
via the operator $L_iQ_j{D}_k^c$ followed by the decay to the 
neutralino decay as shown in \eq{sleptonprod}g.
\beq
u_j{\ol d}_k\ra {\tilde\ell}_i^+\ra\ell_i^++\chi^0_n.
\eeq
The neutralino in turn then decays via the same $\rpv$-operator 
$\mathrm{L_iQ_j{\ol D}_k}$,
\beq
\mathrm{\tilde{\chi}^{0}\ra \{\ell^-_i+u_j+ {\overline d}_k;\; \nu_i+d_j+ 
{\overline d}_k\}}.
\eeq
Since the neutralino is a Majorana fermion it can also decay to the charge
conjugate final states, with equal probability. In spirit, this is
similar to the HERA process considered in \cite{butterworth,Herbi}.

\begin{figure}[htbp]
\includegraphics[angle=90,width=0.45\textwidth,viewport=60 150 410 
550,clip]{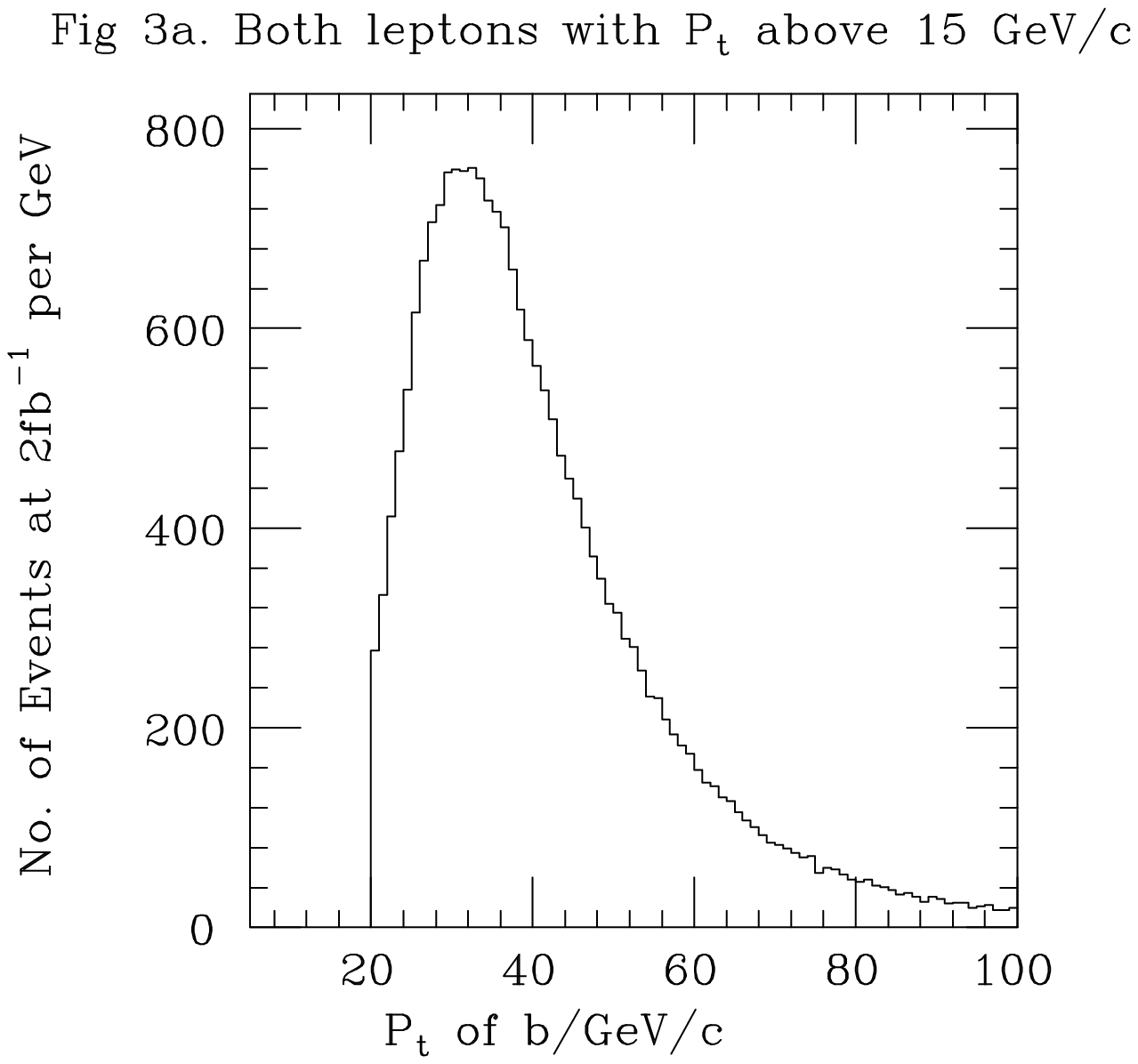}
\hfill
\includegraphics[angle=90,width=0.45\textwidth,viewport=60 150 410 
550,clip]{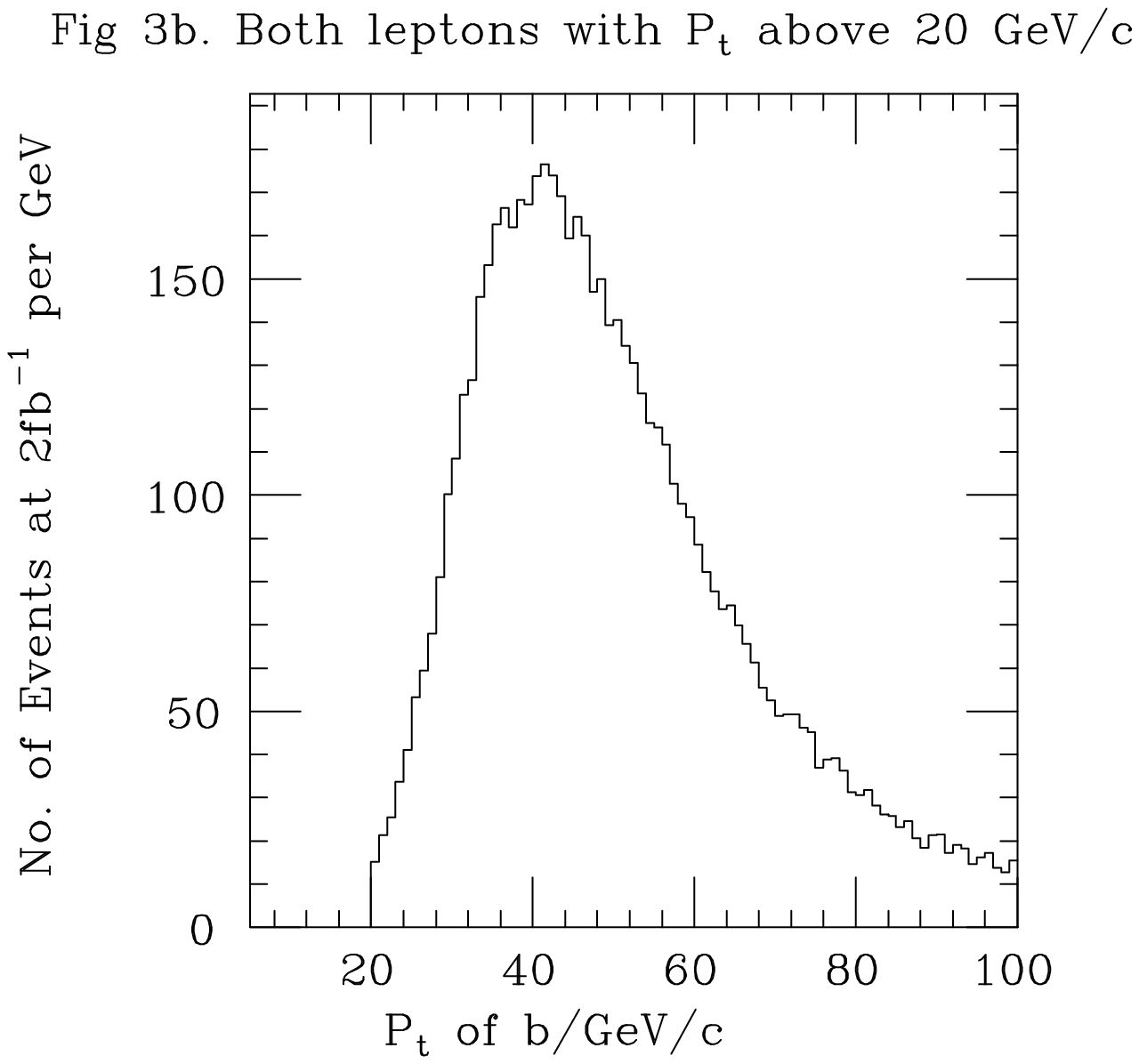}\\
\caption{Effect of the Parton-Level Cuts.\label{fig:parton}}
\end{figure}

The tree-level Feynman diagrams for the slepton production and the
neutralino decay are shown in Fig.\,\ref{fig:cross} and Fig. 
\ref{fig:decay}, respectively. Due to the Majorana nature of the
neutralino, we have a signature of two like-sign charged leptons. In
the following we shall consider only electrons or muons, \ie we focus
on the operators $\mathrm{L_eQ_j{\ol D}_k}$ and $\mathrm{L_\mu 
Q_j{\overline D}_k}$.
We expect these leptons to have high transverse momentum, $p_{\mathrm{T}}$,
and be well isolated whereas the leptons from the Standard Model
backgrounds have lower $p_{\mathrm{T}}$ and are also poorly isolated. We
therefore hope that this signature can be seen above the background if
we apply isolation and $p_{\mathrm{T}}$ cuts.

\begin{table*}[t]
\caption{Summary of the Background Simulation\label{table1}}
\begin{tabular}{|l|ccc|}
Process & Cross section & No. of Events & Expected No. of Events after cuts, \\
& before Cuts /nb & simulated & for $\mathrm{2\,fb^{-1}}$ luminosity\\
\tableline
 $\mathrm{b\bar{b}}$ mixing & $(9.3\pm2.3)\times10^4$ & $8.3\times10^6$ &
 $0.12\pm0.12$ \\
 $\mathrm{t\bar{t}}$        & $6.81\pm0.31$           & $2.0\times10^5$ &
 $0.02\pm0.02$ \\
 single top                 & $1.55\pm0.12$           & $3.5\times10^4$ &
 $0.00\pm0.03$ \\
\tableline
 Total     & & & $0.14\pm0.13$ \\
\end{tabular}
\end{table*}

\begin{figure}[t]
\includegraphics[angle=90,width=0.45\textwidth,viewport=60 150 417 
550,clip]{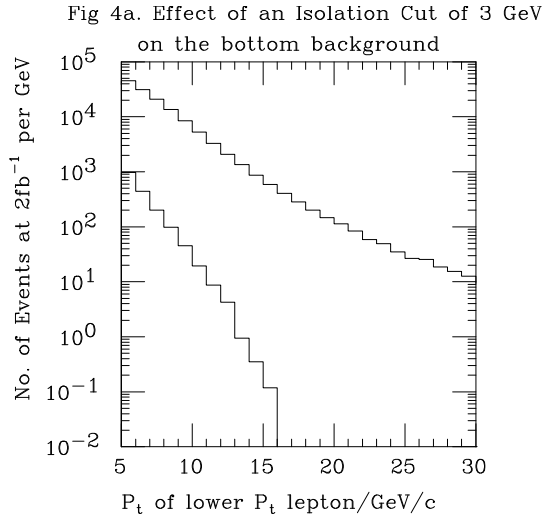}
\hfill
\includegraphics[angle=90,width=0.45\textwidth,viewport=60 150 417 
550,clip]{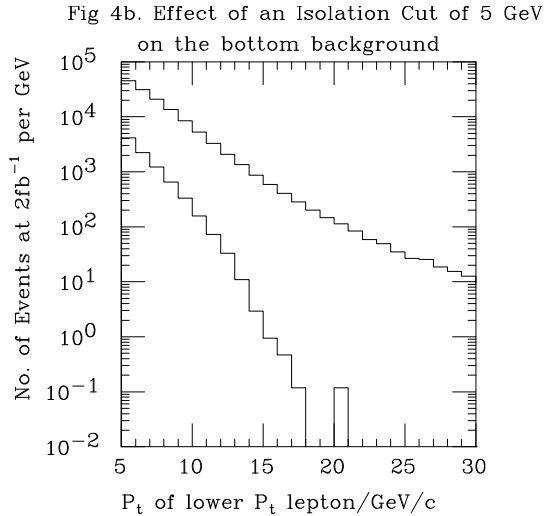}\\
\caption{Effect of the lepton isolation cut on the $\mathrm{b\bar{b}}$
background. The upper curve is the full $\mathrm{b\bar{b}}$ background and the
lower curve is obtained after imposing the isolation cut.
\label{fig:isocut}}
\end{figure}
\begin{figure}[t]
\includegraphics[angle=90,width=0.45\textwidth,viewport=60 150 417 
550,clip]{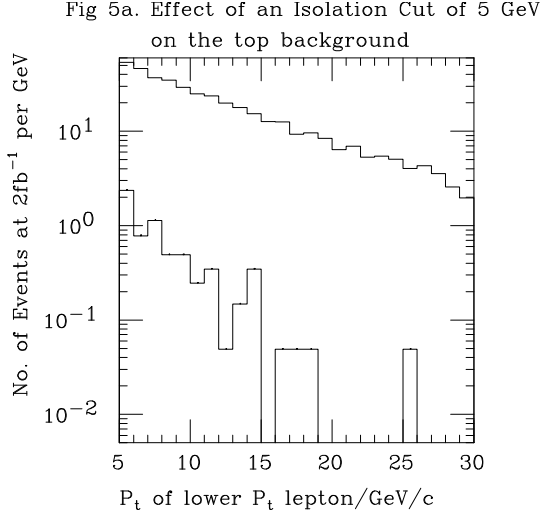}
\hfill
\includegraphics[angle=90,width=0.45\textwidth,viewport=60 150 417 
550,clip]{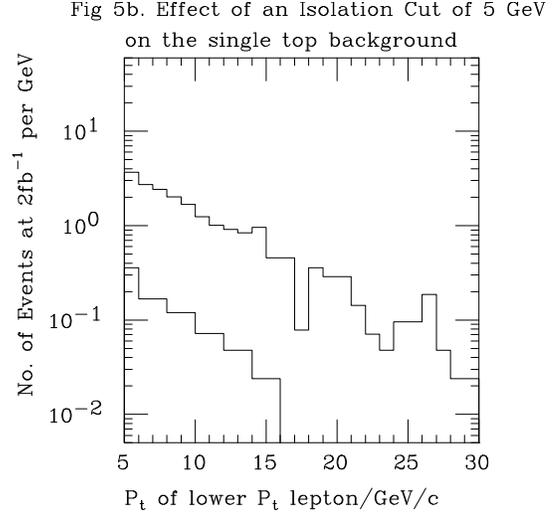}\\
\caption{Effect of the Isolation Cut on the $\mathrm{t\bar{t}}$ and single top 
Backgrounds\label{fig:isocut2}}
\end{figure}
\begin{figure}[t]
\includegraphics[angle=90,width=0.45\textwidth,viewport=50 70 550
700,clip]{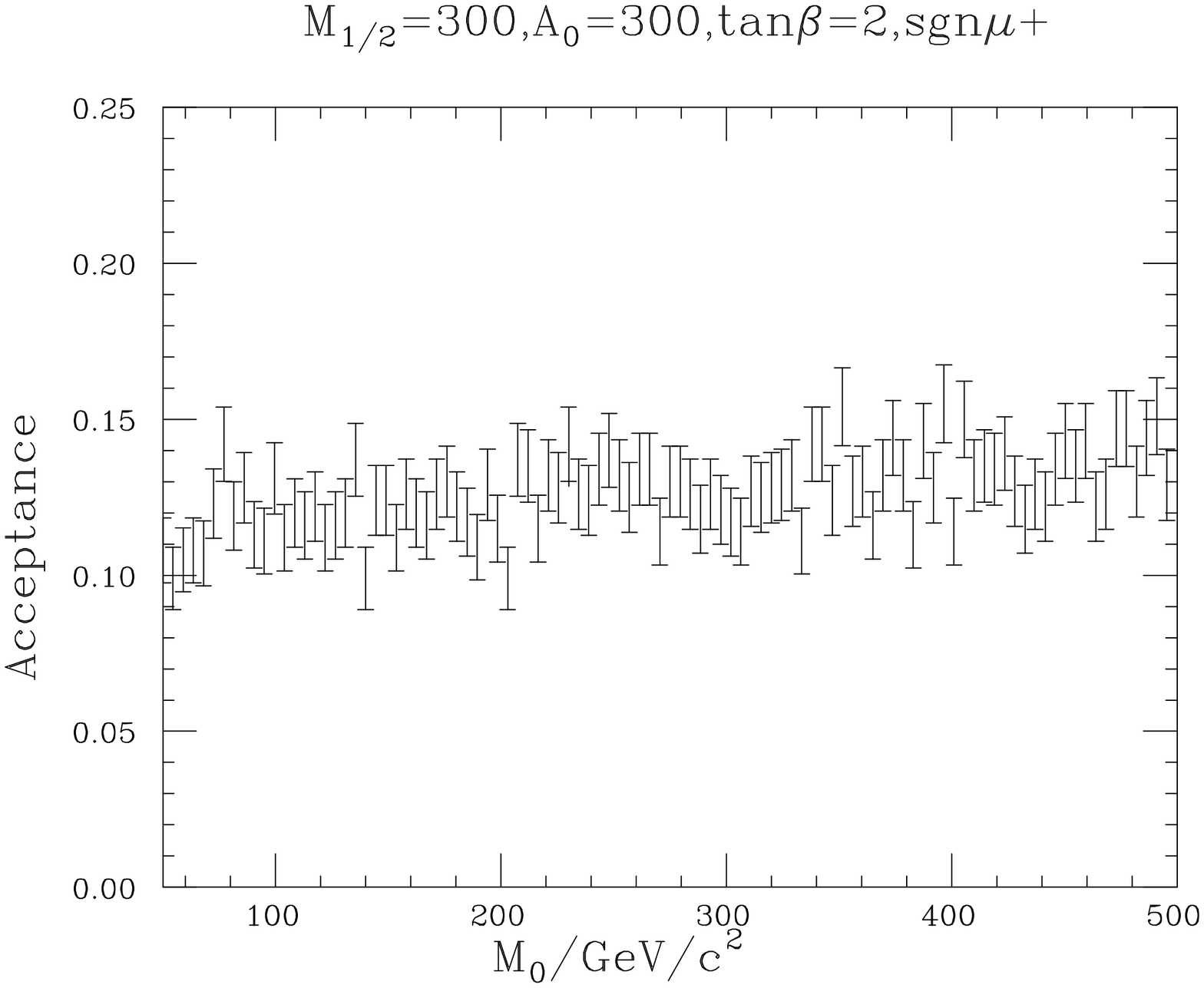}
\hfill
\includegraphics[angle=90,width=0.45\textwidth,viewport=50 70 550 
700,clip]{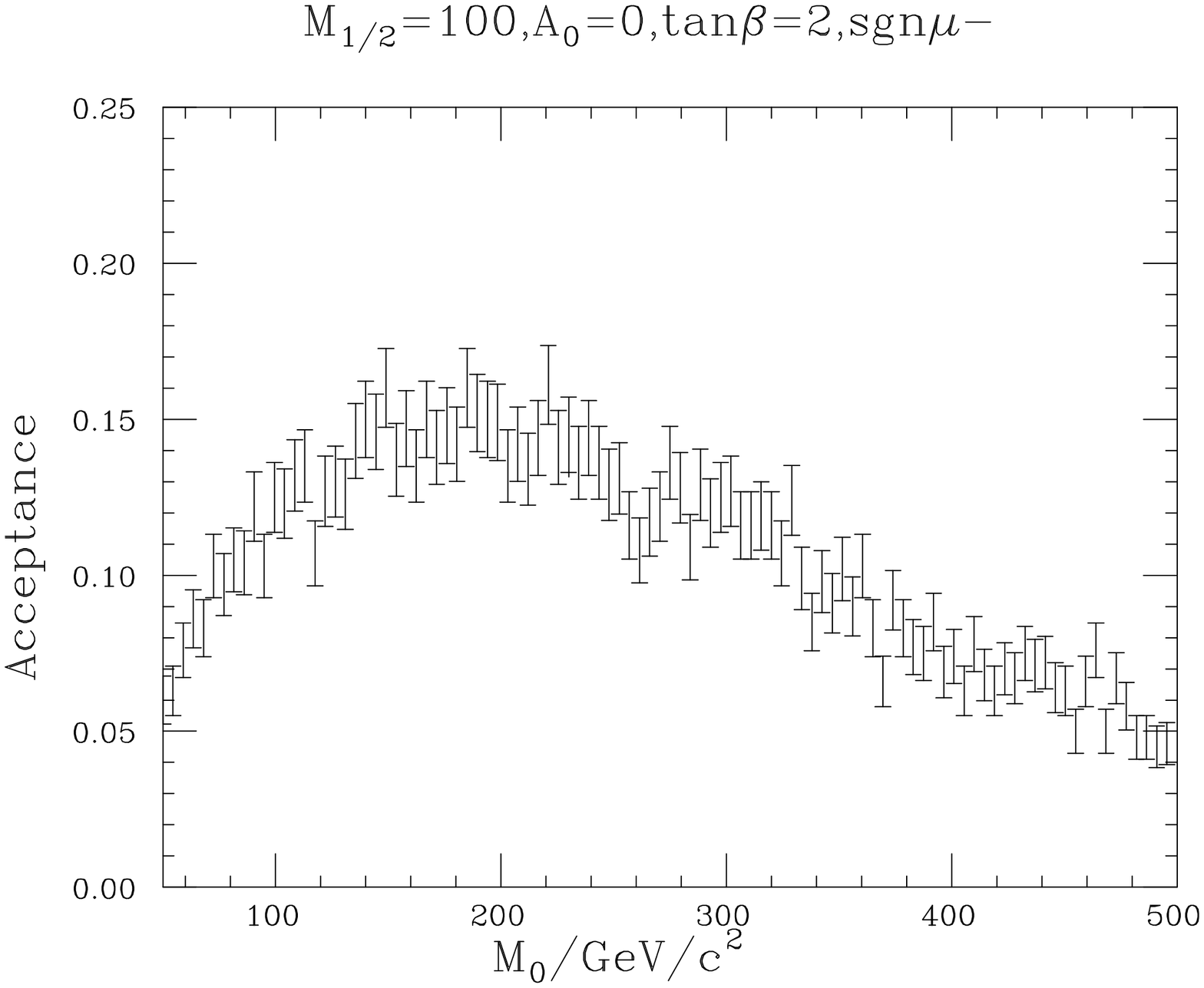}\\
\caption{Acceptance from Monte Carlo Simulation for two different
SUGRA points.\label{fig:eff}}
\end{figure}

\subsection{Backgrounds}
In the following we combine the backgrounds for both electrons and
muons. The main backgrounds to this like-sign dilepton signature 
are as follows
\begin{enumerate}
\item \underline{$\mathrm{b\bar{b}}$ production} followed by the production of
at least one $\mathrm{B^{0}_{d,s}}$ meson, which undergoes mixing. If the two
b-quarks in the event decay semi-leptonically this gives two like-sign
charged leptons.
\item \underline{$\mathrm{t\bar{t}}$ production} followed by 
$\mathrm{t\ra W^{+}b\ra
e^{+} \bar{\nu_{e}} b}$, and $\mathrm{\bar{t}\ra W^{-}\bar{b}\ra q\bar{q}\bar
{b} \ra q \bar{q} W^{+}\bar{c}\ra q\bar{q} e^{+}\bar{\nu_{e}}\bar{c}}$.
\item \underline{Single top production} ($s$ and $t$ channel) followed by
semi-leptonic decays of the top and the B-meson produced after
hadronization.
\item \underline{Non-physics backgrounds} from fake leptons and charge
misidentification. There are also backgrounds due to the production of
weak boson pairs, \ie WZ and ZZ, where at least one of the charged
leptons is not detected \cite{nachtman}. These require a full
simulation including the detector. We do not consider them here.
\end{enumerate}
\begin{figure}[t]
\includegraphics[angle=90,width=0.45\textwidth,viewport=50 70 550 
720,clip]{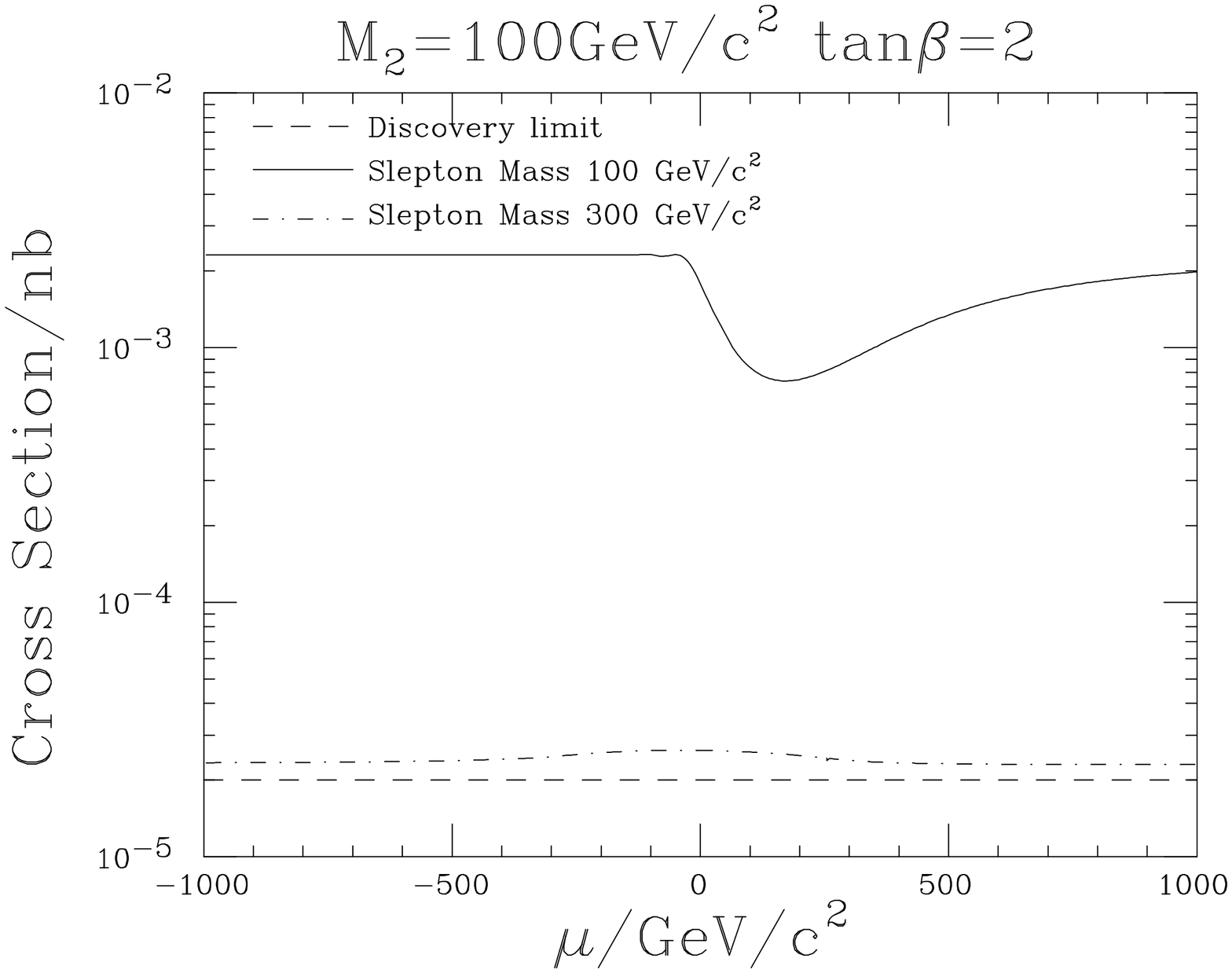}
\hfill
\includegraphics[angle=90,width=0.45\textwidth,viewport=50 70 550 
720,clip]{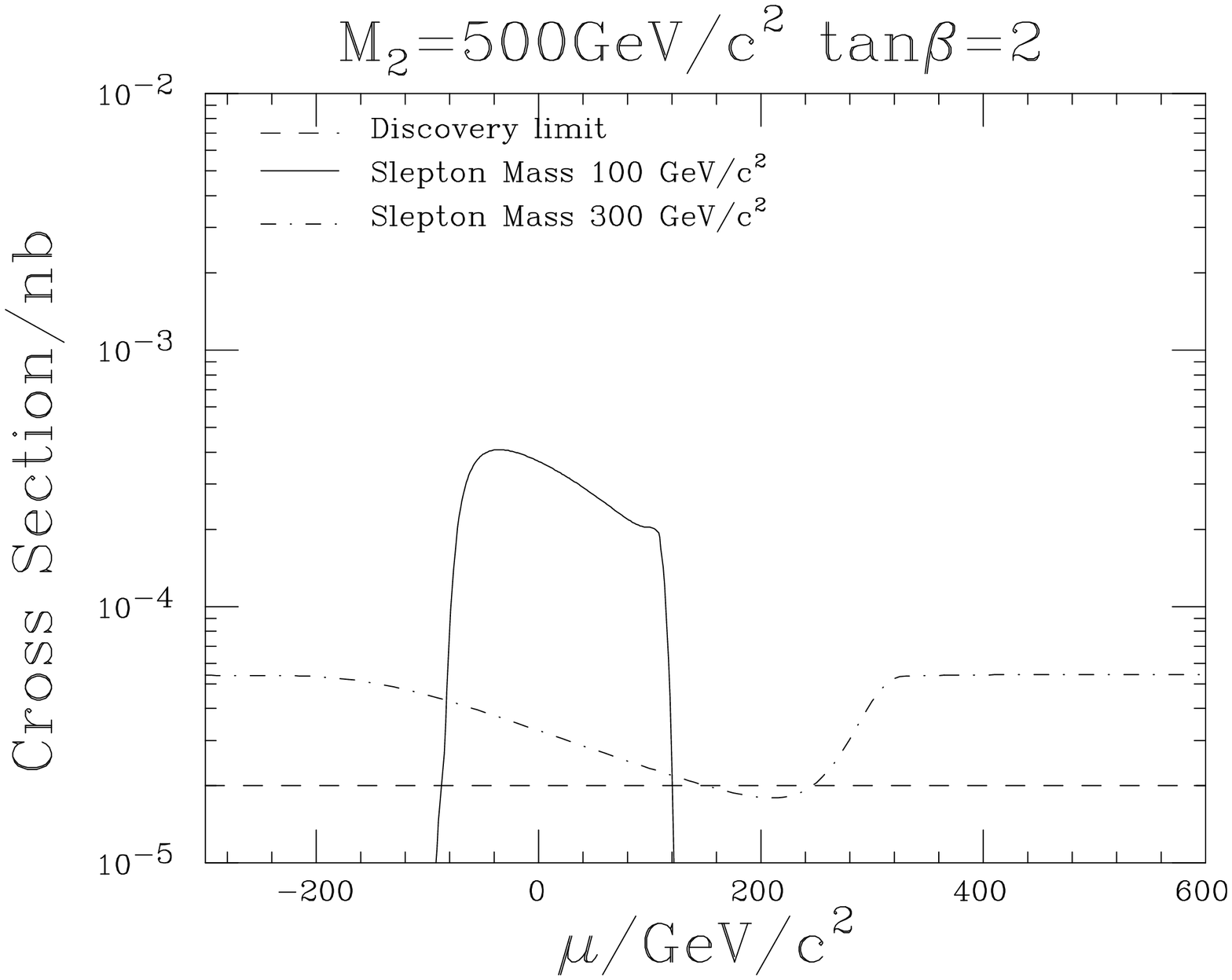}\\
\caption{Cross section for $\tan\beta=2$ and $\lam'_{211}=10^{-2}$. The
long-dashed horizontal line corresponds to the discovery limit of 4
signal events for $\mathrm{2\,fb^{-1}}$. The signal cross section after cuts is
given by the solid and dot-dashed curves for two slepton masses.
\label{fig:Xsect1}}
\end{figure}
\begin{figure}
\includegraphics[angle=90,width=0.45\textwidth,viewport=50 70 550 
720,clip]{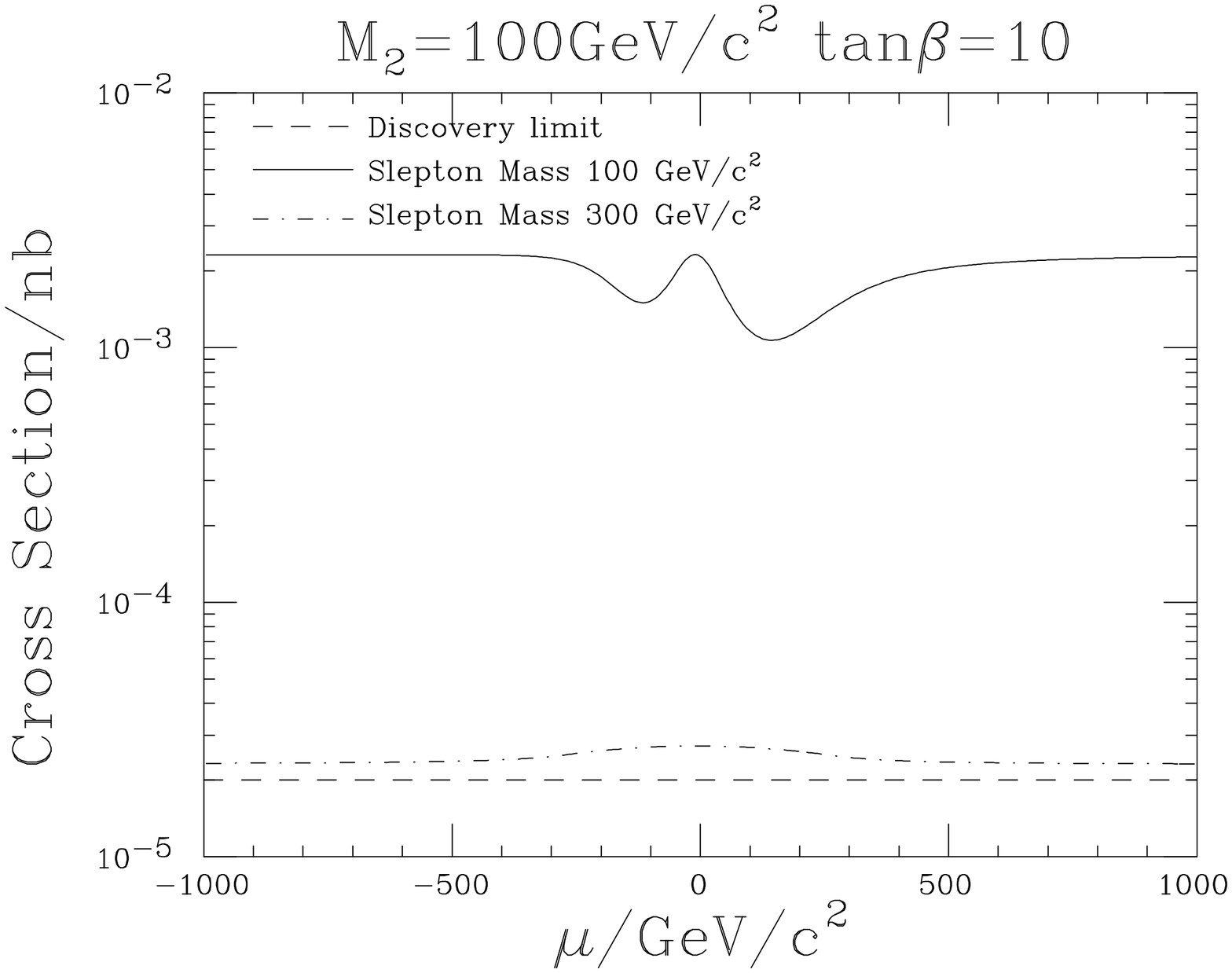}
\hfill
\includegraphics[angle=90,width=0.45\textwidth,viewport=50 70 550
 720,clip]{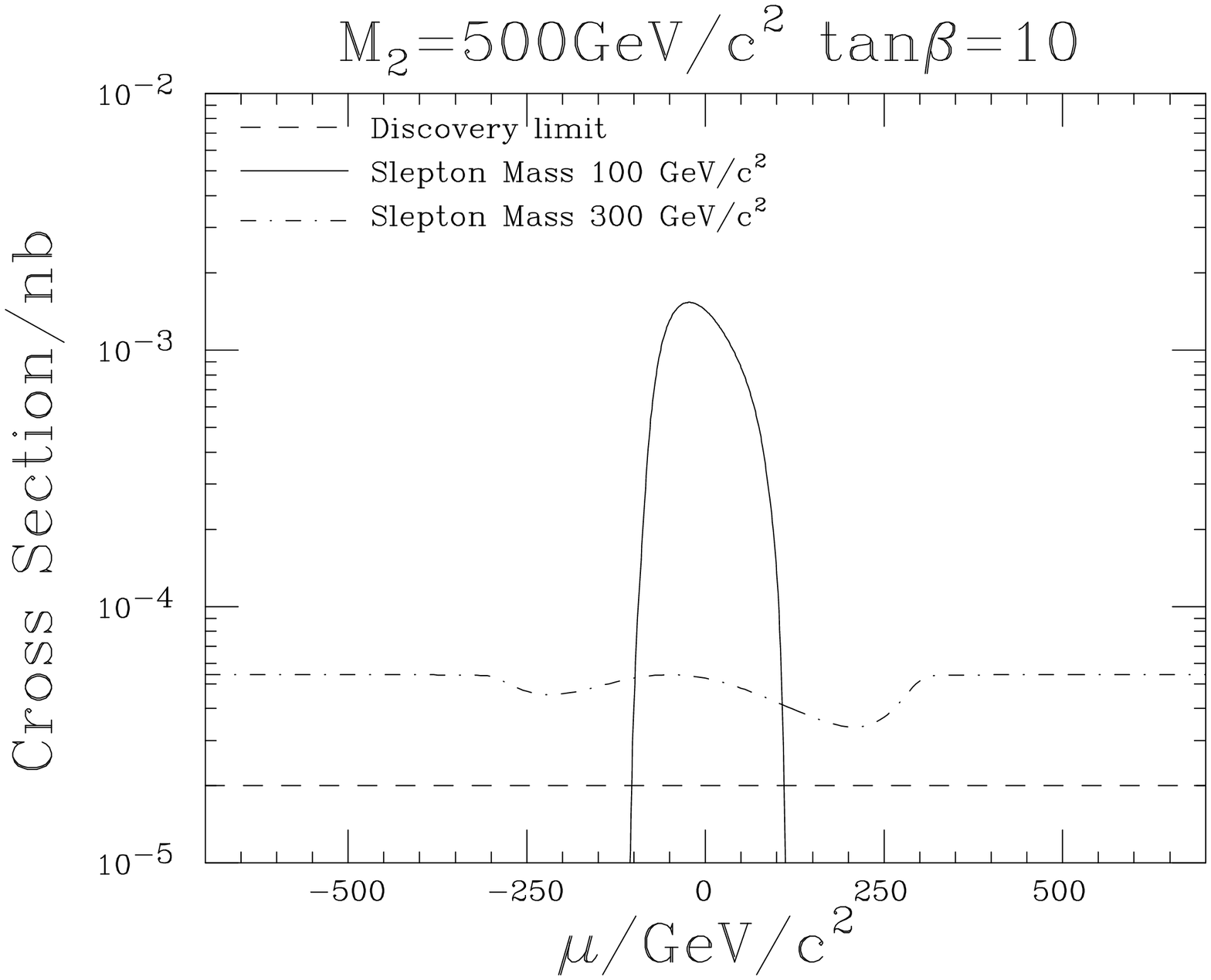}\\
\caption{Cross section for $\tan\beta=10$ and 
$\lam'_{211}=10^{-2}$.\label{fig:Xsect2}}
\end{figure}

We use HERWIG 6.0, \cite{HERWIGA,HERWIGB,HERWIGC}, to simulate these
background processes. The program includes the computation of the
supersymmetric spectrum and the MSSM decay branching ratios from the
ISASUSY program \cite{baertata}.  Due to the high cross section for
the production of $\mathrm{b \bar{b}}$ it was necessary to impose a
parton-level cut of $\mathrm{20\gev}$ on the $p_{\mathrm{T}}$ of the b
and $\mathrm{\bar{b}}$ to enable us to simulate a sufficient number of
events.  In Fig.  \ref{fig:parton} we show the distribution of events
(using the full Monte Carlo simulation) as a function of the
(parton-level) $p_{\mathrm{T}}$ of the bottom quark for two different
values of the lepton $p_{\mathrm{T}}$ cut. We did not simulate any
events for which the $p_{\mathrm{T}}$ of the bottom quark was below
$\mathrm{20\gev}$ since the cross section is too large. If we
extrapolate using Figs.\,\ref{fig:parton}a,\,b to lower b-quark
$p_{\mathrm{T}}$ we can see that for a lepton $p_{\mathrm{T}}$ cut of
$\mathrm{20\gev}$, Fig.\,\ref{fig:parton}b, our approximation should
be good, \ie we expect the area under the curve for $p_{\mathrm{T}}
\mathrm{(b)<20\gev}$ to be negligible. For $p_{\mathrm{T}}\mathrm{(
\ell)>15\gev}$, Fig.\,\ref{fig:parton}a, we would still expect a significant
number of events at $\mathrm{15\gev}<p_{\mathrm{T}}\mathrm{(b)<20\gev
}$. Besides the parton-level cut, we forced the B-mesons to decay
semi-leptonically.  This means we neglect the production of leptons
from the decay of charmed mesons which should also be a good
approximation as we expect the leptons produced from these decays to
be poorly isolated.

Figure\,\ref{fig:isocut} displays the effect of the lepton isolation
cut on the $\mathrm{b\bar{b}}$ background for two different
values. The effect of the isolation cut on the $\mathrm{t\bar{t}}$ and
single top backgrounds is shown in Fig.\,\ref{fig:isocut2}. As can be
seen in Figs.\,\ref{fig:isocut},\,\ref{fig:isocut2}, by imposing an
isolation cut of $\mathrm{5\gev}$ and a cut on the $p_{\mathrm{T}}$ of
the leptons of $\mathrm{20\gev}$ the background can be almost
eliminated.  Table\,\ref{table1} shows the backgrounds with a
$p_{\mathrm{T}}$ cut on the leptons of $\mathrm{20\gev}$ and an
isolation cut of $\mathrm{5\gev}$. We have used the leading-order
cross section for the $\mathrm{b\bar{b}}$ and single top backgrounds
and the next-to-leading order cross section, with next-to-leading-log
resummation, from \cite{cross} for the $\mathrm{t\bar{t}}$ cross
section. In both cases the error on the cross section is the effect of
varying the scale between half and twice the hard scale, and the error
on the number of events is the error in the cross section and the
statistical error from the simulation added in
quadrature. Realistically we cannot reduce these statistical errors
due to the large number of events we would need to simulate. We have
implemented the full hadronization using HERWIG 6.0.

With these cuts and using Poisson statistics, a $5\sigma$ fluctuation
of the total background corresponds to 4 events with an integrated
luminosity of $\mathrm{2\,fb^{-1}}$. Hence we consider 4 signal events to be
sufficient for a discovery of the new $\rpv$\ signal process.

\subsection{Signal}

To simulate the signal and the effect of the cuts, we modified HERWIG
6.0, \cite{HERWIGA,HERWIGB,HERWIGC}, to include the production
process, the MSSM decay of the slepton, and the $\rpv$\ decay of the
neutralino. The decay rate of the neutralino and its branching ratios
were calculated in the code and a matrix element for the neutralino
decay \cite{gondolo,morawitz} was implemented in the Monte Carlo simulation.

We use the program to estimate the acceptance of the signal process,
\ie the fraction of the like-sign dilepton events which pass the cuts
multiplied by the branching ratio to give a like-sign dilepton event.
Fig.\,\ref{fig:eff} shows the acceptance for two different SUGRA
points, with an isolation cut on the leptons of $\mathrm{5\gev}$ and a cut 
$p_{\mathrm{T}}\mathrm{
(\ell)>20\gev}$. As can be seen in Fig.\,\ref{fig:eff}b, the acceptance
drops in two regions. For lower values of $M_0$, the slepton is not
much heavier than the neutralino. The charged lepton from the decay of
the slepton is then quite soft and gets rejected by the $p_{\mathrm{T}}$ cut. 
For
large values of $M_0$ the slepton is much heavier than the
neutralino. The neutralino then gets a significant boost from the
slepton decay. The neutralino decay products are folded forward in
the direction of this boost causing the event to be rejected by the
lepton isolation cut.

To estimate the acceptance properly we need to run a scan of the
SUGRA parameter space using the Monte Carlo event generator. This
still remains to be done. To give some idea of what range of couplings
and masses we may be able to probe instead we assume an acceptance of
10\% using the same cuts as before. We can then estimate the range of
couplings which may be accessible.

As can be seen in Figs.\,\ref{fig:Xsect1},\,\ref{fig:Xsect2} the
production cross section for $\lam'_{211}=10^{-2}$ is sufficient to
produce a signal which is more than $5\sigma$ above the background for
large regions of the SUGRA parameter space. In some regions where the
neutralinos become Higgsino-like the cross section drops. The
cross section also drops as we approach the region where the
neutralino is heavier than the slepton and the resonance becomes
inaccessible.

We focused on the coupling $\lam'_{211}$ because the experimental
bound on $\lam'_{111}$ from neutrinoless double beta decay is very
strict, \cite{klapdor,herbi}, as discussed in Sect. \ref{sec:bounds}.
The bound on $\lam'_{111}$ weakens as the squark mass squared and for
squark masses above about $\mathrm{300\gev}$ (which we expect in the
SUGRA scenario for the heavier slepton masses)
$\lam'_{111}\approx10^{-2}$ is experimentally allowed and our analysis
thus applies for this case as well. $\lam'_{211}\approx10^{-2}$ is
well within the present experimental bounds \cite{herbi}.

In these figures we see that we are sensitive to slepton masses up
to $\mathrm{300\gev}$ for couplings of $10^{-2}$. The production cross section
scales with the square of the coupling. For slepton masses around
$\mathrm{100\gev}$, just above the LEP limits, we can thus probe couplings down
to about $2 \times 10^{-3}$.

\subsection{Conclusion}

We have performed an analysis of the physics background for like-sign
dilepton production at Run II and find that with an integrated
luminosity of $\mathrm{2\,fb^{-1}}$, a cut on the transverse momentum
of the leptons of $\mathrm{20\gev}$ and an isolation cut of
$\mathrm{5\gev}$ the background is $0.14 \pm 0.13$ events. This means
that 4 signal events would correspond to a $5\sigma$ discovery,
although in a full experimental analysis the non-physics backgrounds
must also be considered.

Using a full Monte Carlo simulation of the signal including a
calculation of the neutralino decay rate, its partial widths and a
matrix element in the simulation of the decay we found that the
acceptance for the signal varies but for a reasonable range of
parameter space is 10\% or greater.

When we then look at the cross section for the production of
$\tilde{\chi}^{0}\ell^{+}$ we find that we can probe $\rpv$\ couplings
of $2\times10^{-3}$ for slepton mass of $\mathrm{100\gev}$ and up to
slepton masses of $\mathrm{300\gev}$ for $\rpv$\ couplings of $10^{-2}$,
and higher masses if the coupling is larger.

\medskip

{\bf Acknowledgments:}
The work of P.R. is supported by PPARC, UK.

\section{Single-top-squark production via $UDD$ Couplings}
\label{sec:berger}
\centerline{{\large{\it E.L.~Berger, B. W. Harris, and Z. Sullivan}}}
\vspace{0.7cm}

\begin{figure}[t]
\begin{center}
\epsfxsize= 2.5in  %actual
\leavevmode
\epsfbox{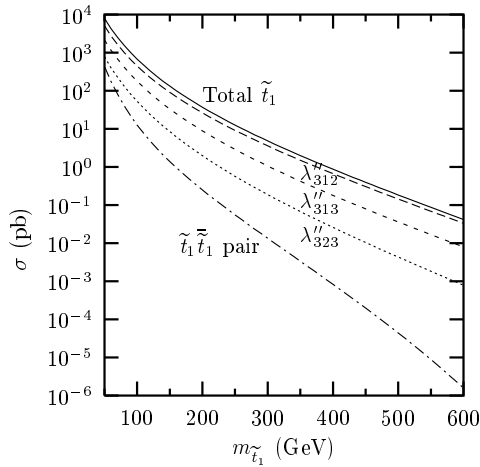}
\end{center}
\caption{Cross section for $R_p$-violating production of a single top-squark 
at Run II of the Tevatron ($\lambda^{\prime\prime}_{3ij}=0.1$) compared with 
the cross section for $R_p$-conserving production of pairs.  Both cross 
sections are computed at leading order.}
\label{bhsfig1}
\end{figure}

In supersymmetric extensions of the standard model, the bounds on
possible $R$-parity-violating couplings are relatively restrictive for
the first two generations of quarks and leptons, but much less so for
states of the third generation ({\it cf.} Sect.\ref{sec:bounds}.  If
$R_p$ is conserved superpartners must be produced in pairs.  In many
models, the squarks and gluinos are relatively heavy, and therefore
their pair production incurs a large phase space suppression at the
energy of the Tevatron. Here, the $s$-channel production of a
{\it{single}} squark through an $\rpv$-mechanism~\cite{esmail} is
considered.  Our motivation is that the greater phase space offsets
the reduced coupling strength in the production.  We focus on the
relatively light top squark $\widetilde{t}_1$ \cite{rudaz} and its
subsequent $R_p$-{\it conserving} decays~\cite{bhs}. The $R_p$
violation penalty is paid only once, in the initial production, and is
offset by the greater phase space relative to pair production.

We choose one clean $R_p$-conserving decay mode, $\widetilde{t}_1
\rightarrow b \widetilde{\chi}^+_1$, with $\widetilde {\chi}^+_1
\rightarrow l^+ + \nu +\widetilde {\chi}^0_1$.  Here, $l$ is an
electron or muon, and the $\widetilde {\chi}^+_1$ and $\widetilde
{\chi}^0_1$ are the chargino and lowest-mass neutralino states of the
MSSM.  For top-squark masses in the range of 180--325 GeV, we simulate
both the signal and standard model background processes and thereby
show that the top squark can be discovered, or the current bound on
the size of the $R_p$-violating coupling $\lambda^{\prime\prime}_{
3ij}$ can be reduced by up to one order of magnitude with existing
data and by two orders of magnitude at the forthcoming run~II of the
Fermilab Tevatron~\cite{bhs}.

In general it is possible to have $R_p$-violating contributions to the
MSSM superpotential of the baryon- or lepton-number violating
type. However, limits on the proton decay rate severely restrict their
simultaneous presence.  We therefore assume the existence of a
baryon-number-violating coupling only of the form~\cite{weinberg}
\begin{equation}
{\cal W}_{\not \! R_p} = \lambda_{ijk}^{\prime\prime}U_i^cD_j^cD_k^c \; .
\end{equation}
Here, $U^c_i$ and $D^c_i$ are right-handed-quark singlet chiral
superfields, $i,j,k$ are generation indices, and $c$ denotes charge
conjugation.  For production of a right-handed top squark via an
$s$-channel diagram $\bar d^j\bar d^k \rightarrow \widetilde{u}^i_R $
the relevant couplings are $\lambda^{\prime\prime}_{312}$,
$\lambda^{\prime\prime}_{313}$, and $\lambda^{\prime\prime}_{323}$.
The current best limits on these couplings come from experiments at
the CERN Large Electron Positron Collider (LEP).  Measurement of
$R_l$, the partial decay width to hadrons over the partial decay width
to leptons of the $Z$ boson, limits the values of $\lambda^{\prime
\prime}_{3ij}$ to be less than approximately 0.5 ({\it cf} Sect. 
\ref{sec:bounds}).

\begin{figure}[htbp]
\begin{center}
\epsfxsize= 2.5in  %actual
\leavevmode
\epsfbox{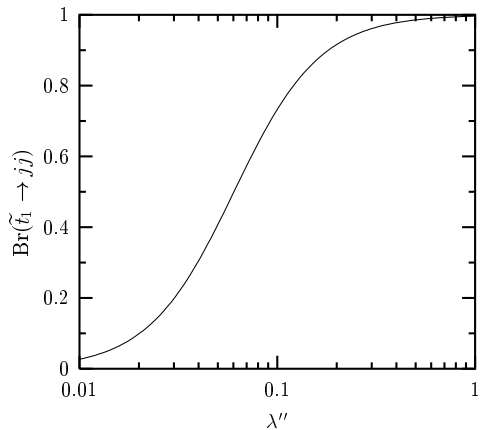}
\end{center}
\caption{Branching ratio for the top squark to decay into two jets
via the $R$-parity-violating coupling $\lambda^{\prime\prime}$
as a function of the coupling. The values of the SUGRA parameters are
given in the text.}
\label{bhsfig2}
\end{figure}

The colour- and spin-averaged partonic cross section for
$q q^{\prime} \rightarrow \overline{\widetilde{t}}_1$ is
\begin{equation}
\hat{\sigma} =
\frac{2\pi}{3} |\lambda^{\prime\prime}_{3ij}|^2
\frac{\sin^2\theta_{\widetilde{t}}}{m^2_{\widetilde{t}_1}} \delta(1 -
m^2_{\widetilde{t}_1}/\hat{s}) \; ,
\end{equation}
where $\sqrt{\hat{s}}$ is the partonic center of mass energy, and
$\theta_{\widetilde{t}}$ relates the right- and left-handed top-squark
interaction states to the mass eigenstates.  The hadronic cross
section depends on the following combinations of incident parton
distribution functions (PDF's): $d \otimes s$, $d \otimes b$, and $s
\otimes b$, where $d$, $s$, and $b$ denote the PDF's of the down,
strange, and bottom quarks, respectively.  We assume 
$\lambda^{\prime\prime}_{312} =
\lambda^{\prime\prime}_{313} = \lambda^{\prime\prime}_{323} \equiv
\lambda^{\prime\prime}$.  Our numerical results represent the sum of
$\widetilde{t}_1$ and $\overline{\widetilde{t}}_1$ production.

\begin{table}
\caption{Cuts used to simulate the acceptance of the detector at the
Tevatron run~II, and run~I (in parentheses if different).
The lepton veto ($E_{Tl}<45$ GeV) is optimized for small top-squark mass.
\label{bhsaccept}} 
\begin{center}
\begin{tabular}{ll} \hline\hline
$|\eta_b|<2$ (1) & $E_{Tb}>40$ GeV \\
$|\eta_l|<2.5$ (1.1) & $E_{Tl}>15$ GeV (20 GeV)\\
$|\eta_j|<2.5$ & $E_{Tj}>20$ GeV \\
$|\Delta R_{bj}|>0.7 $ & $|\Delta R_{jl}|>0.7$ \\
${\not \!E}_{T}>20$ GeV & $E_{Tl}<45$ GeV \\ \hline\hline
\end{tabular} 
\end{center} 
\end{table}

The mass dependences of the cross sections for single and
pair production of top squarks at the Fermilab Tevatron
($\sqrt{s}=2$ TeV) differ significantly, as shown in 
Fig.~\ref{bhsfig1}.  The curves are based on CTEQ4L leading order parton
distribution functions~\cite{CTEQ4} and $\lambda^{\prime\prime}=0.1$,
an order of magnitude below the current limit.  In fact,
$\lambda^{\prime\prime}$ can be smaller by an additional order of
magnitude and the single-top-squark cross section will still exceed
that for pair production if $m_{\widetilde{t}_1} > 100$ GeV.

\begin{figure}[htbp]
\begin{center}
\epsfxsize= 2.7in  %actual
\leavevmode
\epsfbox{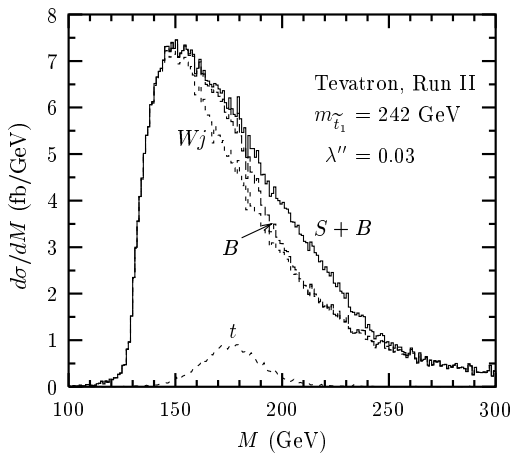}
\end{center}
\caption{The reconstructed-mass $M$ distribution for single-top-squark
production plus background ($S+B$) at the Tevatron ($\sqrt{s}=2$ TeV) for 
a top-squark mass $m_{\widetilde{t}_1}=242$ GeV.  
Also shown is a decomposition of the total background ($B$) into it's 
major components, all W jet modes ($Wj$) and single-top-quark production 
($t$).  The coupling $\lambda^{\prime\prime}=0.03$ produces the minimum 
signal for a $5\sigma$ significance at this mass.}
\label{bhsfig3}
\end{figure}

In the MSSM, the up-type squark $\widetilde u^k$ can decay into
charginos and neutralinos via the two-body processes $\widetilde
u^k\rightarrow d^k+\widetilde{\chi}^+_j$ ($j=1,2$) and $\widetilde
u^k\rightarrow u^k+\widetilde \chi^{0}_j$ ($j=1,2,3,4$).  Three body
modes are possible, including $\widetilde t_1 \rightarrow W^+ + b
+\widetilde \chi_1^{0}$, which is similar to decay into the top quark
(followed by top decay to $W^+ + b$) but softer; and $\widetilde t_1
\rightarrow c + \widetilde \chi^{0}_j$ via a flavor-changing loop
process.  The squark can also decay into a gluino plus a quark if
allowed kinematically.

With baryon-number-violating terms present, the top squark can also
decay into quark pairs via the $\lambda^{\prime\prime}$ couplings.
The branching fraction into two jets is shown in Fig.~\ref{bhsfig2}
for the specific SUGRA point discussed below. The branching ratio is
quite insensitive to the SUGRA parameters over a wide range. If
$\lam^{\prime \prime}$ is large, the decay to quark jets dominates.
However, as shown below, the $R_p$-conserving decay still produces a
measurable and useful cross section.

For the remainder of this study, we focus on the two-body decay
mode~\cite{rosshadro} $\widetilde{t}_1 \rightarrow b +\widetilde
{\chi}^+_1$, with $\widetilde \chi^{+} \rightarrow l + \nu +
\widetilde \chi_1^{0}$.  Here, $l$ denotes an electron or muon, which
usually comes from a $W$.  We expect $\widetilde \chi_1^{0}$ to
undergo an $R_p$-violating decay outside of the detector, so it is
treated as stable for our purposes.

A minimal supergravity model~\cite{sugra} is adopted to obtain the
relevant masses and decay branching fractions.  We begin with common
scalar and fermion masses of $m_0 = 100$ GeV and $m_{1/2} = 150$ GeV,
respectively, at the GUT scale.  We choose a trilinear coupling $A_0 =
-300$ GeV and the ratio of the Higgs vacuum expectation values
$\tan\beta = 4$.  The absolute value of the Higgs mass parameter $\mu$
is fixed by electroweak symmetry breaking and is assumed positive.
Superpartner masses and decay widths are calculated with
ISAJET~\cite{baertata}.  At the weak scale, $m_{\widetilde{t}_1}=$ 183
GeV, $m_{\widetilde {\chi}^0_1}=$ 55 GeV, and $m_{\widetilde
{\chi}^{\pm}_1}=$ 103 GeV.  In order to isolate the effects of the
$\rpv$ sector, we vary $m_0$ and keep the other supersymmetric
parameters fixed.  Since the gaugino masses depend primarily on the
choice of $m_{1/2}$, variation of $m_0$ allows us to vary
$m_{\widetilde t_1}$ without any appreciable change in the masses of
the decay products.

\begin{figure}[htbp]
\begin{center}
\epsfxsize= 2.7in  %actual
\leavevmode
\epsfbox{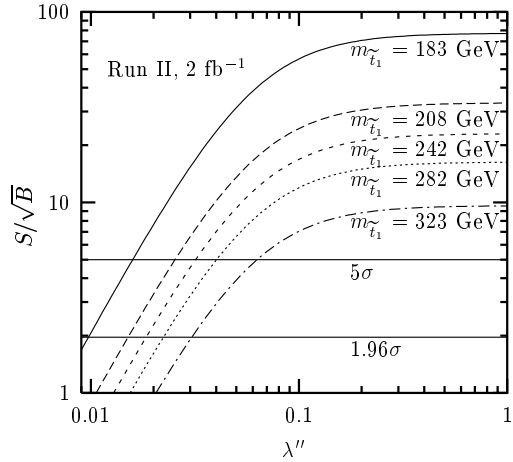}
\end{center}
\caption{Statistical significance of the single-top-squark signal
($S/\sqrt{B}$) in run~II of the Tevatron ($\sqrt{s}=2$ TeV, 2 fb$^{-1}$)
versus $\lambda^{\prime\prime}$ for a variety of top-squark masses.}
\label{bhsfig4}
\end{figure}

For the four-body decay mode of interest, the signal consists of a
tagged $b$-quark jet, a lepton, and missing transverse energy
associated with the unobserved neutrino and $\widetilde{\chi}_1^{0}$.
The dominant backgrounds, in order of importance, arise from
production and decay of the standard model processes $Wc$, in which a
$W$ is produced in association with a charm quark $c$ that is mistaken
for a $b$; $Wj$, in which a $W$ is produced in association with a
hadronic jet that mimics a $b$; $Wb\bar b$; $Wc\bar c$; and
single-top-quark production via $Wg$ fusion.  For these background
processes, we work with tree-level matrix elements obtained from
MADGRAPH~\cite{madgr} convolved with leading-order CTEQ4L
\cite{CTEQ4} parton distribution functions, at a hard scattering scale
$\mu^2 = \hat{s}$.  In an experimental analysis, the $Wj$ background
will be normalized by the data.  To simulate the resolution of the
hadron calorimeter, we smear the jet energies with a Gaussian whose
width is $\Delta E_j/E_j = 0.80/\sqrt{E_j}\oplus 0.05$ (added in
quadrature).

We simulate the acceptance of the detector by using the selections
listed in Table~\ref{bhsaccept}.  The assumed coverage in rapidity
for taggable $b$-quark jets and leptons is smaller for run~I than for
run~II.  However, the signal and background are similar in shape in
these variables, and thus $S/B$ is not sensitive to this cut.  The
lepton must be isolated from any jets, as defined by a cone of radius
$\Delta R$, or it is considered missed.  Similarly, an isolation cut
is used for the $b$-quark jet in order to help identify it.  We assume
a $b$-tagging efficiency of $60\%$ ($50\%$ for run~I) with a mistag
rate of $15\%$ for charm quarks and $0.5\%$ for light
quarks~\cite{tev2000}.  Missing momentum is identified with a {\it
massless} particle whose three-momentum balances that of the $b$-quark
and the lepton.

As expected from the primary decay, $\widetilde{t}_1 \rightarrow b
+\widetilde{\chi}^+_1$, the distribution in the transverse energy
$E_T$ of the $b$ quark is peaked sharply near the maximum value
allowed kinematically.  The spectrum of the background $b$ quark is
soft, and thus we impose a hard cut ($E_{Tb}>40$ GeV) on the minimum
$E_T$ of the $b$-quark jet.  For larger values of
$m_{\widetilde{t}_1}$ this cut would be optimized by raising the
minimum $E_{Tb}$.  The $b$-jet becomes too soft to be detected if
$m_{\widetilde {t}_1} \le m_{\widetilde {\chi}_1} + E_T^{\rm cut}$.
This contributes to a lower limit on $m_{\widetilde {t}_1}$ below
which our proposed search mode is not useful.

The background from single-top-{\it quark} production will produce a
peak in any mass reconstruction.  We utilize the fact that single
top-quarks are often produced with extra hard jets, and we impose a ``jet
veto''.  We require that there be no hard jets in the hadron 
calorimeter ($E_{Tj}>20$ GeV, $|\eta|<2.5$), beyond the one that is $b$-tagged.
After the jet veto, the remaining background is due
almost entirely to misidentified charm and light-quark jets from 
$Wc$ and $Wj$ production.  The transverse energy of the
lepton tends to be relatively soft for the signal at low 
$m_{\widetilde t_1}$, whereas it peaks around 40~GeV when it comes
from the $W$ in the background.  A cut to remove hard leptons, with
$E_{Tl}>45$ GeV, reduces the background by a factor of 2 with little
effect on the signal at low masses.  The final significance for the
signal at run~II is barely changed by this ``lepton veto'', but it is
especially helpful for the run~I data.

\begin{figure}[htbp]
\begin{center}
\epsfxsize= 2.7in  %actual
\leavevmode
\epsfbox{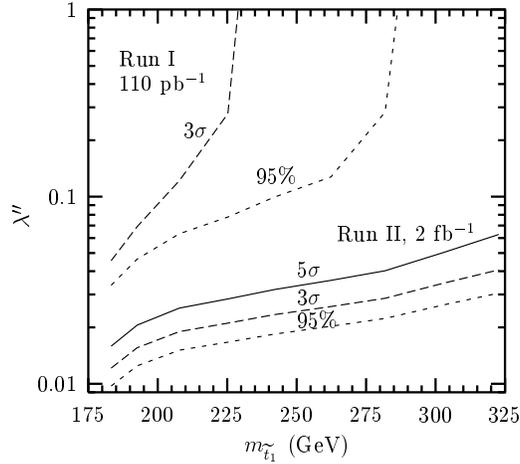}
\end{center}
\caption{Lower limits on discovery ($S/\sqrt{B} = 5$), evidence
($S/\sqrt{B} = 3$), and 95\% confidence-level exclusion ($S/\sqrt{B} =
1.96$) for $\lambda^{\prime\prime}$ versus top-squark mass in Run~I of
the Tevatron ($\sqrt{s}=1.8$ TeV, 110~pb$^{-1}$), and in Run~II
($\sqrt{S}=2$ TeV, 2~fb$^{-1}$).}
\label{bhsfig5}
\end{figure}

Shown in Fig.~\ref{bhsfig3} is an example of the signal and background.
For this case, $m_{\widetilde {t}_1} = 242$ GeV and
$\lambda^{\prime\prime} = 0.03$.  The mass variable is defined as
$M^2=(P_b + P_l + P_X)^2$ where the $P_b$ and $P_l$ are the
four-momenta of the $b$ and lepton.  The four-momentum $P_X$ is
defined such that its three-momentum balances that of the $b$ and
lepton, and $P_X^2 \equiv 0$.  The clearly discernible signal in
Fig.~\ref{bhsfig3} would constitute a discovery at the level of $5\sigma$
with an integrated luminosity of 2~fb$^{-1}$ at $\sqrt{s} = 2$ TeV.
When $m_{\widetilde{t}_1}$ is reduced to 183 GeV, the signal and
background spectra peak at about the same location, and sensitivity to
the signal begins to be lost.

Examination of the structure of the cross section involving $R_p$-{\it
conserving} decay modes reveals that as $\lambda^{\prime\prime}$
grows, the decrease in branching fraction is compensated by the
increase in cross section.  As depicted in Fig.~\ref{bhsfig4},
\begin{equation}
\sigma \propto \frac{|\lambda^{\prime\prime}|^2} 
{|\lambda^{\prime\prime}|^2 + f(R_p)} \;,
\end{equation}
where $f(R_P)$ is a constant times the branching fraction into
$R_p$-conserving modes.  As $\lambda^{\prime\prime} \to \infty$, the
cross section goes to a constant; whereas, when
$\lambda^{\prime\prime} \to 0$ the cross section decreases as
$|\lambda^{\prime\prime}|^2$.  The relationship $S/\sqrt{B} \propto
{|\lambda^{\prime\prime}}|^2/\sqrt{B}$, valid for small
$\lambda^{\prime\prime}$, implies a lower limit on the values of
$\lambda^{\prime\prime}$ that can be probed.  On the other hand, this
relationship highlights an insensitivity to variations in the estimate
of the background.

In Fig.~\ref{bhsfig5}, we show the reach in $\lambda^{\prime\prime}$ for
$180 < m_{\widetilde{t}_1} < 325$ GeV.  With an integrated luminosity
of 2~fb$^{-1}$ at $\sqrt{S} = 2$ TeV, discovery at the level of
$5\sigma$ is possible provided that $\lambda^{\prime\prime} >$
0.02--0.05.  Otherwise, a 95\% confidence level exclusion can be set
for $\lambda^{\prime\prime} >$ 0.01--0.03.  For the lower integrated
luminosity and energy of the existing Run~I data, values of
$\lambda^{\prime\prime} >$ 0.03--0.2 can be excluded at the 95\%
confidence level if $m_{\widetilde{t}_1}=$180--285 GeV.

We conclude that, as long as the $R_p$-conserving decay $\widetilde
{t}_1 \rightarrow b \widetilde {\chi}^+_1 \rightarrow \l \nu
\widetilde{\chi}^0_1$ is allowed, it should be possible to discover
the top squark or to reduce the lower limit on $\lambda^{\prime
\prime}$ by two orders of magnitude at Run~II of the Fermilab
Tevatron, for $180 < m_{\widetilde{t}_1} < 325$ GeV and
$\lambda^{\prime\prime}_{3ij} >$ 0.02--0.05.  Existing data from Run~I
of the Tevatron should allow a reduction of the limit on
$\lambda^{\prime\prime}$ by an order of magnitude.  With such a
reduction, one can establish that $R_p$-violating decay is unlikely
and rule out most of the possible influence of the top squark on
single-top-quark production and decay.

This work was supported by
the U.~S.~Department of Energy, High Energy Physics Division, under
Contract No. W-31-109-Eng-38.

%%%%%%%%%%%%%% Begin References %%%%%%%%%%%%%%%%%%%%%%%%%%%%%%%%%%%%%%%%

%%%%%%%%%%%%%% End of References %%%%%%%%%%%%%%%%%%%%%%%%%%%%%%%%%%%%%%%

\vspace{0.4cm}

\centerline{\Large{\bf PART 2: SParticle Pair Production}}

\section{Gluino and Squark Pair Production with Leptonic Decays}
\label{sec:chertok}
{\centerline{\large{\it M. Chertok}}}
\vspace{0.7cm}

\nn CDF has performed searches for squark and gluino production followed
by $\rpv$-decays using data taken with the Run I detector \cite{det}.
Upgrades of both the detector and Tevatron are currently underway
\cite{tdr}.  These will provide substantial enhancements for these and
other searches in Run II, scheduled to begin in 2000.  During this
run, CDF II will collect roughly 2 \ifb\ of data at $\sqrt{s}=2 \tev$,
corresponding to twenty times the present statistics.  The 10\%
increase in energy corresponds to a 40\% increase in the \ttbar\
yield, and similarly will aid new phenomena searches.

\subsection{$\rpv$ Decays via $LQ{D}^c$: Di-Lepton Events}

Events with a positron and a jet at high $Q^2$ values, detected at the
HERA experiments \cite{hera}, have sparked interest in $\rpv$ since
\cite{herarpv} such events can be explained by the production and
decay of a charm squark (\csquarkl): $\pelp + d \to \csquarkl \to
\pelp+ d$, where \rp\ is violated at both vertices.  For this
scenario, $\csquarkl$ with mass $\mcsl\simeq200 \; \gevcc$ is the
preferred squark flavor, because its associated coupling \lamp\
\cite{choud,choud_prd} is less constrained by experiment than the
others.

We test two $\rpv$\ processes that involve the same \lamp\ coupling 
\cite{cdfrpv}: 
\barr
\ppbar&\to&\gluino\gluino \to(c\,\csquarkl)\,(c\,\csquarkl)\ \Rightarrow
c\, (\pelpm d)\,c\,(\pelpm d) \lab{scharm}\\
\ppbar&\to&\ssb \to (q\none)\,(\bar{q}\none)\ \Rightarrow q\,(dc\pelpm)\,
\bar{q}\,(dc\pelpm)\lab{neutralino}
\earr 
The first we denote the ``charm squark analysis'' and the second the
``neutralino analysis''.  Here, the $\rpv$\ decays are indicated by
``$\Rightarrow$.''  For process \eq{scharm} we assume $\mgluino >
\mcsl=200 \; \gevcc $.  The masses of the other squarks are given in a
MSSM scenario in Ref. \cite{choud}.  For process \eq{neutralino}, we
consider \ssb\ production (5 degenerate squark flavors) and \ttbone\
production separately.  We also make the mass assumptions:
$M(\chione)\,>\, \msquark\,>\,\mchio$, $M(\chione) \approx 2\,\mchio$,
and $\mcone>\mstopo-M(b)$, where the first relation suppresses
$\squark \to \chione$, the second relation arises from gaugino mass
unification, and the third ensures that $Br(\stopo \to c\none)=100\%$
for $\mstopo < M(t)$.  Given the Majorana nature of the gluino and
neutralino, reactions \eq{scharm} and \eq{neutralino} yield like-sign
(LS) and opposite-sign (OS) dielectrons with equal probability.  Since
LS dilepton events have the benefit of small SM backgrounds, we search
for events with LS electrons plus two or more jets.

This analysis requires at least two electrons with \et\ $>$ 15 GeV in
the central electromagnetic calorimeter.  The $\eta$-$\phi$ separation
$\Delta R_{ee} \equiv \sqrt{(\Delta\phi)^2+(\Delta\eta)^2}$ between
two electrons must be greater than 0.4.  Each electron must pass an
isolation cut which requires the total calorimeter \et\ in an
$\eta$-$\phi$ cone of radius $\Delta R=0.4$ around the electron,
excluding the electron \et, to be less than 4 GeV.  Jets are
identified in the calorimeter using cone size $\Delta R = 0.7$ for
clustering and in the range $\modulus{\eta_j}<2.4$.  There must be at
least two jets with \et\ $>$ 15 GeV, $\Delta R_{jj}>0.7$, and $\Delta
R_{ej}>0.7$.  We further require no significant \met\ in the event:
$\met/\sqrt{\sum \et} < 5\ \gev^{1/2}$.  No LS $ee$ events survive our
selection.  The remaining 166 OS $ee$ events are consistent with
expected contributions from SM processes, notably Drell-Yan production
of dielectrons.

Event acceptances are calculated using Monte Carlo samples generated
with ISAJET, CTEQ3L parton distribution functions, 
and passed through the CDF detector simulation program.  
SM backgrounds for this search,
\ttbar\ and $\bbbar/\ccbar$ production, are small.  
We find the total background in 107 \ipb\ is consistent with zero events.  
We set limits on the cross section times branching ratio for the two
processes under study.  For the charm squark analysis, 
the event acceptance is a very weak function of \mgluino\ in the range
of 16.0\% to 16.6\%; we include a 10\% systematic
uncertainty (dominated by the uncertainty on the integrated
luminosity), and 
exclude $\sigma \cdot Br \geq 0.18$ pb independent of \mgluino.
In Figure~\ref{fig:rpv1} we 
plot the results from the charm squark analysis in
the gluino-squark mass plane.  Contours are shown for two
values of the branching ratio $Br(\csquarkl \to ed)$, where we have
compared our results to the NLO $\gluino\gluino$ production cross
section multiplied by the branching ratio to LS $ee$ from
Ref. \cite{choud_prd}. 
\begin{figure}[t]
\begin{minipage}[t]{.50\linewidth}
%\centering\epsfig{file=../eps/rvlim3.eps,width=\linewidth} 
\psfig{figure=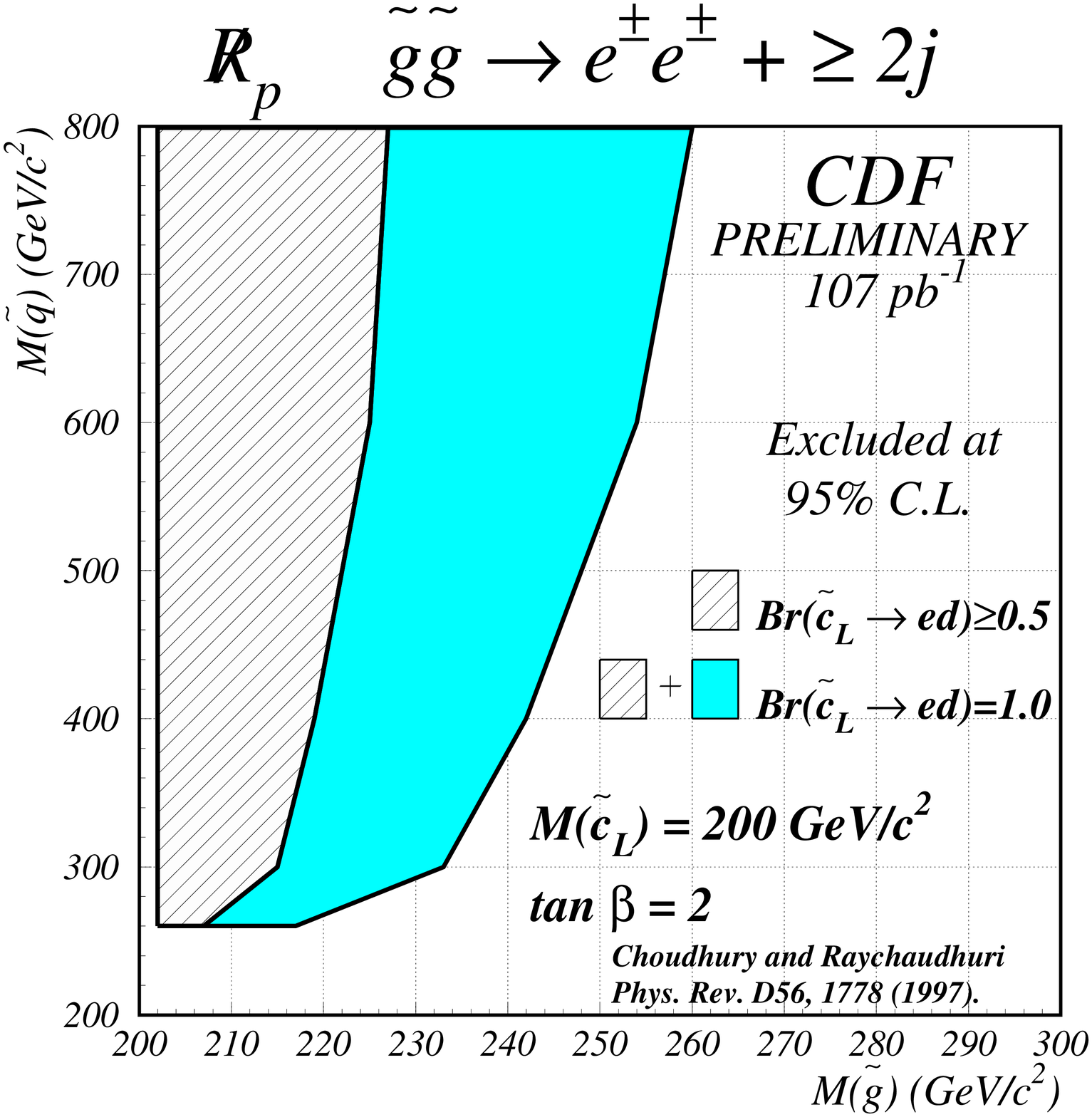,height=6cm,width=7cm,angle=0}
\end{minipage}\hfill
\begin{minipage}[t]{.50\linewidth}
%\centering\epsfig{file=../eps/lvlim1.eps,width=\linewidth} 
\psfig{figure=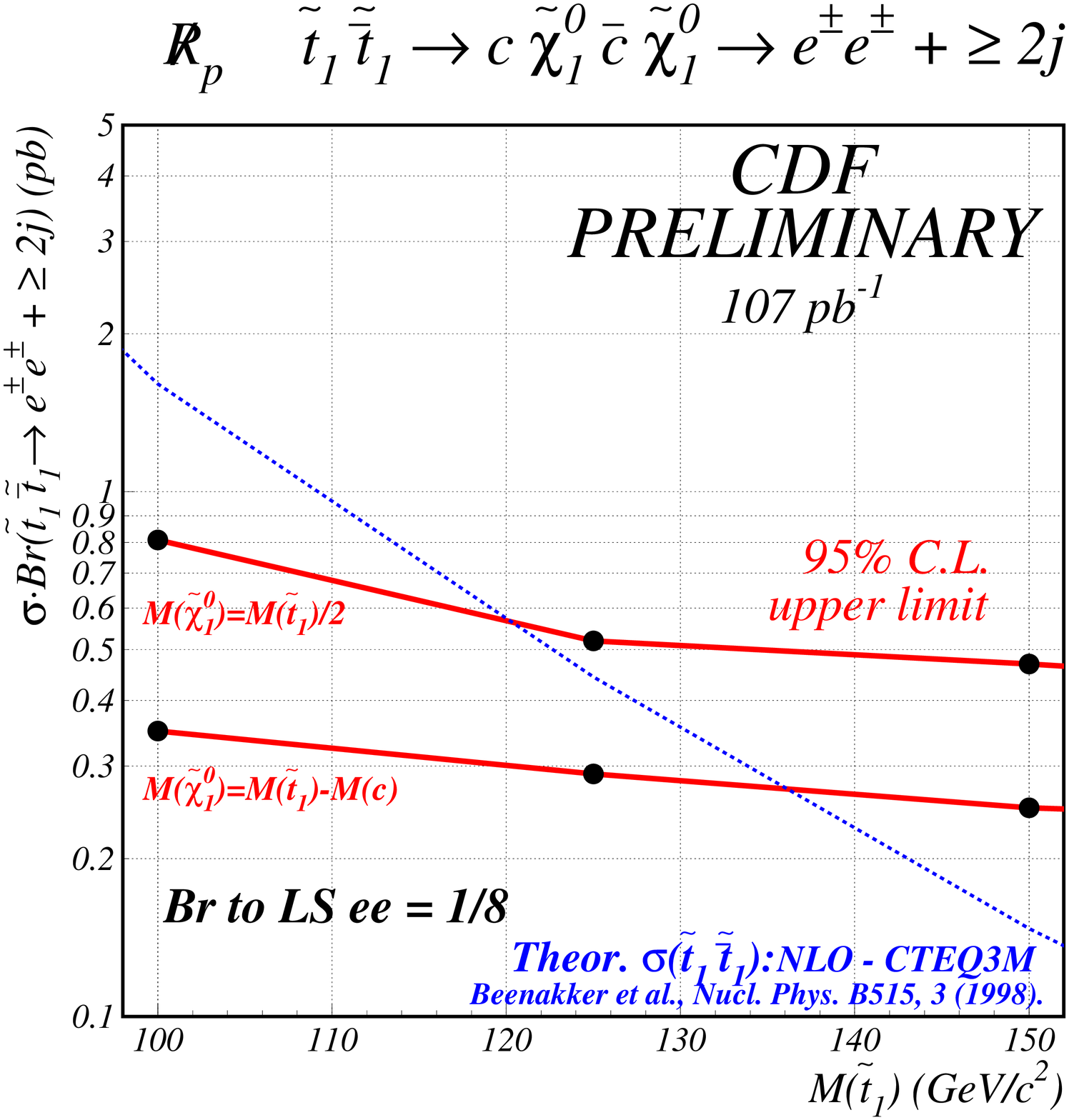,height=6cm,width=7cm,angle=0}
\end{minipage}
  \caption{Exclusion regions in the \msquark-\mgluino\ plane
        (left) and upper limits on cross section times branching ratio
                 for \ttbone\ production (right).}
  \label{fig:rpv1}
\end{figure}

For the neutralino analysis, the acceptance is determined for each
squark and neutralino mass pair and ranges from $3.7\%$ to $15.2\%$.
We obtain $\sigma\cdot Br$ limits which range as a function of the
squark mass from 0.81 pb to 0.26 pb for a light neutralino, and from
0.35 pb to 0.20 pb for a heavy neutralino.  Figure~\ref{fig:rpv1}
shows the results for the neutralino analysis for the case of \ttbone\
production.  Plotted are our 95\% C.L. upper limits along with the
cross section times branching ratio versus \mstopo\ from the NLO
prediction.  The branching ratio $Br(\stopo \to c\,\none)$ is taken to
be 1.0 We also assume $Br(\none\to q\qbar' e)=Br(\none\to
q\qbar'\nu)=1/2$, which need not be the case \cite{butterworth}.  Since
each neutralino decays to $e^+$ or $e^-$ with equal probability, the
branching ratio to LS \ee\ is 1/8.  This analysis excludes $\mstopo$
below 120 (135) \gevcc\ for a light (heavy) neutralino.  The results
for the neutralino analysis for the case of five degenerate \ssb\
production are displayed in Figure~\ref{fig:rpv2}.  Again, plotted is
our cross section times branching ratio limit for two neutralino
masses, along with the NLO prediction which includes a gluino mass
dependent $t$-channel contribution to the cross section.  Thus, we set
gluino and neutralino mass-dependent lower limits on the degenerate
squark mass in the range from 200 to 260 \gevcc.

We include a Run II projection for this search in
Figure~\ref{fig:rpv2} which folds in the increased cross section at
$\sqrt{s}=2 \tev$ along with the ratio of the integrated luminosities.
Assuming event selection and background levels similar to those in Run
I, this search will be sensitive to squark masses up to 380 \gevcc.
The addition of the muon channel will further enhance the reach, as
will the improved lepton identification in Run II.
\begin{figure}[t]
\begin{minipage}[t]{.50\linewidth}
%\centering\epsfig{file=../eps/lvlim2.eps,width=\linewidth} 
\psfig{figure=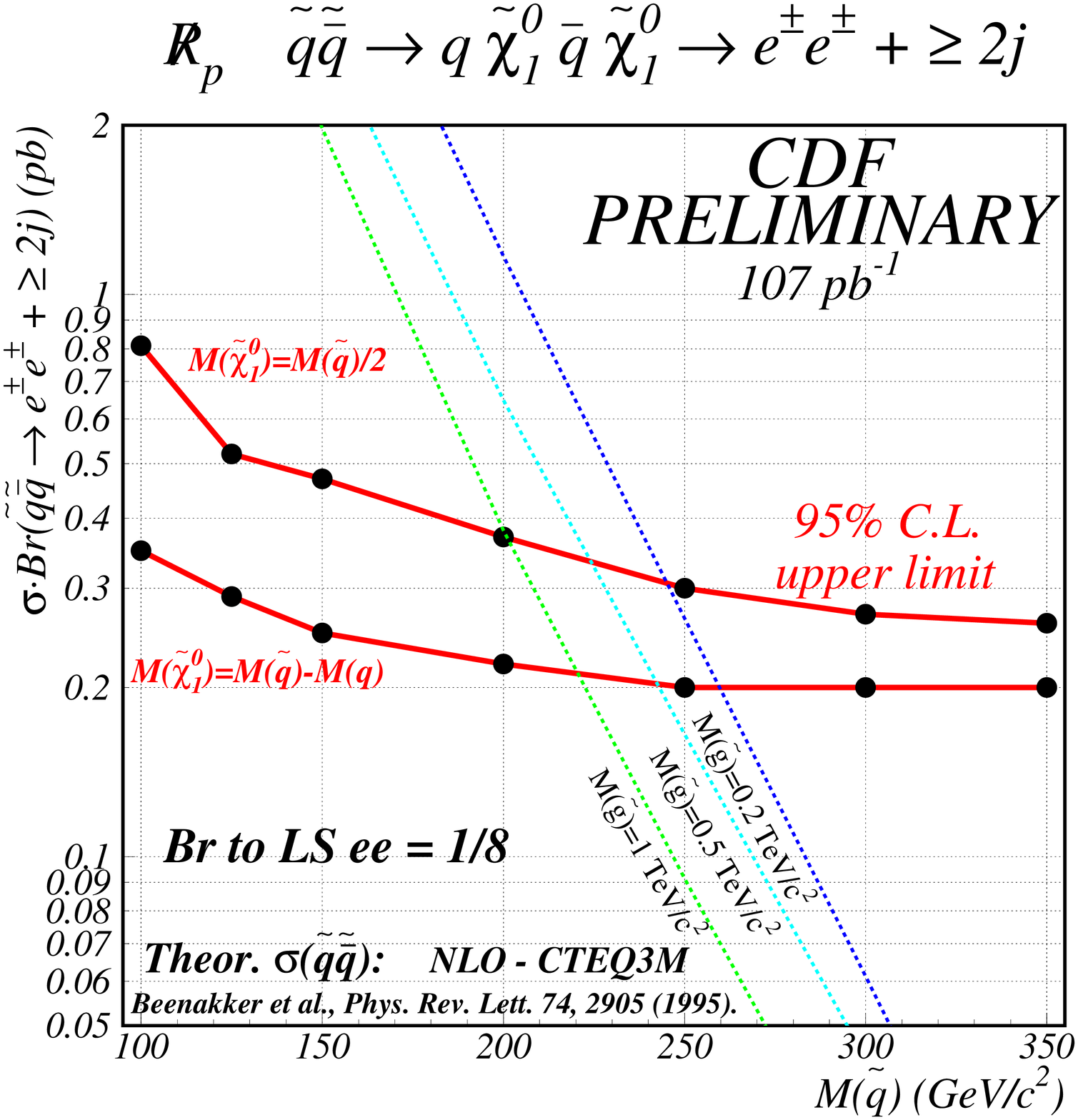,height=7cm,width=7.5cm,angle=0}
\end{minipage}\hfill
\begin{minipage}[t]{.50\linewidth}
%\centering\epsfig{file=../eps/r2lim.eps,width=\linewidth} 
\psfig{figure=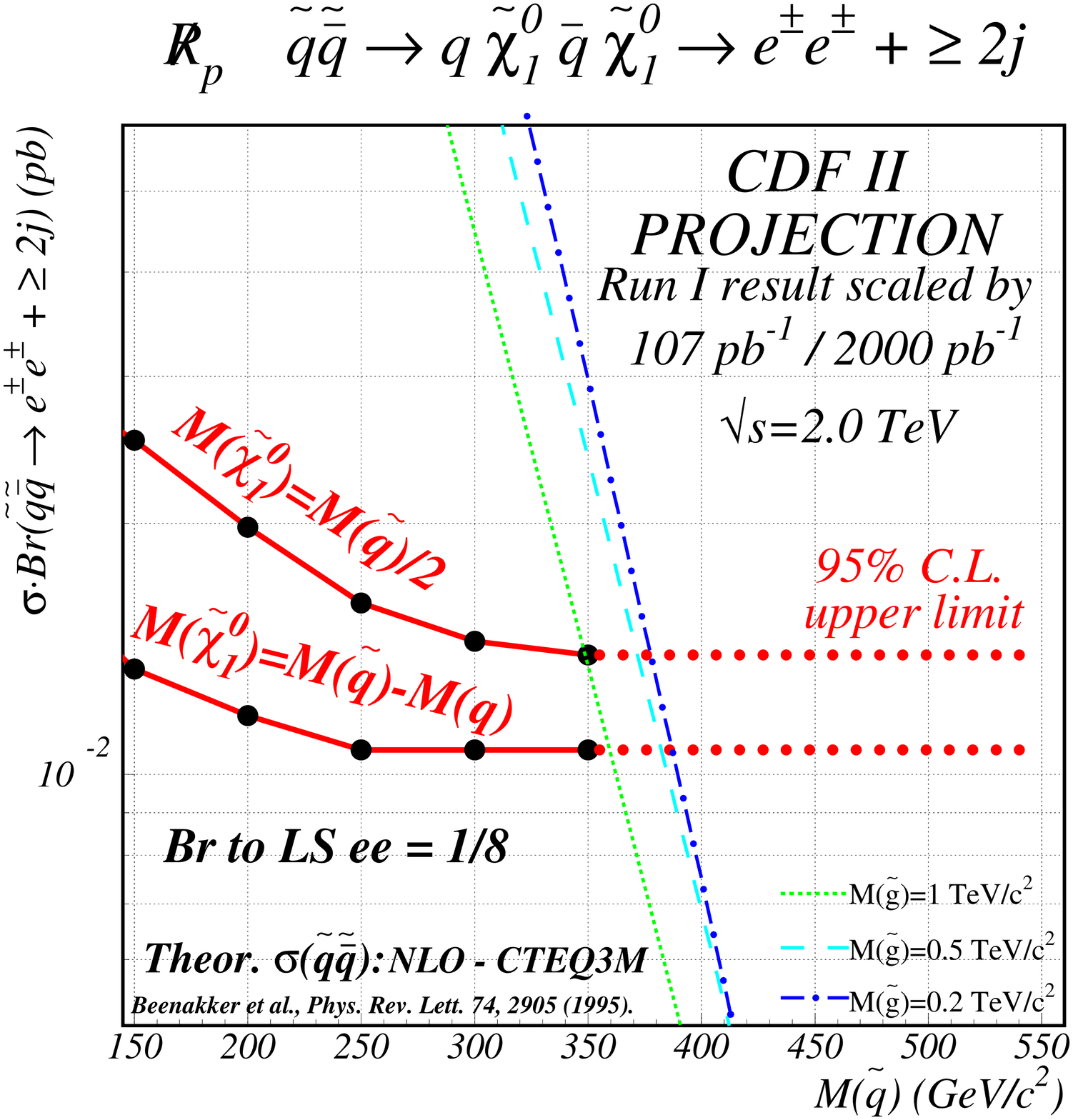,height=7cm,width=7.5cm,angle=0}
\end{minipage}
  \caption{Upper limits on cross section times branching ratio
for five-flavor degenerate \ssb\ production (left) 
and projection for Run II (right).
Also shown are the NLO cross sections
multiplied by a branching ratio to LS $ee$ of 1/8
with the assumption of $Br(\squark \to q \none)$ = 1.0.}
  \label{fig:rpv2}
\end{figure}

%%%%%%%%%%%%%%%%%%%%%%%%%%%%%%%%%%%%%%%%%%%%%%%%%%%%%%%%%%
%%%%%%%%%%%%%%%%%%%%%%%%%%%%%%%%%%%%%%%%%%%%%%%%%%%%%%%%%%
\subsection{$\rpv$ Decays via $LL{E}^c$: Multilepton Events}

\nn If we now assume one of the $LL{E}^c$ operators is non-zero, the
LSP decays as in Eq.\eq{lspdecays}a. For the case of
supersymmetric pair production as in Eq.\eq{prods}, followed by the
cascade decay to the LSP, we then expect events with extra
leptons. CDF has an analysis underway that examines $\lam_{121} \neq
0$ \cite{ccu}. As in \cite{baer}, all production processes (except
scalar top) are considered simultaneously.  The coupling $\lam_{121}$
allows two classes of processes: 1) direct decays of sleptons and
sneutrinos to leptons and neutrinos, and 2), decays of charginos and
neutralinos to leptons and neutrinos as in \eq{lspdecays}a. It is
further assumed that $\lam_{121}$ is large enough so that the $\rpv$\
decays are prompt, \ie within ${\cal O}(1 mm)$ of the interaction
point, but also that $\lam_{121}\ll 1$.  This tends to suppress both
1) and 2), except for decays of the LSP which have no competing MSSM
decays.  Thus, $\lsp \to \mu\bar{e}\nu_e,\;\bar{\mu}e\bar{\nu}_e, \;
e\bar{e}\nu_{\mu} \; e\bar{e}\bar{\nu}_\mu$ each with equal
probability.  Such events therefore contain four (or more) leptons and
\met.

Simulation of the signal is performed using ISAJET
within a SUGRA framework.  The mass parameters \mz\ and \mo\ are
input, while the others are fixed: $A_0=0,\;\tan\beta=2,\;{\rm sgn}\mu=-1$.
The event selection requires four leptons, $\ell=(e,\mu)$, where one lepton
satisfies 
$\et(e)>12\gev,\;|\eta_e|<1.1$ or 
$\pt(\mu)>12\gevc,\;|\eta_\mu|<0.6$,  
while the others must satisfy 
$\et(e)>5\gev,\;|\eta_e|<1.1$ or 
$\pt(\mu)>5\gevc,\;|\eta_\mu|<1.0$.
Furthermore, the leptons must be
separated by $\Delta R>0.4$.  Neither lepton isolation 
nor \met\ requirements are imposed.
SM processes that can yield four lepton events include \ttbar,
\bbbar/\ccbar, and $\zzero\zzero$ production and decay, although the rates are
very low.  Another background under study is that of processes
containing three real leptons plus an additional fake lepton.

Projections for this search in Run II are presented in
Table~\ref{tab:armin}, which folds in the approximate factor of twenty in
statistics and the increase in $\sqrt{s}$.  The background
increase to ${\cal O}(10)$ events in Run II could be controlled by
additional cuts on \met, jets, or lepton isolation.  
Improvements to the detector, for
example the upgraded plug EM calorimeter, will improve the lepton
identification efficiency.  Such contributions are not included in
this estimate.
\begin{table*}
\caption{Multilepton $\rpv$\ search projections for Run II.
The efficiencies are based on the Run I detector simulation.}
\label{tab:armin}
\begin{tabular}{c|c||c|c|c}
\mz\    & \mo\    & $\sigma$ & Efficiency & Four-Lepton\\
(\gevcc)& (\gevcc) & (pb)   &          & Events \\
\hline
50      &190    &0.170         &0.19      &64\\
100     &190    &0.170         &0.12      &41\\
150     &190    &0.166         &0.10      &34\\
200     &190    &0.166         &0.09      &31\\
200     &200    &0.128         &0.10      &25\\
200     &210    &0.102         &0.10      &21\\
200     &220    &0.078         &0.10      &16\\
200     &230    &0.061         &0.10      &12\\
200     &240    &0.049         &0.10      &10\\
\end{tabular}
\end{table*}

%%%%%%%%%%%%%%%%%%%%%%%%%%%%%%%%%%%%%%%%%%%%%%%%%%%%%%%%
%%%%%%%%%%%%%%%%%%%%%%%%%%%%%%%%%%%%%%%%%%%%%%%%%%%%%%%%

\pssilent
\setcounter{topnumber}{5}
\def\topfraction{1.0}
\setcounter{bottomnumber}{5}
\def\bottomfraction{1.0}
\setcounter{totalnumber}{5}
\def\textfraction{0.0}
\def\floatpagefraction{1.0}
\def\met{\mbox{${\hbox{$E$\kern-0.6em\lower-.1ex\hbox{/}}}_T$}} %missing ET
\def\MET{\met}
\def\etal{{\it et al.\/}}

\newpage

\section{D0 Search for SUSY Pair Production followed by
$LQD^c$ LSP Decays}
\label{sec:banerjee}
{\centerline {\large{\it{S.~Banerjee, N.K.~Mondal, V.S.~Narasimham, 
N.~Parua}}}}
\vspace{0.7cm}

\subsection{Physics motivation}
Recent interest in R-parity violating (RPV) SUSY decay modes is
motivated by the possible high-$Q^2$ event excess at HERA~\cite{hera}.
When interpretation of the excess through first-generation leptoquarks
was excluded by the D\O~\cite{D0-lq} and CDF~\cite{CDF-lq}
experiments, it was suggested~\cite{herarpv} that such an effect
could be explained via the $s$-channel production of a charm or top
squark decaying into the $e + jet$ final state. Both the production
and the decay vertices would thereby violate R-parity. Although more
recent data has not confirmed the previous event excess, interest in
RPV signatures has not abated.  The CDF and D\O\ Collaborations have
recently performed searches for RPV SUSY~\cite{cdfrpv,d0rpv}, and
have set new mass limits on the RPV SUSY particles. Both experiments
focussed their searches on the $\lambda'$ couplings, as motivated by
the high-$Q^2$ HERA event excess. The results of the D\O\ searches are
here extended to the Run~II case and the expected sensitivity to the RPV
couplings is discussed.
\begin{figure}[htb]
\vspace*{-0.3in}
\centerline{\protect\psfig{figure=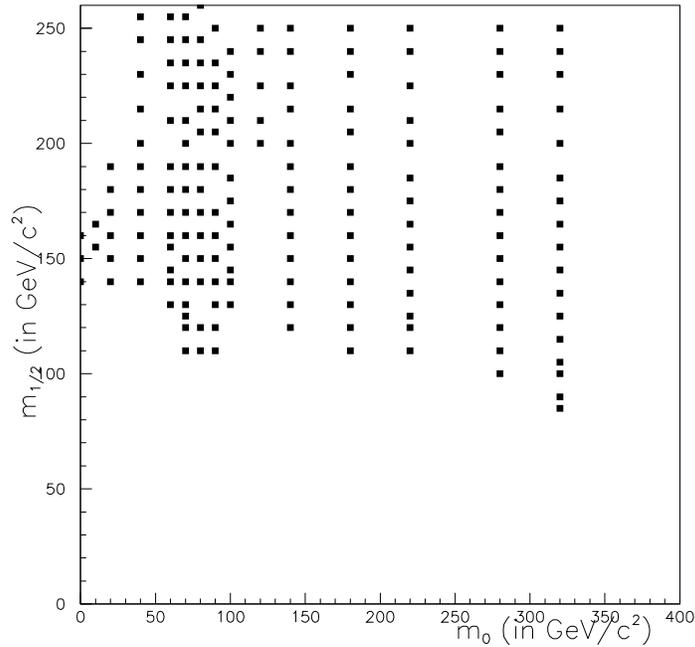,width=4in}}
\caption{Points in the $(m_0,m_{1/2})$ SUGRA parameter space used to 
generate RPV events in the $ee + 4$~jets channel.}
\label{fig:RPV-points}
\end{figure}
\begin{table}[htb!]
\begin{center}
\caption{Efficiency $\times$ BR (\%), signal cross section and the
expected event yield in 2~$fb^{-1}$ of data, at various
$(m_0,m_{1/2})$ parameter space points. }
\begin{tabular}{||c|c|c|c|c||}
\hline
$m_0$ & $m_{1/2}$ & Efficiency $\times$ BR (\%)
      &  Cross section &  $\langle N \rangle$        \\
 (GeV) & (GeV) & & (pb) & (in 2 $fb^{-1}$)           \\
\hline
60    & 235 & $7.9\pm 1.1 $ & 0.16 & $25.2 \pm 3.4 $ \\
60    & 245 & $8.3\pm 1.1 $ & 0.08 & $12.8 \pm 1.7 $ \\
60    & 255 & $8.3\pm 1.1 $ & 0.06 & $10.5 \pm 1.4 $ \\
100   & 220 & $6.1\pm 0.8 $ & 0.10 & $12.2 \pm 1.7 $ \\
100   & 230 & $7.0\pm 1.0 $ & 0.08 & $11.3 \pm 1.5 $ \\
180   & 240 & $7.0\pm 0.9 $ & 0.05 & $7.1  \pm 1.0 $ \\
320   & 240 & $7.1\pm 0.9 $ & 0.05 & $6.9  \pm 1.0 $ \\
%\hline
\end{tabular}
\end{center}
\label{table:RPV-eff}
\end{table}
\begin{table}[hbt!]
\begin{center}
\caption{Expected lower limits on the squark and gluino masses for the
case of no SUSY detection at Run 2.}
\begin{tabular}{||c|c|c|c||}
\hline
~ & Lower limit on $m_{\tilde{q}}$ & Lower limit on $m_{\tilde{g}}$ & 
Limit when
                                       $m_{\tilde{q}} = m_{\tilde{g}}$ \\
~ & ( For any $m_{\tilde{g}}$)  & ( For any $m_{\tilde{q}}$ ) & \\\hline
\multicolumn{4}{||c||}{Electrons}\\
\hline
Run 1 & 252 GeV & 232 GeV & 283 GeV \\
Run 2 (Scenario I) & 430 GeV & 490 GeV & 490 GeV \\
Run 2 (Scenario II) & 520 GeV & 575 GeV & 585 GeV \\\hline
\multicolumn{4}{||c||}{Muons}\\\hline
Run 2 & 560 GeV & 640 GeV & 665 GeV \\
%\hline
\end{tabular}
\end{center}
\label{table:RPV-limits}
\end{table}

\subsection{D\O\ Search for RPV neutralino decays}

The D\O\ search for RPV SUSY considered the case of neutralino LSP
which decays into a lepton and two quarks due to a finite RPV
$\lambda'$ coupling (see Fig.~\ref{fig:decay}). Both the electron and
muon decay channels were considered, corresponding to what is commonly
referred to as $\lambda'_{1ij}$ and $\lambda'_{2ij}$ couplings,
respectively. The corresponding final states contain either $2e $ or
$2\mu $ and at least four accompanying jets. Unlike at HERA, this
search is not sensitive to the value of the RPV coupling, as long as
it is large enough so that the neutralino decays within the D\O\
detector. That corresponds to $\lambda' \geq 10^{-3}$, which gives a
lot of room, given current indirect constraints (see Sect.
\ref{sec:bounds}).

%\begin{figure}[thb]
%\centerline{\protect\psfig{figure=rpv_decay.eps,height=1.8in}}
%\bigskip
%\caption{RPV decay of a neutralino LSP into a lepton and two quarks.}
%\label{fig:RPV-decay}
%\end{figure}

We assume that the neutralino (LSP) pairs are produced in cascade
decays of other supersymmetric particles and use all SUSY pair
production mechanisms when generating signal events.

Signal events were generated within the SUGRA framework with the
following values of SUSY parameters: $A_0 = 0$, $\mu < 0$ and
$\tan\beta = 2$ (the results are not sensitive to the value of $A_0$
.) Center of mass energy of the colliding beams was taken to be 2
TeV. {\footnotesize ISAJET}~\cite{baertata} was used for event
generation. The acceptance and resolution of the D\O\ detector were
parameterized using the following resolutions: $\delta E/E = 2 \%
\oplus 15\%/\sqrt{E}~\mbox{[GeV]}$ (electrons), $\delta (1/p)/(1/p) =
0.018 \oplus 0.008 (1/p)$ (muons), and $\delta E/E = 3\% \oplus
80\%/\sqrt{E}~\mbox{[GeV]}$ (jets) and found consistent with the full
detector simulation based on {\footnotesize GEANT}~\cite{GEANT}.

Figure~\ref{fig:RPV-points} shows the points in the $(m_0,m_{1/2})$
SUGRA parameter space where signal Monte Carlo events were generated
for the electron channel. Similar points were studied for the
muon-decay channel.

\subsection{Selection criteria for the dielectron channel}

A multijet trigger was used for the analysis of Run~1 data. It was
found to be nearly 100\% efficient for the typical RPV signal. Since
the Run~2 trigger list will include a similar trigger, we assume
trigger efficiency of 100\% and do not perform any trigger simulations
for the Run~2 analysis.

The following offline selections  were used: 
\begin{itemize}
\item 
At least two good electrons, the leading one with $E_T(e) > 15$ GeV
and the other one with $E_T(e) > 10$ GeV;
\item 
Rapidity range $\mid \eta \mid \leq 1.1$ (central calorimeter), or
$1.5 \leq \mid \eta \mid \leq 2.5$ (end calorimeters) for all the
electrons;
\item
Energy isolation for the electrons: the EM energy in the R=0.2 cone
about the center of gravity of the EM cluster, subtracted from the
total energy in R=0.4 cone, should not exceed 15\% of the EM energy in
the R-0.2 cone.
\item
At least four jets with $E_T(j) > 15 $ GeV and $\mid \eta \mid < 2.5$;
\item
The dielectron invariant mass ($M_{ee}$) should not be in the Z-mass
interval, ie, $ \mid M_{ee} - M_Z \mid > 15$ GeV/$c^2$.
\end{itemize}

In the present analysis we have dropped the requirement on $H_T = \sum
E_T(e) + \sum E_T(j)$ , but retained all other offline criteria that
were used in the previous analysis of data from Run~I~\cite{d0rpv}.

\subsection{Selection in the dimuon channel}

The following event selection requirements were used for the muon
decay channel:

\begin{itemize}
\item Two muons, the leading one with $p_T >$ 15 GeV, and the other
one with $p_T >$ 10 GeV.
\item Rapidity range $|\eta| < 2.3$ for both muons.
\item Energy isolation requirement for both muons, i.e. the
calorimeter energy accompanying the muon in a ($\eta$ $\phi$) cone of
0.4 should be consistent with that from a minimum ionising particle.
\item At least four jets with $E_T(j) > 15$ GeV and $|\eta| < 2.5$;
\end{itemize}
\begin{figure}[t]
\centerline{\protect\psfig{figure=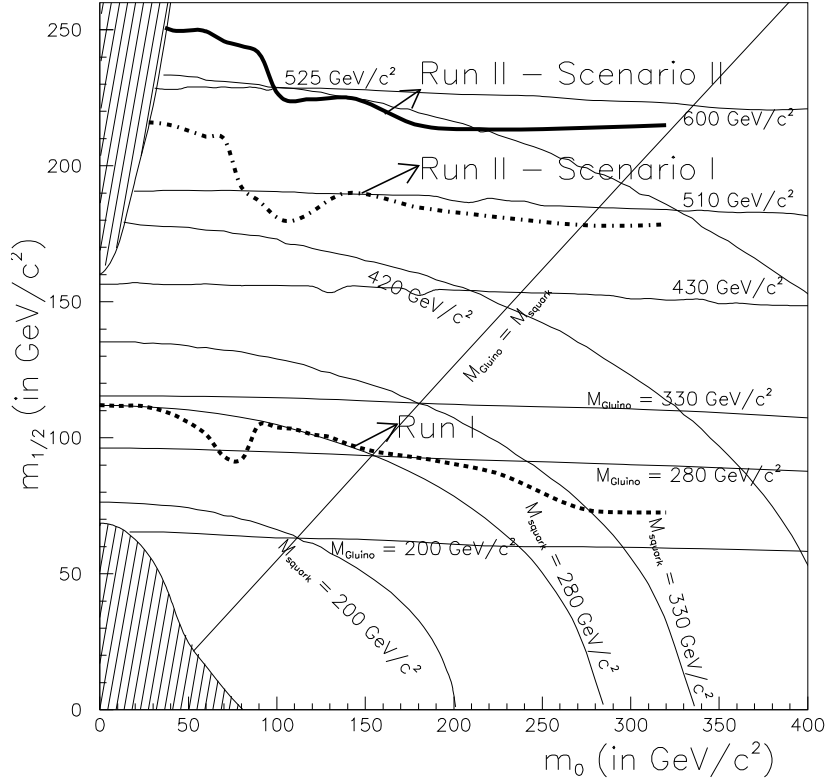,width=12cm}}
\caption{Estimated exclusion contour for Run 2 in the $(m_0,m_{1/2})$
plane for $tan \beta = 2$, $A_0=0$, $\mu <0$, from the $ee + 4$~jets
channel, assuming no SUSY signal is observed. Scenario I corresponds 
to a background of $36\pm 4 \pm 6$ events (direct scaling from Run 1);
scenario II uses the background of $15 \pm 1.5 \pm 1.5$ events 
(scaling, but with improvements in the detector taken into account).}
\label{fig:RPV-e}
\end{figure}
\begin{figure}[t]
\vspace*{0.1in}
\centerline{\protect\psfig{figure=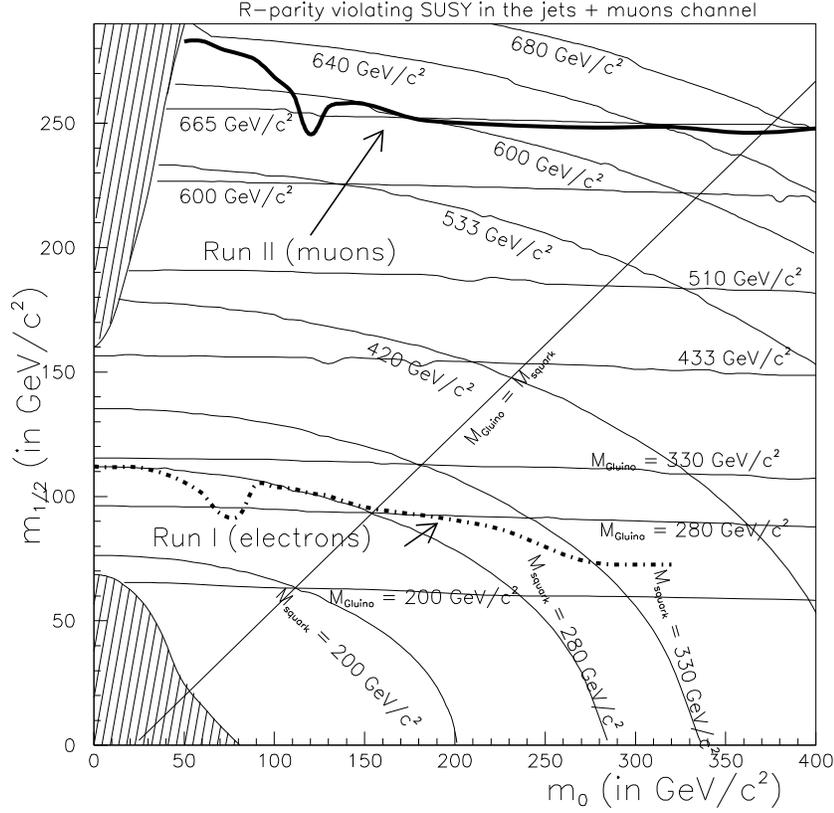,width=12cm}}
\caption{Estimated exclusion contour for Run 2 in the $(m_0,m_{1/2})$
plane for $tan \beta = 2$, $A_0=0$, $\mu <0$, from the $\mu\mu +
4$~jets channel for background of $10 \pm 1.0 \pm 1.0$ (direct scaling
from Run 1).}
\label{fig:RPV-mu}
\vfill
\end{figure}

\subsection{Signal efficiencies} 

The number of signal events expected can be written as: $\langle N
\rangle = \mathcal{L} \cdot \sigma \cdot \epsilon$, where $\langle N
\rangle$ is the expected number of events for luminosity $ \mathcal{L}
$, $\sigma$ is the cross-section, and $\epsilon$ is the overall
efficiency. The efficiency $\epsilon$ can be split into three terms:
$\epsilon = {\epsilon}_{\rm trig} \cdot {\epsilon}_{\rm kin} \cdot
{\epsilon}_{\rm id}$. Here $\epsilon_{\rm trig}$ is the trigger
efficiency for the events that pass the offline cuts ( assumed to be
100\%), $\epsilon_{\rm kin}$ is the efficiency for offline criteria,
which includes kinematic, fiducial and topological requirements, and
$\epsilon_{\rm id}$ is the electron/jet identification efficiency.

The efficiency for identifying jets is very high ($> 95\%$) and is
expected to stay the same in Run 2.

Electron identification efficiencies in Run 1 were $80 \pm 7\%$ in the
central ($|\eta| < 1.1$) and $71 \pm 7\%$ in the forward ($1.5 <
|\eta| < 2.5$) regions~\cite{d0rpv}. These efficiencies were
calculated for electrons with $E_T(e) > 25$~GeV, It drops by about
30\% for electrons with $E_T(e) = 10$~GeV.

The muon identification efficiencies used in Run 1 were $62 \pm 2\%$
in the central ($|\eta| < 1.0$) and $24 \pm 4\%$ in the forward ($1.0
< |\eta| < 1.7$) regions~\cite{rvmu2}. These were calculated for muons
with $p_T > 15$~GeV. For muons with $10~\mbox{GeV} < p_T < 15$~GeV the
efficiencies were 80\% smaller on average~\cite{rvmu3}.

In the present analysis we have taken the overall particle
identification efficiency to be $0.90 \pm 0.09$ in each channel,
independent of lepton $E_T$, primarily due to the expectation of a
better tracker and muon spectrometer for the upgraded D\O\ experiment.

\subsection{Backgrounds}

The main backgrounds are expected to arise from Drell-Yan production
in association with four or more jets, dilepton top-quark events, and
QCD multijet events. The latter is the dominant background for the
electron channel (followed by the Drell-Yan background). In the case
of muons, the background is dominated by the Drell-Yan and top pair
production. We used Monte Carlo to calculate background from the first
two sources, and data to estimate background from QCD jets.

Background for the Run 1 analysis was estimated to be $1.8\pm 0.2\pm
0.3$ (with $1.27 \pm 0.24$ from QCD and $0.42\pm 0.15\pm 0.16$ from
the other processes) for $\sim 100\ pb^{-1}$ of data. To extrapolate
this number to the data set from Run 2, we have simply multiplied it
by the ratio of luminosities to obtain $36\pm 4 \pm 6$
events. However, it is expected that due to the central magnetic field
in the upgraded D\O\ detector, the probability of jets to be
misidentified as electrons will be reduced by a factor of $\sim 2$ in
Run 2. We have therefore considered a second scenario with the smaller
expected background of $15 \pm 1.5 \pm 1.5$ events.

For the muon channel, the expected background has been scaled directly
from the Run 1 analysis. We expect $10 \pm 1 \pm 1$ background events
in Run 2.

\subsection{Results}

In order to obtain the sensitivity of Run 2 in to RPV decays, we
calculated the efficiency for signal for all the mass points shown in
Fig.~\ref{fig:RPV-points}. Typical efficiencies, the signal cross
section in the $ee + 4$~jets channel, and the expected event yield in
2~$fb^{-1}$ of data, for several representative $(m_0,m_{1/2})$
points, are given in Table~\ref{table:RPV-eff}. Similar numbers are
obtained for the muon channel.

We use these efficiencies to obtain exclusion limits in the
$(m_0,m_{1/2})$ plane at 95\% CL, assuming that no excess of events
will be observed above the predicted background. The exclusion
contours for the electron and muon channel are shown in
Fig.~\ref{fig:RPV-e} and \ref{fig:RPV-mu}, respectively. Numerical
values of the limits are summarized in Table~\ref{table:RPV-limits}.

It's worth mentioning that our analysis provides a conservative
estimate of the sensitivity achievable in Run~2, since no formal
optimization of the signal vs. background has been performed. We
expect that a formal optimization can improve the sensitivity in the
mass reach by 15--20\%.

\section{Sparticle Pair Production with Baryonic Decays:\\
Difficult Scenarios for SUSY Detection at the Tevatron}
\label{sec:baer}
\centerline{{\large{\it H. Baer, K. Cheung, J. Gunion}}}
\vspace{0.7cm}

%\renewcommand\linebreak{\unskip\break} %% breaks line & still justifies
%%%%%%%%%%%%%%%%%%%%%%%%%%%%%%%%%%%%%%%%%%%
%\newlength{\captsize} \let\captsize=\small % use \let\normalsize=\captsize
%\newlength{\captwidth}                     % just before \caption{ ...

%\title{
\subsection{Motivation}

\nn GUT scale boundary conditions (\eg\ the 
mSUGRA universal boundary conditions) in which the gaugino
soft-SUSY-breaking
masses are assumed to have a common value at $\mgut$ lead 
to ratios of 7:2:1 (at low energy) for $\mgl :\mcpmone :\mcnone$
due to $M_3:M_2:M_1=\alpha_3:\alpha_2:\alpha_1$.
In particular, the LSP $\cnone$ is 
bino-like and thus substantially lighter than
the wino-like $\cpmone$. However, there are well motivated
models in which $M_1>M_2$. 
\bit
\item 
In the context of string models, if the dilaton $F$ component
does not acquire a large vacuum expectation value (the most natural
result in many models), then the leading contributions to
gaugino masses will arise from loop corrections. 
This was first noted in the context of the
the O-II model of Ref.~\cite{ibanez}
in the limit where supersymmetry breaking is 
ascribed entirely to the overall size
modulus with none arising from the dilaton field. Including the possibility
of a Green-Schwarz mixing term, parameterized by $\delgs$, 
one has (at $\mgut$):
\beq
M_3:M_2:M_1\stackrel{\scriptstyle O-II}{\sim}-(3+\delgs):(1-\delgs):
({33\over 5}-\delgs)\,.
\label{oiibc}
\eeq
At low energies,
one finds $M_2:M_1=2(1-\delgs):(33/5-\delgs)$. For $\delgs$ in the preferred
range from 0 to $-4$, $M_2<M_1$ and both the $\cpmone$ and $\cnone$
are wino-like and very degenerate. 
More recently \cite{murayama,randall} the above boundary conditions
have been re-discovered (at least in the $\delgs=0$ case) in a more
general context. If the soft masses are dominated by
the conformal anomaly, one obtains
(to all orders in perturbation theory)
$M_i={\beta_i(g_i^2)\over 2 g_i^2}\mth$.
Going to low energy scales this translates into
\bea
M_1= {11\alpha\over 4\pi\cos^2\thetaw}\mth\,,~~
M_2= {\alpha\over 4\pi\sin^2\thetaw}\mth\,,~~
M_3= -{3\alpha_s\over 4\pi}\mth\,,
\label{mureq}
\eea
as consistent with the evolved result of Eq.~(\ref{oiibc}) when $\delgs=0$.
\item
$M_2<M_1$ at low energies can also arise if SUSY is broken by an
$F$-term that is not an SU(5) singlet \cite{sm96nonuniv}.  Recall that
the $M_a$ arise from a chiral superfield $\Phi$ that appears linearly
in the gauge kinetic function with $\langle F\rangle\neq0$ for its
auxiliary component.
\begin{equation}
\call= \int d^2\theta W_a W_b {\Phi_{ab}\over \mplanck} + h.c.
\sim  {\langle F_{\Phi} \rangle_{ab}\over \mplanck}\lam_a\lam_b\,,
\end{equation}
where $\lam_{a,b}$ ($a,b=1,2,3$) are the gaugino fields. In the general case,
$\Phi,F_{\Phi}\in ({\bf 24}{\bf \times} 
{\bf 24})_{\rm symmetric}={\bf 1}\oplus {\bf 24} \oplus {\bf 75}
 \oplus {\bf 200}$.
Only $F_{\Phi}$ components that are `neutral' under
SU(3), SU(2), U(1) can have a vev if these
groups are unbroken after SUSY breaking. As a result,
$\langle F_{\Phi} \rangle_{ab}=c_a\delta_{ab}$, with $c_a$ depending
on the representation. $c_a$ is independent of $a$ for 
the ${\bf 1}$ representation (universal $M_a$ at $\mgut$) but
$M_2<M_1$ for the ${\bf 200}$ representation.
\eit
To give some specific results, at $\mgut$ ($\mz$) the O-II
model with $\delgs=-4$ yields $M_3:M_2:M_1=1:5:53/5$ ($6:10:53/5$)
while the ${\bf 200}$ model yields $1:2:10$ ($6:4:10$).
Even though $M_2$ is only slightly
less than $M_1$ at low energies in the O-II and {\bf 200} models,
$\dmchi\equiv \mcpmone- \mcnone<$ a few GeV is typical.
Note that in both models $\mgl$ is close to the common $\cpmone,\cnone$ mass.

The phenomenology for a model in which $M_2<M_1$ and R-parity is
conserved was explored in \cite{cdg1,cdg2} (see also \cite{gmrunii}).
There it was noted that at an $\epem$ collider it will be virtually
impossible to see $\cpone\cmone$ pair production since the potentially
visible particles in $\cpmone\to \cnone+\ldots$ decays will be
{\it very} soft. One can evade this problem by tagging the $\cpone\cmone$
event with a photon, \ie\ look for $\epem\to\gam\cpone\cmone$.
Sensitivity is substantially reduced compared to normal mSUGRA
boundary conditions.  At hadron colliders, if $\mcpmone\simeq\mcnone$
one loses the 
tri-lepton signal from $\cntwo\cpmone\to \ell^+\ell^-\ell^{\prime\,\pm}+\ptmiss$
since the $\ell^{\prime\,\pm}$ is too soft to be detected.  Further,
the like-sign di-lepton signal from $\gl\gl$ production (for instance)
deriving from 
$\gl\gl\to 4j+\cpmone\cmone\to 4j+\ell^{\pm}\ell^{\prime\,\pm}+\ptmiss$
will also be undetectable.  Still, as summarized
in more detail later, in an R-parity conserving model
the jets+$~\ptmiss$ signal will remain and allow reasonable
(although somewhat reduced compared to mSUGRA) 
sensitivity to SUSY particle production \cite{cdg2}.

\subsection{Baryon Number Violation}
This leads us to the worst-case scenario in which we allow for
baryonic R-parity violation. We consider the case where the leptonic
R-parity violating terms are zero ($\lam_{ijk},\lam^\prime_{ijk},
\kap_i=0$) and 
only $\lampp$ is non-zero.  We note that $\lampp_{pnm}$ must be
antisymmetric in the $n,m$ family indices (due to the implicit
antisymmetric colour coupling). For example, converting to a notation
where $p,n,m$ are given in terms of the quark type, we might imagine
$\lampp_{uds}=-\lampp_{usd}\neq 0$.

The most obvious implication of baryonic R-parity violation is the
possible absence of any specifically supersymmetric source of missing
energy.  The decay products of the $\cnone$-LSP\footnote{We do not
have space here to consider the \glsp\ case that is also possible for
O-II model boundary conditions when $\delgs\sim-3$.}  would be jets
that would be visible in the detector provided the decays are prompt.
Combining baryonic R-parity violation with $\cpmone-\cnone$ degeneracy
provides some very unusual and, in some cases, very difficult
scenarios for detecting supersymmetry.

Of course, there are various means for placing bounds on some of the
$\lampp_{ijk}$ (see, \eg, \cite{ck} and \cite{france}).
Perturbativity up to the GUT scale requires that all $\lampp$'s be $<1$.
The experimental limit on
the $\Delta B=\Delta S=-2$ decay $^{16}O\to$$^{14}C$+$K^+$+$K^+$
\cite{goity,barbierimasiero} leads to
$
%\beq
|\lampp_{usd}|<5.6\times 10^{-7}\left({200\mev \over \wtil
\Lambda_{QCD}}\right)^{5/2}
\left({\mgl \over100\gev}\right)^{1/2}\left({\msq \over 1\tev}\right)^2 \,.$
%\label{usdnum}
%\eeq
Here, $\wtil\Lambda_{QCD}$ is a scale that dimensional analysis
suggests should be of order $\Lambda_{QCD}\sim 200\mev$.  Constraints
on $\lampp_{ubd}$, $\lampp_{tsd}$ and $\lampp_{tbd}$ from $N\lra\anti
N$ oscillation limits (see Ref.~\cite{ck}) vary roughly as $1/M_{L-R}
^4$, where $M_{L-R}^2$ is the $L-R$ mixing entry in the relevant
squark mass-squared matrix.  We assume that $L-R$ mixing can be
neglected, implying that $M_{L-R}^2$ is small and that these bounds
are not relevant. $2\to 2$ processes such as $u+d\to \wtil d^*\to
\anti d\cnone$ disturb the standard baryogenesis picture {\it if
sphalerons are present} and if any $|\lampp|$ were larger than $\sim
5\times 10^{-7}(\msq/1\tev)^{1/2}$, a very severe constraint. However,
these constraints do not apply if sphalerons are not present at the
appropriate epoch \cite{rosscosmo}.  Bounds that can be placed on
various products of the $\lampp$'s are not relevant to our analysis
since we shall consider only one coupling to be non-zero at a time.

\subsection{Phenomenology}

\noindent\underline{\boldmath $\cnone$ decays}

The expected $c\tau$ for the $\cnone$ is of considerable importance.
If the $\cnone\to u_Rd_Rd_R^\prime$ (family indices implied) decay
is prompt, discovering supersymmetry at a hadronic
collider in this model is far from easy.  However, if this decay
has a significant path length, our ability to search for supersymmetry 
would be greatly enhanced
by virtue of being able to look for a decay vertex.
In the limit of a very long path length, we would revert
to the jets+$~\ptmiss$ signature. This was studied in Refs.~\cite{cdg1,cdg2}
for the present class of models. Since $\mcpmone\simeq\mcnone$ and
$\mgl$ is not that much larger than $\mcpmone$
supersymmetry signals are weaker than
for mSUGRA boundary conditions. Still, the LHC would be guaranteed to
find a signal for $\mgl\lsim 1\tev$; but at the Tevatron 
the weaker signals would imply a fairly limited discovery reach.

A result for the path length is easily given in the limit where
the sfermions are much heavier than the $\cnone$. Assuming also
that the relevant $u$-type and $d$-type squarks are more or less degenerate
with mass $\mtil$ and neglecting $L-R$ squark mixing,
one obtains 
\beq
c\tau=.03~\mu{\rm m}\left({0.1\over N_{11} }\right)^2 \left({1\over
\lampp}\right)^2\left({100\gev\over\mcnone}\right)^5\left({\mtil\over
1\tev}\right)^4\left[1
-{4\over 5}\,\left({\mcnone^2\over\mtil^2}\right)+\ldots\right]\,,
\label{ctauform}
\eeq
where $N_{ij}$ are the neutralino mixing matrix entries with $i$ refering
to $\wtil\chi^0_{i=1,2,3,4}$ and $j$ refering to $B,W_3,H_1^0,H_2^0$.
Eq.~(\ref{ctauform}) neglects the tiny contributions from
the Higgsino components of $\cnone$.  In the
absence of $L-R$ squark mixing there is
no contribution to the RPV decay of the $\cnone$ coming from its primary
wino component (of size $N_{12}$). (This is because the RPV
$\lampp$ superpotential term only involves right-handed superfields.)
In the model class being considered, $N_{11}\sim 0.1$ is typical and
$\mtil> 1\tev$ is preferred (although this could be
avoided in a `no-scale' type model). For $\mtil\sim 1\tev$,
$\lampp< 0.017$ ($<0.0043$) is required for 
$c\tau> 100~\mu$m if $\mcnone\sim 100\gev$ ($\sim\mt$), respectively.
For $\mtil\sim 300\gev$, the corresponding $\lampp$ values must
be smaller by a factor of .09 for $c\tau>100~\mu$m.
Given that typical Yukawa couplings for the heavier `light' quarks are
$gm_s/(2\mw)\sim 0.0008$ and $gm_c/(2\mw)\sim 0.006$, the RPV
analogues could easily be this small.

\begin{figure}[ht]
\leavevmode
\begin{center}
\epsfxsize=3.8in
\hspace{0in}\epsffile{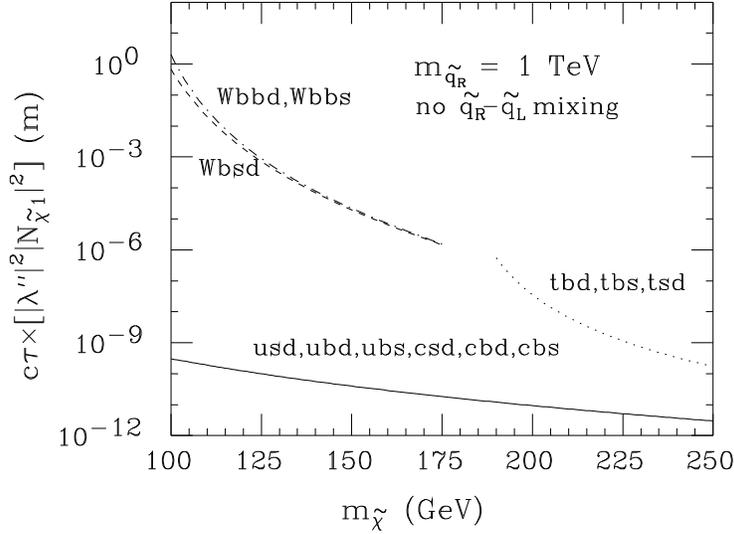}
\end{center}
\caption[]{We plot $c\tau\times[|\lampp|^2|N_{11}|^2]$
as a function of $\mcnone$ for a variety of final states,
including those arising from virtual top intermediate states.}
\label{chidecay} 
\end{figure} 

For the constraint given earlier on $\lampp_{usd}$,
the $\cnone$ would have a very long path length if
all other $\lampp$'s are zero. However, it seems likely
that the higher generation $\lampp$'s would be larger than
$\lampp_{usd}$. The largest $\lampp$'s
might be those associated with the 3rd family.
However, since the mass of the $\cnone$
is typically not very large in the models we consider, a top
quark in the final state might have to be virtual, implying
substantial phase-space suppression. 
The values for $c\tau\times[|\lampp|^2|N_{11}|^2]$
for $t^{(*)}bd$, $t^{(*)}bs$ and $t^{(*)}sd$
are plotted in Fig.~\ref{chidecay} in comparison to that
obtained in any one of the $usd$, $ubd$, $ubs$, $csd$, $cbd$ and $cbs$
channels.  At $\mcnone=100\gev$ ($175\gev$),
one of the three virtual top channels would dominate
over one of the other six channels if $\lampp$ for the former is
a factor of $\sim 5\times 10^4$ ($\sim 500$) larger than for the latter.
Keeping in mind that Yukawa couplings differ by ratios of $\sim \mt/m_u\sim
5\times 10^4$ and $\sim \mt/m_s\sim 350$, 
such large ratios should not be regarded as impossible.

Typical $c\tau$ values for the various decays are
obtained from  Fig.~\ref{chidecay} by multiplying by $1/|N_{11}|^2\sim
100$ and by $1/|\lampp|^2$. If $\cnone$
decay is dominated by a virtual top channel and the relevant $\lampp\sim 1$,
a path length of order $100$~m ($100~\mu$m) is to be expected
for $\mcnone\sim 100\gev$ ($\sim\mt$). For smaller $\lampp$, $c\tau$
is even larger.  However, if one of the $\lam_{usd,\ldots}$ is substantial,
the $\cnone$ decay will be very prompt. 

\noindent\underline{\boldmath $\cpmone$ decays}

%\begin{figure}[ht]
%\leavevmode
%\begin{center}
%\epsfxsize=4.25in
%%\epsfysize=5.25in
%\hspace{0in}\epsffile{user$1:[jfgucd.drees]lifebrs_new.ps}
%\end{center}
%\caption[]{Lifetime and branching ratios for the $\cpmone$
%as a function of $\dmchi$. Both are very nearly
%independent of other parameters such as $\mcnone$ itself, $\tanb$, and $\mu$.}
%\label{lifebrs} 
%\end{figure} 
%

In the models being considered, 
$\dmchi\equiv\mcpmone-\mcnone<1\gev$ is typical and the
$c\tau$ for $\cpmone\to \cnone\ldots$ (see \cite{gmrunii})
can be so large that the $c\tau$ for
direct RPV decay of the $\cpmone$ (values being comparable to those
plotted for the $\cnone$) can easily be shorter (\ie\ the width
larger).  In such scenarios, direct $\cpmone$ decays to the $\cnone$
might be quite rare. Phenomenologically, however, it would hardly make
a difference since the decay products of the $\cpmone$ decay are so
soft for the small expected values of $\dmchi$ anyway.

\noindent\underline{\boldmath $\gl$ decays}

The normal SUSY $\gl\to q\anti q\cnone$ amd $\gl\to q^\prime
\anti q\cpmone$ decays compete with the RPV $\gl\to \anti q \wtil
q_R^*\to 3\anti q$ decays. For our boundary conditions,
the former SUSY modes can be sufficiently phase space suppressed
that the RPV decays could dominate if one or more of
$\lampp_{usd}$, $\lampp_{ubd}$, $\lampp_{ubs}$,
$\lampp_{csd}$, $\lampp_{cbd}$, or $\lampp_{cbs}$ were large.
However, this would result in little change in the phenomenology since
the extra 2 jets coming from the SUSY $\gl$ chain decays would
generally be quite soft and not easily reconstructed at a hadron collider.
At an $\epem$ collider, $\gl$ production (via $\epem\to q\anti q\gl\gl$)
has a small cross section and $\gl$ decays are not an important issue.

\subsection{\boldmath SUSY detection at a hadron collider}

Various possible scenarios could emerge. Suppose first
that the relevant $\lampp$ is sufficiently small that the $\cnone$
is effectively stable in the detector. Then, the R-parity conserving
results of Ref.~\cite{cdg2} for $\mgl\sim\mcpmone\simeq\mcnone$
apply. Since the lepton from $\cpmone\to
\ell^{\pm}\cnone$ decay is very soft, 
standard SUSY signals relying on leptons are no longer viable.
This includes the like-sign dilepton signal for $\gl\gl$ production 
in which $\gl\gl\to 4j+\cpone\cpone + X\to 4j+\ell^+\ell^++\etmiss$ 
(or the charge conjugate) and the trilepton signal from
$\cpmone\cntwo\to\ell^\pm\ell^{\prime\,+}\ell^{\prime\,-}$.
One must rely entirely on the jets plus missing energy signal
for $\gl\gl$ production, which
is weakened by the fact that the jets in both
$\gl\to q^\prime \anti q\cpmone$ and $\cpmone\to
q^\prime\anti q\cnone$ decays are no longer hard enough to pass typical cuts
used by CDF and D0 to define a visible jet. The only energetic high-$p_T$
jets present are those arising from initial state radiation.
Due to the smaller number of jets passing cuts, 
discovery reach at the Tevatron
is substantially reduced compared to the normal mSUGRA
scenario, even if the relatively soft (as appropriate for
the scenario being considered) Run I CDF and D0 cuts are employed.
This is illustrated in Table~\ref{maxmgl}.

\begin{table}[hbt]
\caption[fake]{Maximum $\mgl$ values that can be probed
using D0 and CDF Run I cuts in the $jets+\etmiss$ final state
for different integrated luminosities, $L$, at the Tevatron and Tev33
using the O-II model boundary conditions with
$\delgs=-4.5$. Observability is defined by $S/\sqrt B\geq 5$ and $S/B\geq 0.2$.
Also given are the maximum $\mgl$ values for which $\gl\gl$
production can be observed in the case of mSUGRA universal boundary conditions
for both Run I cuts and also stronger cuts optimized for Run II/TeV33.
The results of this table are for $\tanb=2$ and $\mu<0$.}
\begin{center}
\begin{tabular}{|c|l|rrrrr|}
\hline
Cuts & 
\ \ \ \ \ $L=$ & $8\pbi$ & $19\pbi$ & $100\pbi$ & $2\fbi$ & $25\fbi$ \\
\hline
D0 & $\delgs=-4.5$ & ---\ \ \ \ \  & $80\gev$ & $110\gev$ & $150\gev$ & $150\gev$ \\
RunI & mSUGRA & ---\ \ \ \ \  & $140\gev$ & $200\gev$ & $200\gev$ & $200\gev$ \\
\hline
CDF & $\delgs=-4.5$ & $100\gev$ & $130\gev$ & $140\gev$ & $160\gev$ & $160\gev$ \\
RunI & mSUGRA & $150\gev$ & $170\gev$ & $210\gev$ & $240\gev$ & $240\gev$ \\
\hline
Strong & mSUGRA & \multicolumn{3} {c} {$\mgl$ excluded
by $\mcpmone\leq 47\gev$} & $250\gev$ & $300\gev$ \\
\hline
\end{tabular}
\end{center}
\label{maxmgl}
\end{table}

Next, suppose $\lampp$ is of `moderate' magnitude such that $\cnone$ decay
is visible in the detector, \ie\ has a path length leading to
a visible vertex in the inner vertex detector or tracker or a
recognizable delayed decay in the outer portions of the detector.
The combined requirement of many jets and a clear vertex or delayed
decay should leave very little QCD background.  Supersymmetry
discovery would be largely limited by event rate
and discovery reach would exceed even that typical of mSUGRA
boundary conditions.

The most difficult case for supersymmetry discovery would be that
in which $\lampp$ is sufficently large that the RPV $\cnone$
decay is prompt and  the $\lampp$'s involving the top/stop are sufficiently
small that the virtual or real top-quark decay modes of the $\cnone$
are not important and we must consider $\cnone\to 3j$ decays. In this case,
$3j=uds$, $udb$, $usb$, $cds$, $cdb$, and/or $csb$
are all, in principle, possible. There would be no specifically SUSY
source of missing energy in sparticle-pair production.
The most characteristic feature of $\gl\gl$ production would be
the presence of many energetic jets in most events.  
To explore the characteristics of such events in more detail, we have 
considered
a sample case defined by the following O-II (GUT scale) model parameters:
\beq
\delgs=-4.1,~~m_0=1800\gev,~~\mhalf=10\gev,~~A_0=0,~~\tanb=3,~~\mu>0\,,
\label{params1}
\eeq
leading to
\bea
&&\mgl\sim 102\gev,~~\mcnone\simeq\mcpmone\sim 100\gev,~~\dmchi\sim 0.23\gev,
~~\mcntwo\sim 120\gev,\nonumber\\
&&\mcpmtwo\sim\mcnthree\sim\mcnfour\sim 1140\gev,~~m_{\wtil \ell,\wtil q}\sim
1.4-1.8 \tev.
\label{masses1}
\eea
Note the very low mass scales for all gauginos, including the $\gl$,
and the high mass scales (typical for the O-II boundary conditions)
of all sfermions and higgsinos.

\begin{figure}[ht]
\leavevmode
\begin{center}
\epsfxsize=6.in
\hspace{0in}\epsffile{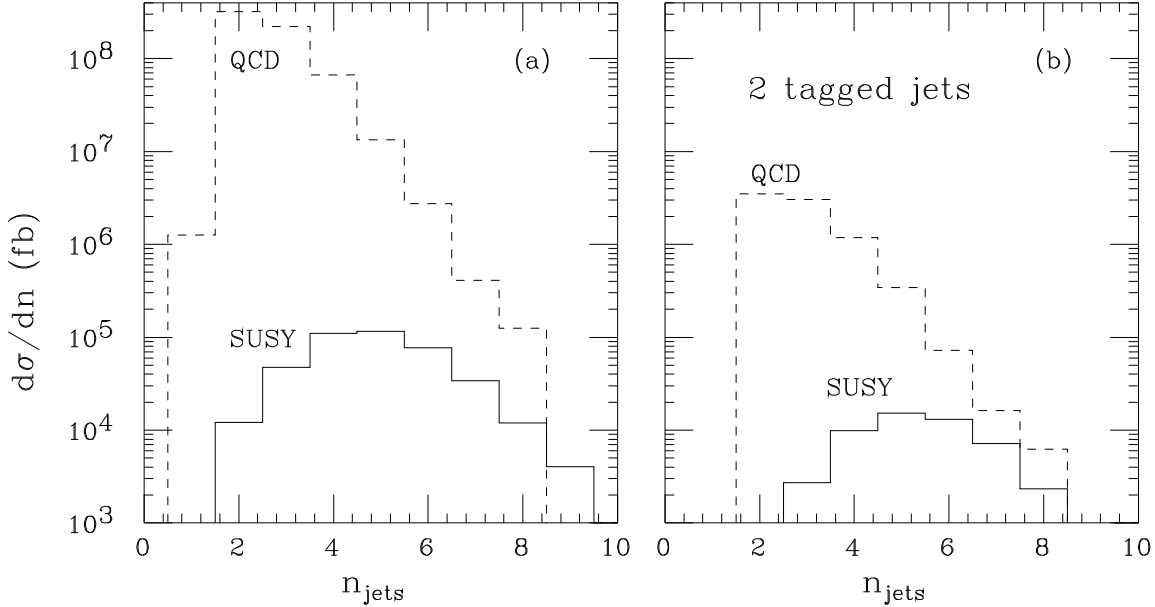}
\end{center}
\caption[]{The number of events is plotted as a function
of the number of jets for the QCD background (dashes) and for SUSY
production processes (solid) (a) without tagging (b) after
requiring that at least two jets be tagged (using the CDF
JP charm tagging algorithm).
O-II model parameters as in Eq.~(\ref{params1}).}
\label{njets} 
\end{figure} 

We have performed a Monte Carlo simulation of signal and background
(using ISAJET \cite{baertata}). We define jets using $E_T>15\gev$,
$|\eta|<4$, and a cone size of $\Delta R=0.5$.  In
Fig.~\ref{njets}(a), we plot the event rate for the QCD background as
compared to that for supersymmetric particle pair production (all
channels included, but mainly $\gl\gl$) as a function of the number of
jets, $n_j$, before any tagging requirement, and before any other
cuts.  The signal has larger $n_j$ on average, but the QCD background
is still bigger by a factor of 10 even if we require $n_j\geq
8$. Nonetheless, if we compute the nominal $S/\sqrt B$ for the $n_j=8$
bin, assuming $L=2\fbi$ ($30\fbi$), we get $S/\sqrt B=45$ ($170$).
Thus, if the systematic uncertainty for the $n_j=8$ background could
be reduced to below 10\% we would achieve an observable signal without
tagging. This might be possible by normalizing using the $n_j\leq 7$
bins once a reliable calculation of relative rates for multi-jet
events with $n_j$ ranging from low values up to $n_j=8$ becomes
available. (The ISAJET shower model results for relative rates are
probably not accurate at the 10\% level.)

However, the very strong limit on $\lampp_{usd}$ means that one (or
more) of the $ubd$, $ubs$, $csd$, $cbd$, $cbs$ channels will dominate
$\cnone$ decay. All of these contain at least one heavy quark.  Since
SUSY pair production events contain at least two $\cnone$'s, we can
then use double-tagging to reduce the background relative to the
signal. For all but the $cds$ decay channel, at least two of the jets
in the final state will be $b$-jets.  If $cds$ decays dominate, we
would have two $c$-jets.  Since $c$-tagging will have lower efficiency
than $b$-tagging, dominance of $\cnone$ decays by the $cds$ channel
results in the most difficult scenario for supersymmetry detection.

To explore this `worst case', we assume $\cnone\to csd$ exclusively.
Fig.~\ref{njets}(b) gives the $n_j$ distributions for signal
and background after
requiring that every accepted event contain at least two tagged jets.
For this plot we use the CDF `JP'-tag algorithm (designed for $c$ quark
tagging). We require that a tagged jet have
$|\eta|<2$~\footnote{This is the rapidity coverage
for Run II.} and $p_T>25\gev$. For such a jet, the JP-tag yields~\cite{regina}:
%\beq
$\eps_b\geq 0.5\,,~~\eps_c\geq 0.3\,,~~{\rm for}~\eps_{q,g}\sim 0.05\,,$
%\label{effs}
%\eeq
where $q=u,d,s$.~\footnote{
For $p_T\geq  50\gev$, $\eps_b\sim 0.6$
and $\eps_c\sim 0.35$ for $\eps_{q,g}\sim 0.05$. 
If the L00 inner layer is constructed, $\eps_b$ ($\eps_c$) will be larger by 
a factor of 1.12 (1.25).} From Fig.~\ref{njets}(b), we see that
after tagging it is best to focus on the $n_j=7$ bin, with $S=1.4\times 10^4,
B=3.2\times 10^4$ ($S=2.1\times 10^5,B=4.8\times 10^5$) for
$L=2\fbi$ ($L=30\fbi$), leading to $S/\sqrt B=78$ ($S/\sqrt B=303$)
with $S/B\sim 45\%$ (a big improvement).
Thus, for low-mass SUSY scenarios [see Eq.~(\ref{masses1})]
of the type being considered, double-tagging leads to a very clear excess
(beyond Standard Model expectations) of events with a large number of jets.

At the low $\mgl$ value considered above, additional cuts in the
variables $E_T^{\rm tot}\equiv \sum_j|E_T^j|$ and 
$H_T\equiv\sum_j|E_T^j|+\etmiss$
are not particularly helpful.  However, in order to reduce the QCD
background when searching for gluinos of substantially 
higher mass, cuts requiring high values for
one or both of these variables are essential. For an appropriate cut,
we would again examine the $n_j$ distribution after requiring two tagged
jets. We anticipate that a signal will be observable so long
as the basic production rate for $\gl\gl$ pairs is substantial.
The exact mass reach (in $\mgl$) is under study \cite{inprogress}.

\subsection{Discussion and Conclusions}

We have shown that observation of a signal from $\gl\gl$ pair production 
is likely to be possible at the Tevatron even in the extreme of
O-II model boundary conditions and RPV $3j$ decays of the $\cnone$,
so long as the gluino mass is not too large. 
However, the high-jet-number signal upon which we focused
does not allow us to determine the strength of the dominant
RPV $\lampp$ coupling. 
There are two possibilities for directly determining $\lampp$.
If $\lampp$ is large, RPV-induced single squark production
cross sections are also typically substantial, and, if a
signal can be isolated \cite{inprogress}, 
the cross section size gives a measure
of $\lampp$. If $\lampp$ is small, the decay path length
for the $\cnone$ might be observable and would again provide
a measure of $\lampp$. However, our preliminary estimates \cite{inprogress}
indicate that there is a region of intermediate
$\lampp$ at higher $m_{\wtil q_R}$ for which
neither $c\tau(\cnone)$ nor $\sigma(\wtil q_R)$ will be measurable. 
In this region, determination of $\lampp$ would only be possible if
an RPV decay mode of the $\cpmone$ is competitive
with its standard SUSY decay modes and these can
be separated from one another (which is not at all certain); 
the relative size of the branching
ratios would then provide a measure of $\lampp$. Parenthetically, 
we note that at an $\epem$ collider the only means for measuring
$\lampp$ would be via the $\cnone$ decay length; there
are no sources of quarks or antiquarks as needed for squark production
via baryonic RPV couplings.

To summarize, even if there is baryonic R-parity violation, implying absence
of missing energy in SUSY events, the very large number of jets
expected for a typical SUSY event will allow us to isolate the SUSY
signal so long as gluino (and/or squark) masses
are sufficiently low that SUSY event rates are substantial. Further,
there is at least a decent chance that we will be able to measure
the strength of the RPV coupling.

\medskip

\centerline{\bf Acknowledgments}

This work was supported by the Department of Energy and by the Davis Institute
for High Energy Physics.

\vspace{0.5cm}
\centerline{{\Large{\bf PART 3: Spontaneous R-Parity Breaking}}}

\section{Decays of the Top-Quark 
and the Top-Squark} 

{\large{\it F.~de Campos, M.A.~D\'\i az, O.~J.~P.~Eboli, M. B. Magro,
L.~Navarro, W.~Porod, D.~A.~Restrepo, and J.W.F.~Valle}}

\subsection{Introduction}
\nn In this section we study the case of spontaneous R-parity breaking.
This is a modification of the previous studies in that an extra
superfields is added: $\nu_R$, which is a gauge singlet and
corresponds to a right-handed neutrino. Here we focus on the case of 
bilinear R--Parity Violation 
(BRPV)~\cite{epsrad,BRPVtalk,BRPhiggs,others}, where $\lam,\,\lam',\,
\lam''=0$ in Eq.\eq{superpot} and $\kap_i\not=0$\footnote{Throughout
the rest of this section we shall denot the parameter $\kap$ as
$\epsilon$.}. This includes as an additional feature a vev for the
sneutrinos. These models are well-motivated theoretically as they
arise as effective truncations of models where R--Parity is broken
spontaneously \cite{SRpSB} through {\it right handed} sneutrino vacuum
expectation values (vev) $\vev{\tilde\nu^c}=v_R \neq0$.
They open new possibilities for the study of the unification
of the Yukawa couplings~\cite{epsbtaunif}. In particular it has been
shown that in BRPV models bottom-tau unification may be
achieved at any value of $\tan\beta$. 
From a phenomenological point of view these models predict a plethora
of novel processes \cite{BRPVtalkphen} that could reveal the existence
of SUSY in a totally different way, not only through the usual missing
momentum signature as predicted by the MSSM.
They provide a very predictive approach to the
violation of R--Parity, which renders the systematic study of
R-parity violating physics \cite{BRPVtalkphen} possible.

We consider the simplest superpotential which violates R-Parity
\begin{equation} 
W_{R_p \hspace{-3.4mm} /} = W_{MSSM} 
+\epsilon_i \widehat L_i \widehat H_u \,,
\label{eq:Wsuppota}
\end{equation}
assuming that TRPV terms are absent or suppressed, as would be
the case if their origin is gravitational~\cite{BJV}.  The
$\epsilon_i$ terms violate lepton number in the $i$th generation,
respectively. Models where R--Parity is broken
spontaneously \cite{SRpSB} through a vev of the
right handed sneutrinos $\vev{\tilde\nu^c}=v_R \neq0$ generate only
BRPV terms. The $\epsilon_i$ parameters are then identified as a product of
a Yukawa coupling and $v_R$. This provides the main
theoretical motivation for introducing explicitly BRPV in the MSSM
superpotential.
For simplicity we set from now on $\epsilon_1=\epsilon_2=0$.  Thus we
have tau--lepton signatures and we only test the case where
tau--lepton number is violated. The MSSM--BRPV has the following
superpotential
\begin{equation} 
W_{R_p \hspace{-3.4mm} /}=\varepsilon_{ab}\left[
 h_t\widehat Q_3^a\widehat U_3\widehat H_u^b
+h_b\widehat Q_3^b\widehat D_3\widehat H_d^a
+h_{\tau}\widehat L_3^b\widehat R_3\widehat H_d^a
+\mu\widehat H_u^a \widehat H_d^b
+\epsilon_3\widehat L_3^a\widehat H_u^b\right]\,,
\label{eq:Wsuppot}
\end{equation}
where the first four terms correspond to the MSSM. The last term violates
tau--lepton number as well as R--Parity.

\begin{figure}[t]
\begin{minipage}[t]{.50\linewidth}
%\centering\epsfig{file=../eps/rvlim3.eps,width=\linewidth} 
\psfig{figure=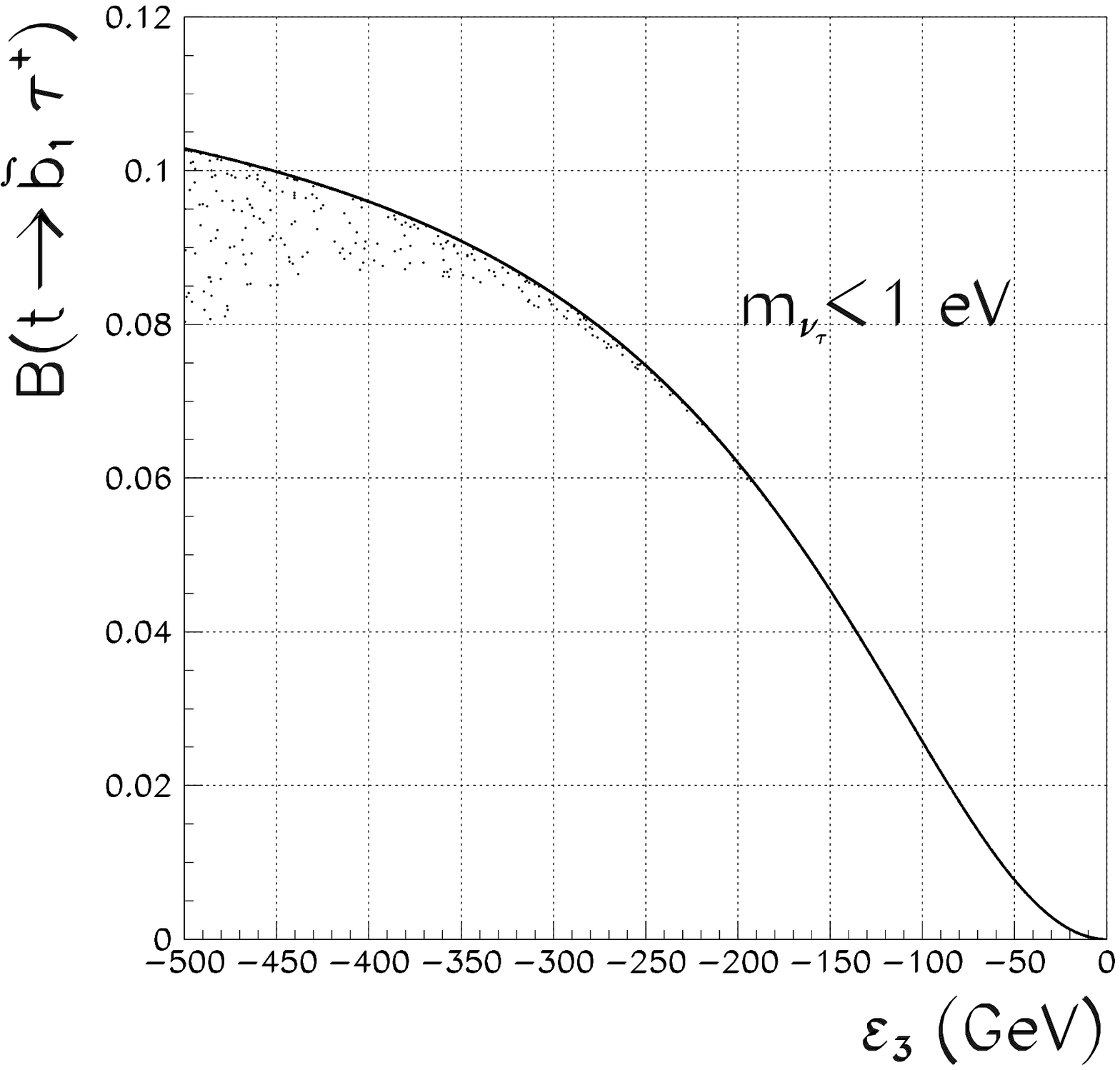,height=6cm,width=7cm,angle=0}
\end{minipage}\hfill
\begin{minipage}[t]{.50\linewidth}
%\centering\epsfig{file=../eps/lvlim1.eps,width=\linewidth} 
\psfig{figure=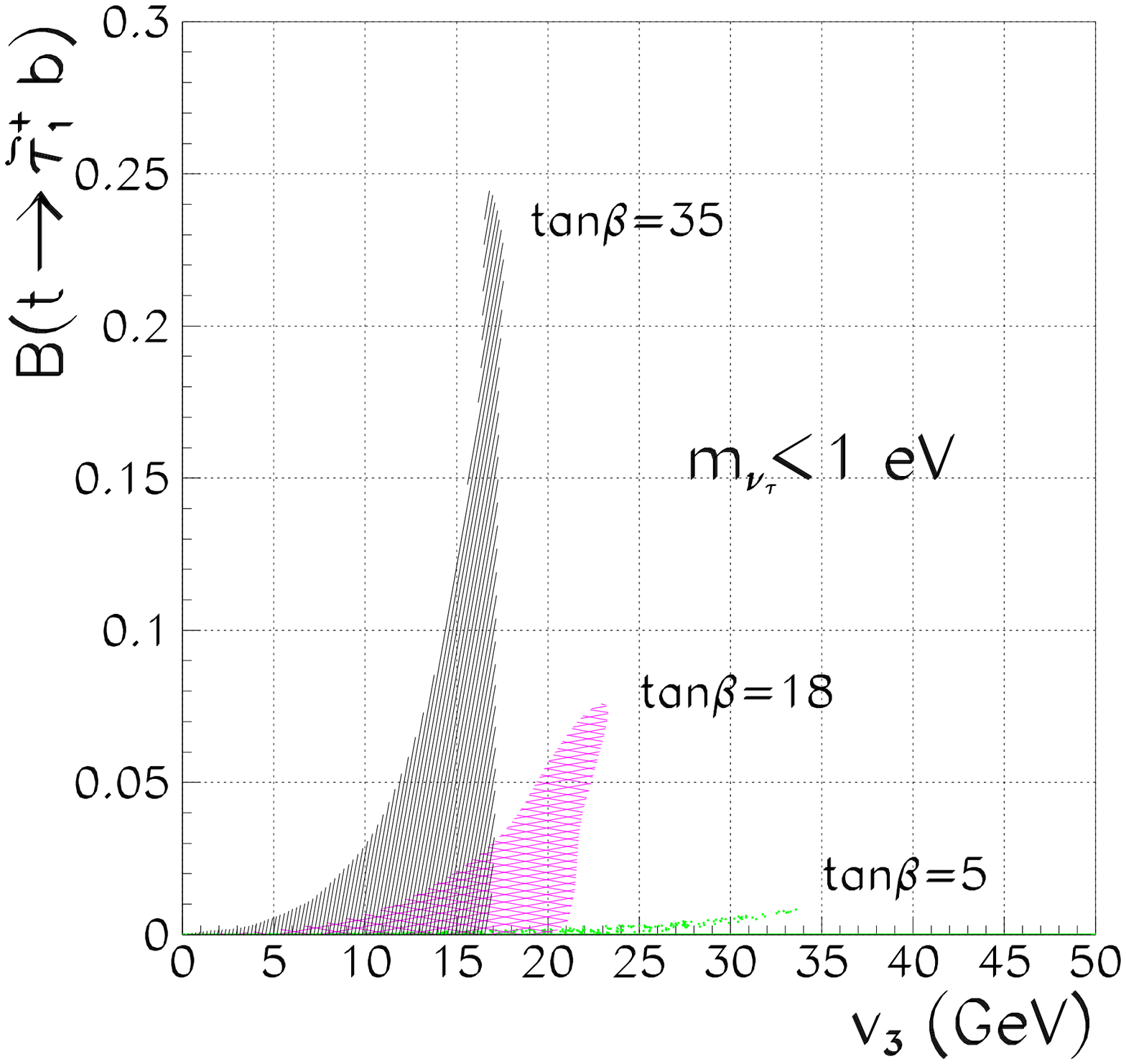,height=6cm,width=7cm,angle=0}
\end{minipage}
  \caption{(a) Branching Ratios for $t \to {\tilde b}_1 \, \tau^+$ as a
         function of  $\epsilon_3$. (b) Branching Ratios for 
         $t \to {\tilde \tau}^+_1 \, b$ as a
         function of  $v_3$ for different values of $\tan \beta$.
In both cases the parameters are:
         $M = 180$~GeV, $\mu = 200$~GeV, $\tan \beta = 35$,
         $M_{E_3} = 285$~GeV, $A_\tau = 280$~GeV, $M_Q = 285$~GeV,
         $M_U = 180$~GeV, $M_D = 190$~GeV, $A_t =320$~GeV,
         $A_b = 120$~GeV, $B = 50$~GeV, $-500$~GeV $< \epsilon < 0$~GeV,
         $0$~GeV $< B_2 < 200$~GeV,  $1$~GeV $< v_3 < 50$~GeV.}
  \label{fig:valle1}
\end{figure}

In this model the top-quark and the top-squark get additional
decay modes, e.g. $t \to {\tilde \tau}^+_1 \, b$ or ${\tilde t}_1 \to
\tau^+ + b$.  We study these decays in view of the Tevatron (for top decays
in TRPV models see \cite{dreiner}). We show that existing Tevatron
data give additional constraints on the parameter space and
extrapolate to Run II.

\subsection{Top Decays}

\nn One of the major successes of the Tevatron has been the discovery of the
top-quark \cite{Topdiscovery}. The large top mass implies a relatively
small production cross section at the Tevatron. Therefore, the sum of
all branching ratios of the top decays except $t \to W^+ \, b$ is only
restricted to be smaller than approximately 25 \% \cite{TopEx}.  In
the MSSM the top can decay according to: $t \to W^+ \, b$, $t \to H^+
\, b$, $t \to {\tilde \chi}^0_1 \, {\tilde t}_1 $, $t \to {\tilde
\chi}^+_1$. \footnote{The mode ${\tilde b}_1 $ is 
practically ruled out by existing LEP2--data \cite{LepTalk}.} In the
BRPV model the charginos mix with the charged leptons, the neutralinos
with neutrinos, and the charged sleptons with the charged Higgs boson
\cite{epsrad,BRPVtalk,BRPhiggs}. Therefore, the top can have
additional decay modes:
\begin{equation}
t \to {\tilde \tau}^+_1 \, b \, , \hspace{4mm}
t \to \nu_\tau \, {\tilde t}_1 \, , \hspace{4mm}
t \to \tau^+ \, {\tilde b}_1 \, .
\end{equation}
As an illustrative example, we show in Fig.~\ref{fig:valle1}a the
branching ratio for $t \to \tau^+ \, {\tilde b}_1 $ as a function of
a) $\epsilon_3$ and b) $v_3$. We have randomly chosen 10,000 points
imposing the following experimental constraints: $m_{\nu_\tau} < 18$~MeV, $m_{
{\tilde t}_1}, m_{ {\tilde b}_1} > 80$~GeV, $min(m_{H^+}, m_{ {\tilde
\tau}_1}) > 70$~GeV, and $m_{{\tilde \chi}^+_1} > 85$~GeV. The
parameters are listed in the figure caption. We find a strong
correlation between the R-parity decay branching ratio $BR(t \to
\tau^+ \, {\tilde b}_1)$ and the magnitude of $\epsilon_3$. 
This can be understood in the following way: in the chargino mass
matrix the mixing between the leptons and the charginos disappears if
one does the following rotation in the superfields: $\widehat H_1 \to
N (\mu \widehat H_1 - \epsilon_3 \widehat L_3)$ and $\widehat L_3 \to
N (\mu \widehat L_3 + \epsilon_3 \widehat H_1)$ (N being the
normalization). In this basis the coupling between $t$, $\tau$, and
${\tilde b}_1$ is proportional $N \, h_b \,\epsilon_3$ leading to this
feature.  

In Fig.~\ref{fig:valle1}b we show the branching ratio for $t \to {\tilde
\tau}^+_1 \, b$ as a function of $v_3$. As in Fig.~\ref{fig:valle1}a, 
results in Fig.~\ref{fig:valle1}b are displayed for different values of
$\tan \beta$ and the other parameters are also the same. The
dependence on $\tan \beta$ is a result of: (i) The stau - charged
Higgs boson mixing is proportional to the R-parity breaking parameters
$\epsilon_3$ and $v_3$ (ii) The decay width depends on the bottom
Yukawa coupling which increases with $\tan \beta$.  As can be seen
from the figure, there is a strong correlation between the magnitude of
the R-parity breaking branching ratios and the mixing between the stau
and the charged Higgs boson.

We have performed a similar scan for small $\tan \beta$ for both of
BRPV decay channels discussed above.  These are suppressed in this case
and can not exceed 2\% or so, i.e. $\sum BR(t \to b \, X) < 1-2\% \,
(X\ne W)$), because their decay widths are in all cases proportional
to the bottom Yukawa coupling squared.  In the case of $t \to \tau^+
\, {\tilde b}_1$ this is clear from the discussion of
Fig.~\ref{fig:valle1}a.  In the case of $t \to {\tilde \tau}^+_1 \, b$
one has to note that the stau mixes mainly with the charged component of
the down-type Higgs multiplet $H_1$ ($\tilde \tau_L$ and $H_1$ have
the same gauge quantum numbers) and the $H_1 t b$ coupling is
proportional to $h_b$.
 
In every case the various decay modes lead to the cascade decays
\beq
t \to {\tilde \tau}^+_1 \, b\to \left\{ 
          \begin{array}{ll}
              \tau^+ \, \nu_\tau \, b & \\
             \tau^+ \,{\tilde \chi}^0_1 \, b&\to \left\{
                 \begin{array}{l}          
                      \tau^+ \, f \, \bar{f} \, \nu_\tau \, b \\
                      \tau^+ \, f \, \bar{f}' \, \tau^\pm \, b 
                 \end{array} \right.\\
             \nu_\tau \, {\tilde \chi}^+_1 \, b &\to \left\{
                  \begin{array}{l} 
                    \nu_\tau \, f \, \bar{f}' \, \nu_\tau \, b \\
                    \nu_\tau \, f \, \bar{f} \, \tau^+ \, b 
                  \end{array} \right. \\
              c \, s \, b & 
          \end{array} \right.,
 \qquad
t \to \tau^+ \, {\tilde b}_1 \to\left\{
  \begin{array}{ll}
  \tau^+ \, \nu_\tau \, b & \\
           \tau^+ \, {\tilde \chi}^0_1 \, b \to \left\{
                     \begin{array}{l}
                         \tau^+ \, f \, \bar{f} \, \nu_\tau \, b \\
                         \tau^+ \, f \, \bar{f}' \, \tau^\pm \, b 
                     \end{array}\right.
\end{array}\right. 
\eeq
In nearly all cases there are two $\tau$'s and two $b$-quarks in the
final state plus the possibility of additional leptons and/or jets.
Therefore, $b$-tagging and a good $\tau$ identification are important
for extracting these final states. Moreover there is in general a
large multiplicity of charged particles in the final state which
should be helpful in reducing the background. The background will come
mainly from the production of one or two gauge bosons plus
additional jets.
The conclusion in similar cases \cite{LeCompte98} has been that in its
next run the Tevatron should be sensitive to branching ratio values up
to $10^{-3} - 10^{-2}$ depending on the mode. Therefore, the possible
observation of one of these additional decay modes at the Run II of the
Tevatron should give a strong hint on the underlying parameters.

\subsection{Top-Squark Decays}

\nn Top-squark physics is a very interesting part of supersymmetric
theories, because the lighter top-squark might be the lightest charged
SUSY particle \cite{rudaz}.  
In the kinematic region accessible to the Tevatron the light
top-squark has the following MSSM decay modes: $ {\tilde t}_1 \to
{\tilde \chi}^+_i + b$, $ {\tilde t}_1 \to {\tilde \chi}^0_1 + c$, $
{\tilde t}_1 \to {\tilde l}^+_i + \nu_l + b$, $ {\tilde t}_1 \to
{\tilde \nu}_l + l^+ + b$, $ {\tilde t}_1 \to {\tilde \chi}^0_1 + W^+
+ b$, and $ {\tilde t}_1 \to {\tilde \chi}^0_1 + H^+ + b$ (for a
discussion see e.g. \cite{StopDecays} and references therein).  In
BRPV models the top-squark has an additional and phenomenologically
very interesting decay mode \cite{stop}:
\begin{equation}
 {\tilde t}_1 \to \tau^+ + b
\end{equation}
In the following we have concentrated on scenarios where only the
two-body decay modes are possible. We adopt the framework of
Supergravity unification as in \cite{epsrad} in order to reduce the number
of free SUSY parameters. However, we keep $\epsilon_3$ and $v_3$ as
free parameters for the moment. In Fig.~\ref{fig:valle2}a we show the
areas in the $m_{{\tilde t}_1}$-$m_{{\tilde \chi}^0_1}$ plane where
the branching ratio ${\tilde t}_1 \to \tau^+ + b$ is larger than 90\%
for different values of $\epsilon_3$ and $v_3$. We restrict to the
range $|\epsilon_3|, |v_3|<1$ GeV, and vary randomly the MSSM
parameters keeping $m_{\tilde t_1}<m_{{\tilde \chi}^{\pm}_1}+m_b$.
This demonstrates that one can get a dominance of the R-Parity
violating decay mode even for relatively small values of the R-parity
breaking parameters.  The upper--left triangular region corresponds to
$m_{\tilde t_1}<m_{{\tilde \chi}^0_1}+m_c$ and thus $BR(\tilde t_1 \to
b \tau)=1$. In the lower--right triangular region $m_{\tilde
t_1}>m_{{\tilde \chi}^+_1}+m_b$ and therefore $\tilde t_1 \to b
\, {\tilde \chi}^+_1$ is open. In the central region the top-squark
has the two decay modes $\tilde t_1 \to b \, \tau$ and $\tilde t_1 \to
c \, {\tilde \chi}^0_1$. The solid lines, defined by the maximum value
of $|\epsilon_3|$ and $|v_3|$, are the boundary of the regions where
$BR(\tilde t_1 \to b \tau)>0.9$ such that points at the left of the
boundary satisfy that condition.

Since BRPV models allow the decay $({\tilde t}_1 \to \tau^+ + b)$ we
can interpret the top squark as a third generation leptoquark.
Therefore we can use the limits obtained from leptoquark searches
\cite{D0-lq,CDF-lq} to derive limits on the top-squark for this case. In
Fig.~\ref{fig:valle2}b we show an exclusion plot in the
$m_0$-$m_{1/2}$ plane. The nearly horizontal dashed lines are chargino
mass contours and the lines forming radial patterns are the top-squark
mass contours.  The upper to the lower radial curves corresponds to
$m_{\tilde{t}_1}=120, 100$ and 80. The region limited by the
dotted-dashed line is defined by $m_{\tilde{t}_1}\;<\;m_{{\tilde
\chi}^+_1}$.  The analysis rules out $m_0$ and $m_{1/2}$ points in the
dark hashed region. In the lower hashed region no points with
radiative electroweak symmetry breaking can be found. We have taken
$\tan \beta= 3$, $A_0=-650$ GeV and $\epsilon_3 / \mu = -0.5$ and
verified that in this region $BR(\tilde{t}_1 \to b \tau)=1$.  The
Tevatron limits can not be directly applied when $A_0>-500$ GeV,
because in this case $m_{\tilde{t}_1}\;>\;m_{{\tilde \chi}^+_1}+m_b$.
The regions in the $m_0$--$m_{1/2}$ plane where 
$m_{\tilde{t}_1}\;<\;m_{{\tilde \chi}^+}$ are excluded if $-650 < A_0
< -500$ GeV and $|\epsilon_3 / \mu|$ is sufficiently large so that the
three-body decays are negligible.
Fig.~\ref{fig:valle2}a shows that the region where
$m_{\tilde{t}_1}\;<\;m_{{\tilde \chi}^+_1}+m_b$ and $BR({\tilde t}_1
\to \tau^+ + b) \approx 1$ is practically ruled out by experiment.
For this particular choice  of SUSY parameters there is only
a little window still to explore at the Run II of the Tevatron. However for
other choices of SUSY parameters, e.g. $A_0=-900$ GeV the dark-hatched
region fills up only about half of the allowed region where
$m_{\tilde{t}_1}\;<\;m_{{\tilde \chi}^+_1}+m_b$ and would therefore
be open for investigation at the next run.

The MSSM three-body channels could be competitive with the BRPV one if
$|\epsilon_3 / \mu|$ is very small and $m_0 \ll m_{1/2}$. In this case
the condition $BR({\tilde t}_1 \to \tau^+ + b) \approx 1$ no longer
holds and our analysis is not applicable.  If $|\epsilon_3 /
\mu|<10^{-3}$ (which leads to tau neutrino mass in the 10$^{-2}$eV
range in the mSUGRA model) for the same value of $\tan\beta$ we find 
\cite{staumad} that the decay mode into $c\, {\tilde \chi}^0_1$ is 
competitive with the $b \tau$ channel.
The $\tilde{t}_1 \to c  {\tilde \chi}^0_1$ channel becomes more
important for large $\tan \beta$ and $m_{\tilde{t}_1}\;<\;m_{{\tilde
\chi}^+_1}$. In this case one needs $|\epsilon_3 / \mu|>10^{-2}$ in 
order to get a negligible $BR(\tilde{t}_1 \to c {\tilde \chi}^0_1 )$.

\vspace*{-0.5cm}
\nn
\begin{figure}[htbp]
\centerline{
\psfig{figure=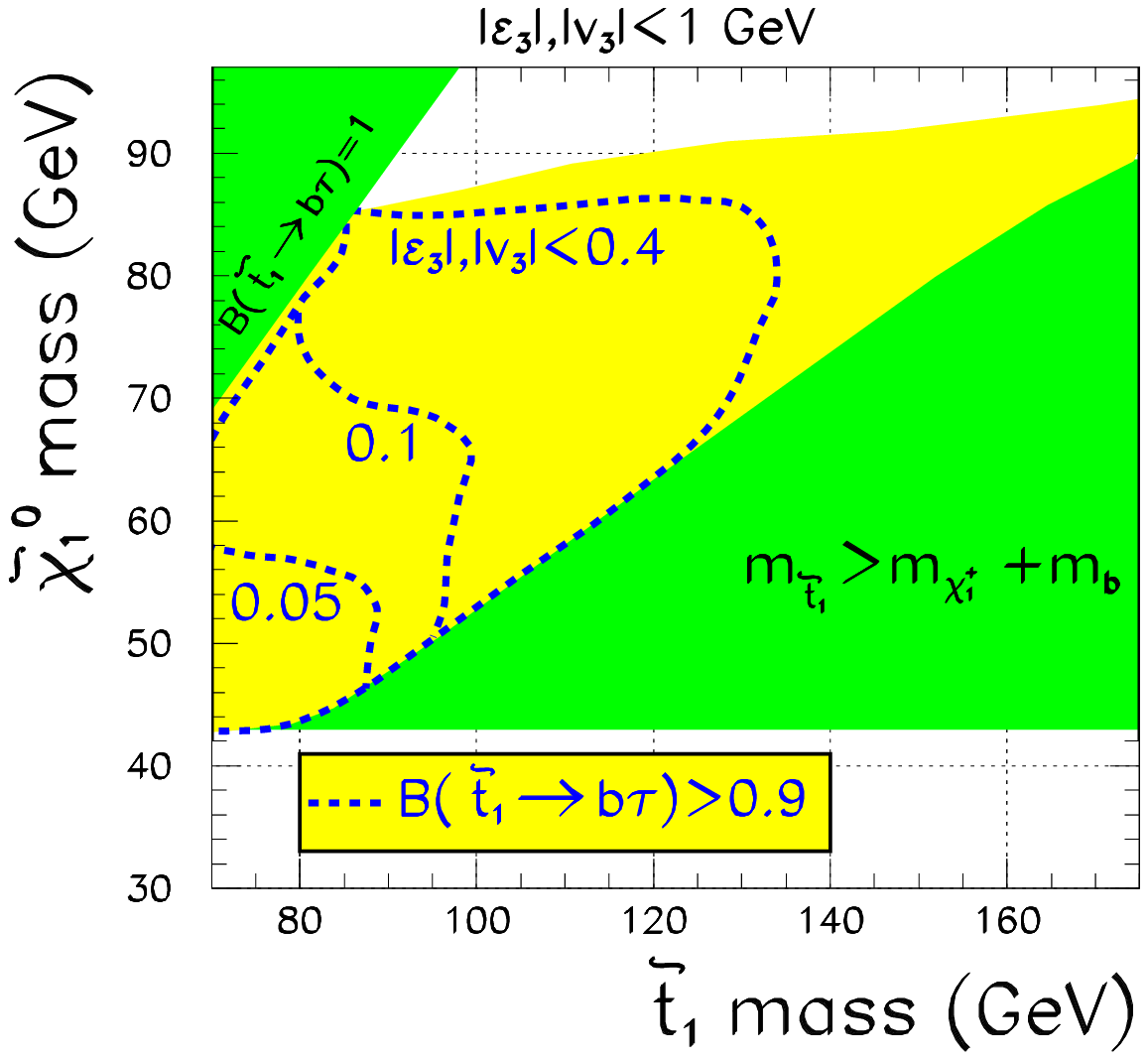,height=6cm,width=8cm,angle=0}
\hspace*{-5mm}
\psfig{figure=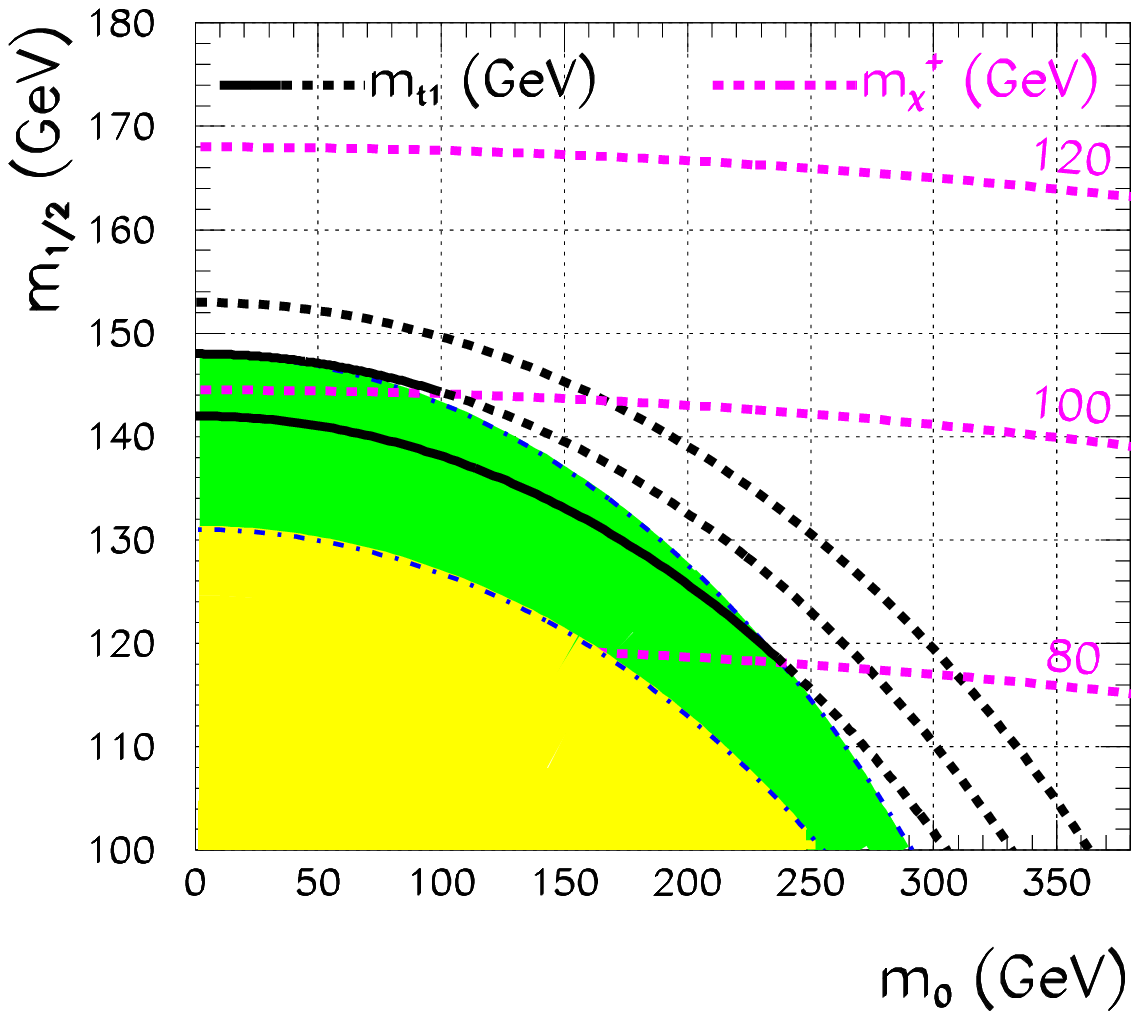,height=6cm,width=8cm,angle=0}
}
\vspace*{0.6cm}
\caption{(a) Contour-lines for BR$(\tilde t_1 \to b \tau)>0.9$ in the 
$m_{{\tilde t}_1}$--$m_{{\tilde\chi}_1^0}$
plane. The gray region shows the area where only those two decay modes
are open. We consider $|\epsilon_3|,|v_3|<1$ GeV, and the MSSM
parameters are varied randomly such that $m_{\tilde t_1}<m_{{\tilde
\chi}^{\pm}_1}+m_b$.  The lines are defined by the maximum value of
$|\epsilon_3|$ and $|v_3|$ and delimit the regions where $BR(\tilde
t_1 \to b \tau)>0.9$. (b) Exclusion contour in the
$m_0$--$m_{1/2}$ plane. The nearly horizontal dashed lines are
chargino mass contours while the radial-like dashed lines are the
top-squark mass contours. These change from solid to dashed when the
top--squark becomes heavier than the lightest chargino.  The radial
curves correspond to $m_{\tilde{t}_1}=120, 100$ and 80, respectively,
from upper to the lower. The region limited by the dotted-dashed line
has $m_{\tilde{t}_1}\;<\;m_{{\tilde \chi}^+}$.  The dark hashed region
is excluded by experimental data while the lower light-hashed region
is disfavoured by theory. We have fixed $\tan \beta= 3$, $A_0=-650$
GeV and $\epsilon_3 / \mu = -0.5$.}
\label{fig:valle2}
\end{figure}
\vspace*{-.2mm}

\subsection{Summary}

We have studied top-quark and top-squark decays in a supersymmetric
model with bilinear R-parity breaking. We have found that in both
cases there exist additional top and stop decay modes leading to novel
phenomenological implications with respect to those of the MSSM. In
the top-quark case the new decay modes are $t \to {\tilde
\tau}^+_1 \, b$, $t \to \nu_\tau \, {\tilde t}_1$, and $t \to \tau^+
{\tilde b}_1$.  We have shown that existing data on non-W top decay
from the Tevatron are already sensitive to the BRPV parameters, adding
both sbottom and stau decay channels. 

In this model the top-squark has the additional channel ${\tilde t}_1
\to \tau^+ + b$. This channel will be 100\% if the stop is the lightest 
SUSY particle, which is possible in the BRPV model. Moreover, we have
demonstrated that this decay can be dominant even when the lightest
neutralino below the stop and the R-parity breaking parameters
$|\epsilon_3|$ and $|v_3|$ are well below a GeV, as long as the
R-parity conserving chargino decay mode is kinematically closed,
i.e. for $m_{\tilde{t}_1}\;<\;m_{{\tilde \chi}^+_1}$. We have studied
scenarios in a SUGRA model with universality of the soft breaking
terms at the unification scale and we have found that the Tevatron
data on third generation lepto-quark data rule out scalar-top masses
below 80-100~GeV, depending on the parameters. 

\medskip

{\bf Acknowledgments:}
{\small This 
work was supported by DGICYT under grant PB95-1077 and Acci\'on
Integrada Hispano-Austriaca HU1997-0046, by the TMR network grant
ERBFMRX-CT96-0090 of the European Union, by CNPq and FAPESP (Brazil),
by Programa de Apoio a N\'ucleos de Excel\^encia (PRONEX), by a
CSIC-CNPq exchange agreement. D.~A.~R.~was supported by Colombian
{COLCIENCIAS fellowship.  W.~P.~was supported by the ''Fonds zu}
F\"orderung der wissenschaftlichen Forschung'' of Austria, project
No. P13139-PHY. L.~N. was supported by spanish CSIC fellowship.}

\vspace{3cm}

{\bf Note Added in Proof:} After this report was submitted a further
slepton production process via R-parity violation was discussed in the
literature \cite{bkp}.

% ---- Bibliography ----

\end{document}